\documentclass[useAMS,usenatbib,fleqn]{mnras}
\usepackage{graphicx,color,amsmath,amssymb,amsfonts}
\usepackage{subfig}
\usepackage[toc,page]{appendix}
\usepackage{hyperref} 
\usepackage{etoolbox}
\usepackage{natbib}
\usepackage{rotating}
\usepackage{float,lscape}
\usepackage[normalem]{ulem}

\newcommand{\be}{\begin{equation}}
\newcommand{\ee}{\end{equation}}
\newcommand{\bea}{\begin{eqnarray}}
\newcommand{\eea}{\end{eqnarray}}

\newcommand{\F}[1]{{Fig. \ref{#1}}}
\newcommand{\T}[1]{{Table \ref{#1}}}
\newcommand{\Eq}[1]{{Eq. \ref{#1}}}
\newcommand{\tb}[1]{{\boldsymbol{#1}}}
\newcommand{\nest}{\textsc{MultiNest}}
\newcommand{\me}{\text{e}}

\title[SFRD below the radio detection threshold]{A deep radio view of the evolution of the cosmic star-formation rate density from a stellar-mass selected sample in VLA-COSMOS}

\author[Malefahlo ~et~al.]{Eliab D. Malefahlo\thanks{eliabmalefahlo3@gmail.com}$^1$, Matt J.~Jarvis$^{2,1}$,  Mario G. Santos$^{1,3}$,  Sarah V. White$^{4}$, \and Nathan J. Adams$^{2}$, Rebecca A.A. Bowler$^{2}$\\\\
$^{1}$Department of Physics \& Astronomy, University of the Western
Cape, Private Bag X17, Bellville, Cape Town, 7535, South Africa\\
$^{2}$Astrophysics, University of Oxford, Keble Road, Oxford, OX1 3RH, UK\\ 
$^{3}$South African Radio Astronomy Observatory (SARAO), 2 Fir Street, Observatory, Cape Town, 7925, South Africa\\
$^{4}$Department of Physics and Electronics, Rhodes University, PO Box 94, Makhanda, 6140, South Africa\\
}

\begin{document}



\maketitle

\label{firstpage}
\begin{abstract}
We present the 1.4~GHz radio luminosity functions (RLFs) of galaxies in the COSMOS field, measured above and below the $5\sigma$ detection threshold, using a Bayesian model-fitting technique. The radio flux-densities from VLA-COSMOS 3-GHz data, are extracted at the position of stellar mass-limited near-infrared (NIR) galaxies. We fit a local RLF model, which is a combination of active galactic nuclei (AGN) and star-forming galaxy (SFG), in 10 redshift bins with a pure luminosity evolution (PLE) model. We show that the evolution strength is similar to literature values up to $z\sim 1.6$. Beyond $z\sim 2$,  we find that the SFG RLF exhibits a negative evolution ($L^*$ moves to lower luminosities) due to the decrease in low stellar-mass sources in our stellar mass-limited sample at high redshifts. From the RLF for SFGs, we determine the evolution in the cosmic star-formation-rate density (SFRD), which we find to be consistent with the established behaviour up to $z\sim 1$. Beyond $z\sim 1$ cosmic SFRD declines if one assumes an evolving infrared--radio correlation (IRRC), whereas it stays relatively higher if one adopts a constant IRRC. We find that the form of the relation between radio luminosity and SFR is therefore crucial in measuring the cosmic SFRD from radio data. We investigate the effects of stellar mass on the total RLF by splitting our sample into low ($10^{8.5} \leq M/\mathrm{M}_{\odot} \leq 10^{10}$) and high ($M>10^{10}\,\mathrm{M}_{\odot}$) stellar-mass subsets. We find that the SFRD is dominated by sources in the high stellar masses bin, at all redshifts. 
\end{abstract}


\section{Introduction} \label{sec:intro}

Understanding the evolution of star formation (SF) in galaxies over the history of the Universe is a key aspect of galaxy-formation studies. It has the potential to tell us how, when and where, star formation happened from the onset of the first galaxies within the epoch of reionisation, through to the present day. Measuring the star formation rate (SFR) in galaxies can be done at a variety of wavelengths \citep[see e.g.][]{Kennicutt-1998}. The most sensitive tracer of young massive stars within star-forming regions of galaxies comes from rest-frame ultraviolet (UV) observations, where the depth that can be reached with current telescopes means that very low star-formation rates can potentially be reached to the highest redshifts \cite[e.g][]{McLure-2013,Bouwens-2015A, Adams-2020, Bowler-2020}.

However, the rest-frame ultraviolet is readily absorbed by dust, both within and along the line of sight to distant galaxies, resulting in the SFR measurements made at these wavelengths being lower limits. The dust that absorbs this ultraviolet radiation is heated and re-radiates the energy at far-infrared wavelengths with a spectrum close to a blackbody. The combination of UV, through to the far-infrared emission can therefore provide measurements of the total SFR in galaxies, both unobscured and obscured \citep[e.g.][]{Burgarella-2005,dacunha-2008,Berta-2013,Smith-2018}. 

Unfortunately, far-infrared observations are generally limited in their spatial resolution. For example, the {\em Herschel Space Observatory} has a resolution of 18\,arcsec at 250\,$\mu$m, leading to imaging surveys that are generally limited by source confusion \citep[e.g.][]{Oliver-2012}. ALMA can detect this dust emission at much higher angular resolution, but these surveys are limited in area \citep[e.g.][]{Dunlop-2017, dudzevi-2020, Yamaguchi-2020, Franco-2020, Gruppioni-2020}, or rely on pointed observations of pre-selected samples \citep[e.g.][]{Zavala-2019,boogaard-2019,Simpson-2020}. Thus, it is unsurprising that over the past few years, alternative tracers of star-formation rates of galaxies have been considered at other wavelengths \citep[e.g.][]{Ouchi-2010, Drake-2013,Schober-2015,Aird-2017}. Possibly the most promising one is using deep radio continuum observations at GHz frequencies.

 The radio SFR estimate relies on the Far-infrared (FIR) SFR through the Far-infrared--radio correlation (FIRC). This is a tight correlation between the radio luminosity and the total infrared luminosity of galaxies \citep[e.g.][]{vanderKruit1971,deJong-1985,Condon-1991, Jarvis2010, Delhaize-2017}. The correlation spans over five orders of magnitude and its existence has been attributed to young massive stars. After their short life-span, of a few Myr, the massive stars reach a catastrophic end in a supernova explosion, which accelerates electrons that then emit synchrotron radiation observed in the radio. During their short lifetimes these same massive stars emit optical and ultraviolet radiation that is then absorbed and re-radiated into the IR by surrounding dust. Thus resulting in a correlation between the radio synchrotron emission and the dust continuum emission. In recent years, it has become apparent that the form of the FIRC may also depend on other properties of the galaxy \citep[e.g.][]{Read2018,Molnar2018,Delvecchio-2020, Smith2020}. These dependencies could be due to excess radio emission due to AGN activity, or that the far-infrared emission is not fully accounting for the total star formation rate in some galaxies. Indeed, using a total star-formation rate from full spectral energy distribution modelling or by combining UV and far-infrared emission may alleviate some of these concerns, or possibly complicate them further for certain types of galaxy \citep[e.g.][]{Gurkan2018}.

However, in all these studies, the most reliable estimates for the star-formation rate based on radio emission tend to be for those where the contribution from a central accreting black hole is thought to be negligible. One way to do this, is to select galaxies based on their optical properties, rather than use a radio selected sample. This mitigates against the inevitably bias for ``normal" galaxies with low-level AGN-related radio emission to be boosted above the flux-density limit of the radio survey. Whereas, those galaxies with the same SFR but no AGN-related emission fall below the same flux limit.

Therefore, in this work we measure the radio-luminosity function (RLF) of near-infrared selected galaxies below the nominal flux-limit by applying the technique developed by \citet{Zwart-2015}, extended upon in \citet{Malefahlo-2020}, and used in a similar way to measure the H{\sc i} mass function \citep{Pan-2020}. 

We use a set of (SFG and AGN) models for the RLF and fit directly to the radio data using a full Bayesian approach. In Section~\ref{sec:data} we describe the radio, optical and near-infrared survey data we use, along with the photometric redshifts and the derived stellar masses. In Section~\ref{sec:method} we provide a description of the Bayesian methodology we use to model the RLF below the nominal detection threshold as a function of redshift. In Section~\ref{sec:results} we present the results of the various RLF model forms and in Section~\ref{sec:CSFRD} we use the most appropriate RLF models to calculate the evolution of the cosmic star formation rate density, and compare this with other studies in the literature. Section~\ref{sec:conc} summarises our conclusions.

Throughout the paper we use the following $\Lambda$CDM cosmology, with $H_0 = 70$ km$^{-1}$ Mpc$^{-1}$, $\Omega_\Lambda = 0.7$ and $\Omega_{\rm M} = 0.3$. All quoted optical and near-infrared magnitudes are in the AB system \citep{Oke-1983}. We assume a  spectral index, defined as $\alpha \equiv \log(S/S_0)/ \log(\nu/\nu_0)$, with $\alpha=-0.7$, when converting flux density to luminosity and one reference frequency to another. 

\section{Data}\label{sec:data}
\subsection{Near-infrared data}
In order to select the galaxies for this study, we use the near-infrared (NIR) imaging in $Y, J, H$ and $K_s$ bands taken with the VIRCAM \citep{EmersonEmerson-Sutherland-2010} as part of the ultra deep survey on the VISTA telescope, UltraVISTA \citep{McCracken-2012} and the deep optical data from Canada-France-Hawaii-Telescope Legacy Survey (CFHTLS). The HyperSuprimeCam Strategic Survey Programme \citep[HSC; ][]{Aihara-2018a,Aihara-2018b} over the Cosmic Evolution Survey (COSMOS; \citealt{Scoville-2007}) field. {Additionally, we use mid-infrared data from {\it Spitzer}/Infrared
Array Camera (IRAC, \citealt{Sanders-2007,Steinhardt-2014,Ashby-2018}).} In the fourth data release (DR4) the survey covers a total area of $\sim 1.9$\,deg$^2$ which is reduced to an effective area of $\sim 1.8$\,deg$^2$ when masked regions (saturated by stars, regions of high noise) are excluded \footnote{http://ultravista.org/release4/dr4\_release.pdf}. The overlapping effective area between DR4, IRAC and CFHTLS or HSC is $\sim 1.45$ deg$^2$. The flux densities were extracted from a $2''$-diameter aperture in each band using the $K_{s}$-band as the detection image (a rough proxy for stellar mass over the redshift range we are interested), and extracting the flux at these positions across the other near-infrared and optical data \citep[following ][]{Bowler-2020, Adams-2020}. The catalogue has a minimum $5\sigma$ detection threshold of $K_{s} = 24.5$.

\subsubsection{Photometric redshifts and Stellar Masses}
The photometric redshifts are the same as those used by Adams et al 2020b (in prep) and are measured by fitting the multi-band data available in the COSMOS field to a synthetic library of galaxy templates using {\sc LePHARE} \citep{Arnouts-2002,Illbert-2009}. In summary they follow \citet{Ilbert-2013} using several synthetic galaxy multi-band templates from \cite{Bruzual_Charlot-2003} and \cite{Polletta-2007} [generated using the Stellar Population Synthesis model of \cite{Bruzual_Charlot-2003} assuming a \citet{Chabrier-2003} initial mass function (IMF)] to compare with the observed photometry. 
 
Comparing the photometric redshifts to the spectroscopic redshifts available in the literature \citep{Lilly-2009,Coil-2011,Cool-2013,LeFevre-2013,Alam-2015,Hasinger-2018}, Adams et al 2020b (in prep) reports an outlier rate of 828/19 752 (4.2 per cent) and a normalised mean absolute deviation (NMAD) of 0.0312.

The stellar masses are determined again by using {\sc LePHARE}  to fit the multi-band data with templates but with the redshift fixed at the best-fit photometric redshift.  $\chi^2$ minimisation is  used to find the best fit template from the \cite{Bruzual_Charlot-2003} models on the total flux measurements in each filter.

\subsubsection{Sample}
Our goal in this paper is to measure the evolution of the RLF and thus the cosmic star-formation rate density (SFRD) using SFGs, which means removing contamination from stars and emission from active galactic nuclei (AGN). Sources are classified as a star if: (1) The best-fit star template has higher probability than the best-fit galaxy template and, (2) The source does not meet the $BzK$ colour--colour selection criteria from (\citealt{Daddi-2004}, which combines the B and $z$ optical bands with NIR K band to identify stars). Passive galaxies (predominantly AGN) are traditionally identified using colour--colour plots with ($U - V$) vs ($V - J$), usually referred to as $UVJ$ \citep[e.g.][]{Wuyts-2007,Williams-2009}. However, several studies have shown that $\sim 10 - 20$ per cent of passive sources have significant SF in the host galaxy \citep[e.g.][]{Belli-2017,Merlin-2018,Leja-2019}.

Therefore,  we choose not to separate the galaxies in our sample into quiescent and SF, and instead aim to detect radio emission from star formation for all galaxies that lie above our flux/mass limit. 

\subsubsection{Completeness \label{sec:complete}}
\begin{figure}
    \centering
     \includegraphics[width=0.5\textwidth]{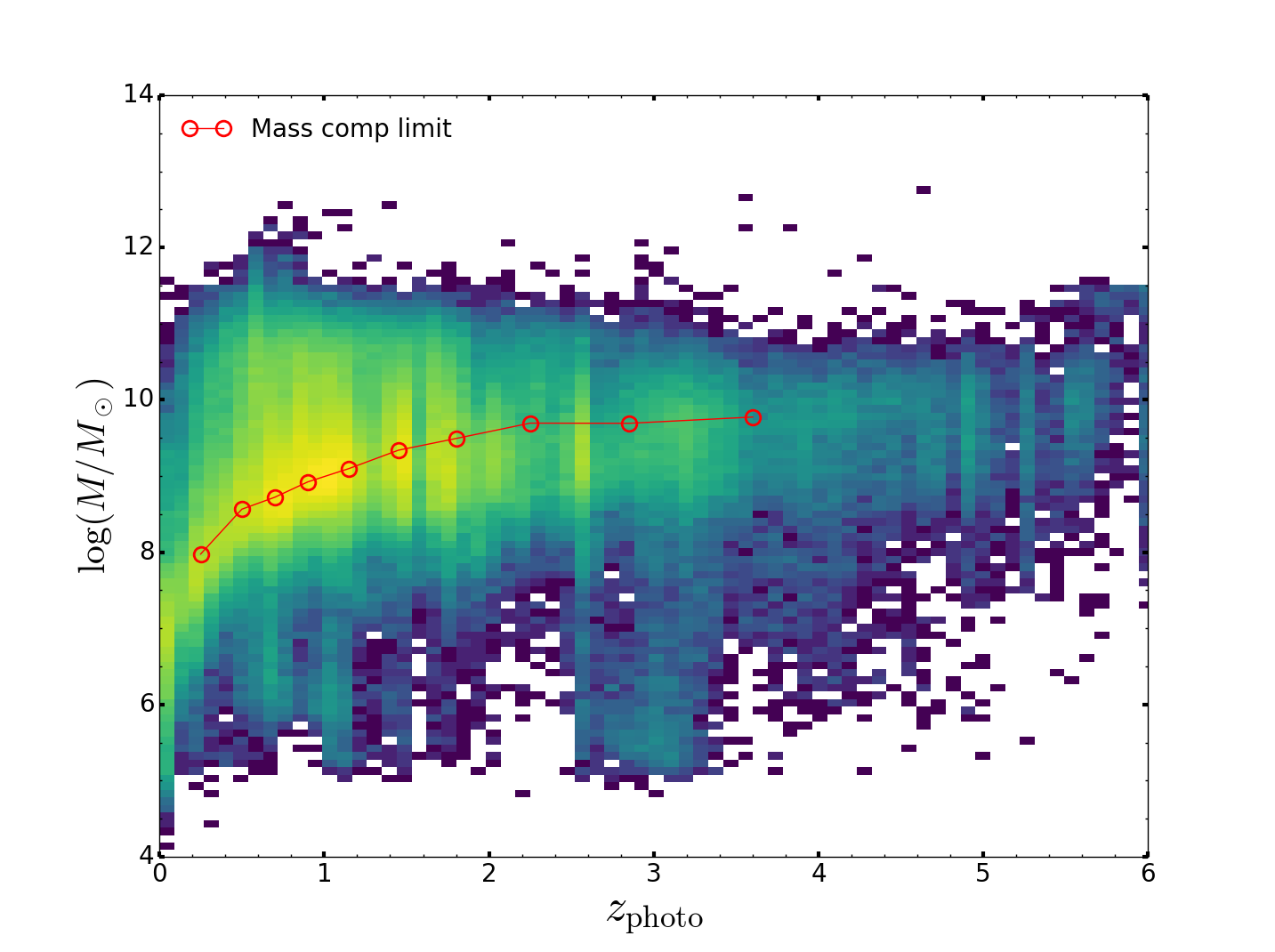}
    \caption{The stellar mass of the galaxies in the UltraVISTA DR4 sample as a function of photometric redshifts. The red circles connected by lines represents the stellar mass completeness limit.}
    \label{fig:stellar_mass}
\end{figure}

In a magnitude-limited survey, the stellar mass completeness is a function of the mass-to-light ratio (which depends on a galaxy template) and redshift. In light of this we divide the data into ten redshift bins (from $z=0.1$ to $z=4$, ensuring that the redshift bins are large enough to not be compromised by photometric-redshift uncertainties [i.e. the photometric redshift uncertainty $\ll$ redshift bin width])  and estimate a conservative stellar-mass completeness limit ($M_{\rm{lim}}$) at each redshift. To estimate this stellar-mass completeness limit we follow \cite{Ilbert-2013}. We start by computing the stellar-mass limit ($M_{\rm{min}}$) for each galaxy, and the limit is the stellar mass that a galaxy at a certain redshift with stellar mass ($M$) would have if it was observed at the $5\sigma$ flux limit ($K_s=24.5$),
\begin{equation}
\log(M_{\rm{min}} ) = \log (M) + 0.4(K_S - 24.5).
\end{equation}
The completeness limit is then given by the stellar-mass that is above 90 per-cent of the stellar-mass limits in the redshift bin Table~\ref{table:stellar_mass}, Fig~\ref{fig:stellar_mass}. This stellar mass completeness limit takes into account the different galaxy templates (and their corresponding mass-to-light ratio) which then ensures that not more than 10 per cent of the low-mass galaxies are missing in our sample.
Applying the stellar-mass cut results in a total galaxy sample size of 171 621, over the redshift range $0.1<z<4$.

\begin{table*}
\caption{This table shows the redshift bins along with the median redshift of the data in each bin. $N_{\rm{Tot}}$ is the total number of galaxies in each bin. We show the stellar-mass completeness limit that contains 90 per cent of the galaxies stellar mass completeness. We also present the number of galaxies with stellar mass above the stellar-mass completeness limit ($N$, our sample) and the number of sources in our sample that have VLA-COSMOS 3-GHz counterparts 
($N_{\rm{VLA}}$).}
\label{table:stellar_mass}

\begin{tabular}{cccccc} \hline
Redshift bin &$z_{\rm{Med}}$ & $N_{\rm{Tot}}$  & $\log(M_{\rm{lim}}/M_\odot)$ & $N$  &$N_{\rm{VLA}}$\\ \hline

$ 0.1 < z < 0.4 $& 0.32 &$ 30127 $& 8.0 & 17816 & 504  \\
$ 0.4 < z < 0.6 $& 0.53 &$ 25927 $& 8.6 & 17412 & 587  \\
$ 0.6 < z < 0.8 $& 0.7 &$ 36373 $& 8.7 & 23775 & 690  \\
$ 0.8 < z < 1.0 $& 0.9 &$ 39404 $& 8.9 & 26498 & 778  \\
$ 1.0 < z < 1.3 $& 1.12 &$ 46335 $& 9.1 & 26586 & 896  \\
$ 1.3 < z < 1.6 $& 1.45 &$ 31832 $& 9.3 & 15845 & 645  \\
$ 1.6 < z < 2.0 $& 1.74 &$ 36235 $& 9.5 & 14021 & 838  \\
$ 2.0 < z < 2.5 $& 2.24 &$ 15995 $& 9.7 & 5319 & 364  \\
$ 2.5 < z < 3.2 $& 2.82 &$ 22144 $& 9.7 & 6332 & 286  \\
$ 3.2 < z < 4.0 $& 3.42 &$ 9845 $& 9.8 & 2805 & 49  \\

\hline 
\end{tabular} 

\end{table*} 

\subsection{Radio data}
We use radio data from the VLA-COSMOS 3-GHz survey \citep{Smolcic-2017}. The survey covers 2.6\,deg$^2$ with a resolution of $0.75''$ and a rms sensitivity with a median value of $2.3~\mu$Jy. 10,830 detected sources were extracted in the central 2\,deg$^2$ using {\sc BLOBCAT} (\citealt{Hales-2012}) with 67 found to be multi-component. The multi-component sources are visually confirmed and most of them are galaxies with resolved structures such as jet/lobe/core. A small portion of the multi-component sources are SFGs with disk-like structures (\citealt{Smolcic-2017}).

\subsection{Flux-density extraction}
For sources that lie significantly above the noise limit of the data, flux densities are usually extracted by running a source finder that identifies a source lying significantly above the noise and then aims to quantify the integrated flux-density of the sources, either using the fitting of multiple Gaussians \citep[e.g. {\sc PyBDSF}; ][]{PyBDSF} or by flood filling to a certain level above the background noise [e.g. {\sc Blobcat} \citep{Hales-2012} and {\sc ProFound} \citep{Robotham-2018,Hale-2019}].

The challenge here is that most of the NIR sources do not have a radio counterpart above the detection threshold. The simplest approach would be to use aperture centred at the NIR position and measure total flux density by summing the individual flux densities per pixel, accounting for the beam area. 

The size of the aperture plays an important role because if it is too big compared to the projected  size of the galaxy then there will be increased contribution from noise, and there is a greater probability that the measured flux will also include a contribution from nearby objects. If the size is too small then the flux density of the galaxy might be underestimated. The extraction ``stamp" should therefore be as close as possible to the expected size of the galaxies. In this paper we use a square with size of $7\times 7$ pixels ($1.4 \times 1.4$~arcsec), which is large enough to contain the average size of galaxies, based on several studies on radio-continuum sizes of $\mu$Jy galaxies \cite[e.g.][]{Murphy-2017,Guidetti-2017,  Bondi-2018,Cotton-2018,Andrade-2019}, and small enough to avoid contamination from background sources.  We note that as the flux density in the images is per beam area, we note that unresolved (or marginally resolved) galaxies will have all of their flux accounted for using this box size.

\begin{figure*}
\centering
\includegraphics[width=0.9\textwidth]{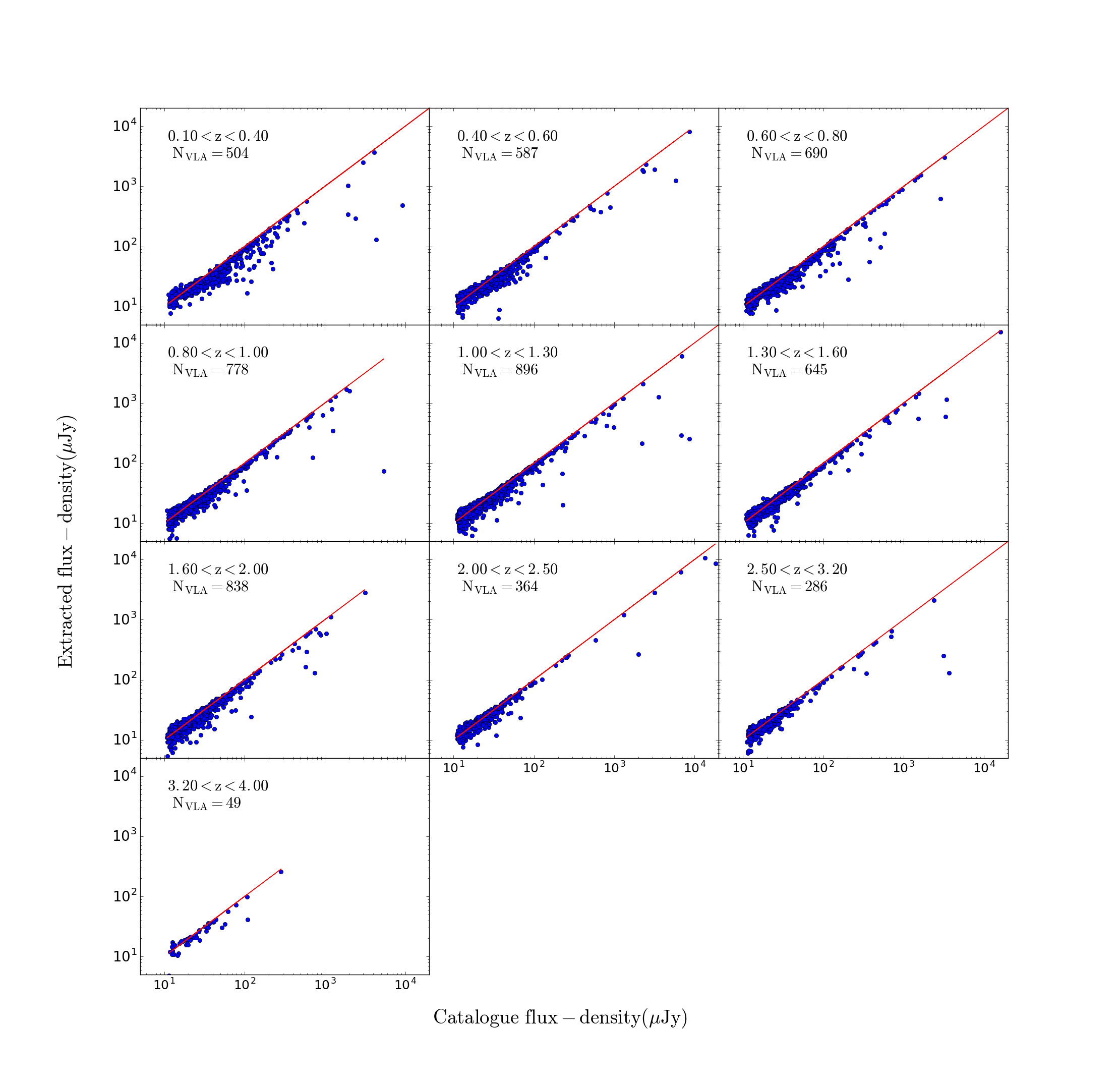}
\caption{Comparison between the VLA-COSMOS 3-GHz flux densities extracted around the NIR position and VLA-COSMOS 3-GHz-COSMOS2015 matched flux densities from
 \protect\cite{Smolcic-2017b}, represented by the blue points.
  We matched the UltraVISTA sources with \protect\cite{Smolcic-2017b} sources using their COSMOS2015 ID.}
\label{fig:smolcic}
\end{figure*}

In Fig~\ref{fig:smolcic} we compare our extracted flux densities at the positions of our stellar-mass selected galaxy sample, with the detected VLA-COSMOS 3-GHz \cite{Smolcic-2017} sources. 
Our measured flux densities scatter uniformly (in log-scale) around the 1-to-1 line at faint flux densities (i.e. $S_{\rm{3~GHz}} < 30~\mu$Jy). Above $S_{\rm{3~GHz}} \sim 40~\mu$Jy our extracted flux densities underestimate the VLA-COSMOS 3-GHz flux by $\sim 6$ per cent. The underestimation is due to resolved sources larger than our aperture (predominantly large low-redshift star-forming galaxies and more distant extended AGN). 

\begin{figure*}
\centering
\includegraphics[width=0.9\textwidth]{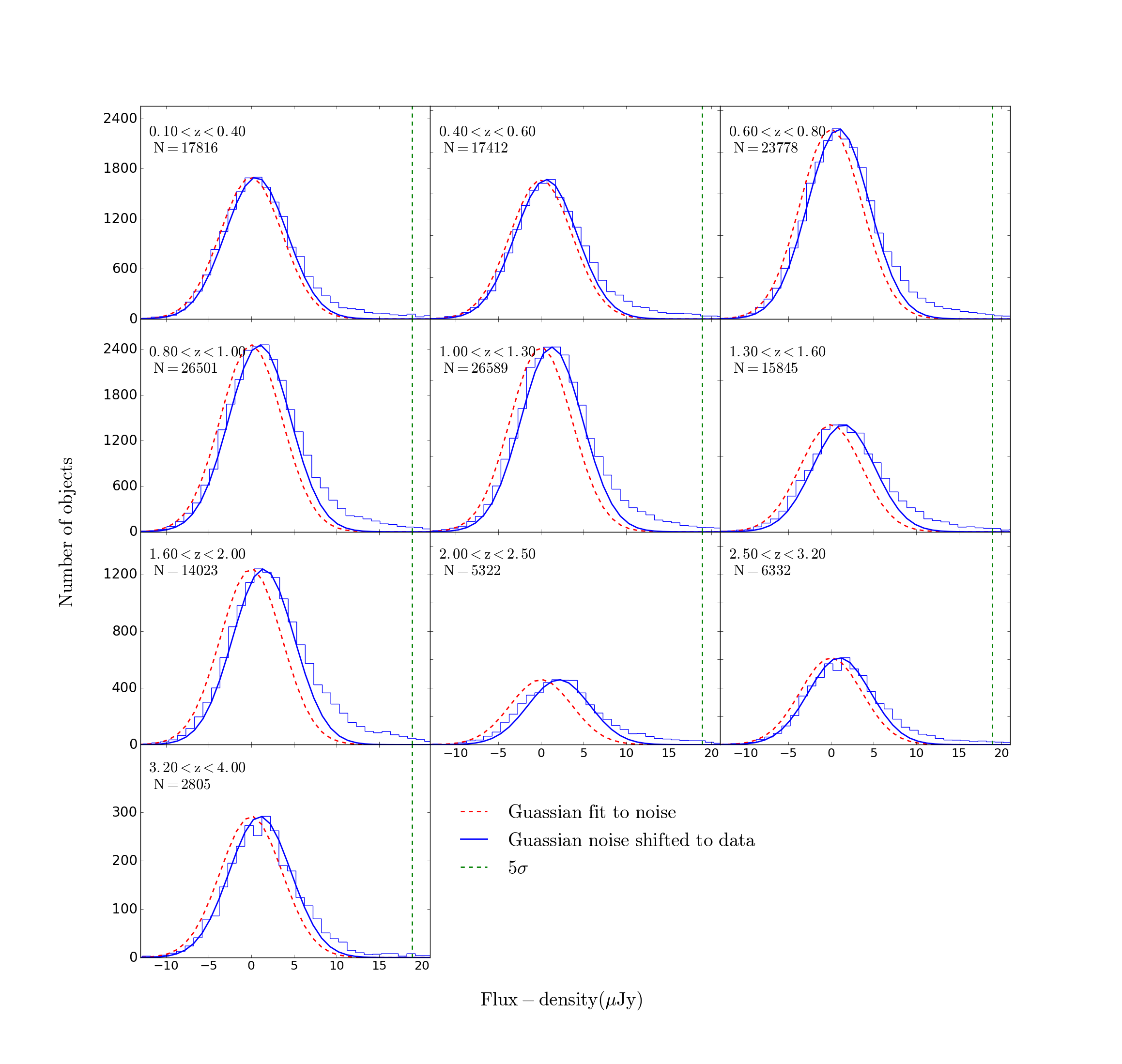}
\caption{The histograms of the VLA-COSMOS 3~GHz integrated flux-density extracted from boxes ($7\times 7$ pixels) centred at the NIR positions. The red dashed curve centred at zero is a Gaussian fit to the flux-densities extracted from boxes centred $50''$ from the NIR positions in each redshift bin. The Gaussians have mean $\sigma = 3.67 \pm 0.025 \mu$Jy over all the redshift bins. The blue line represents a shift in the red dashed line to fit (by-eye) the Gaussian part of the source flux-densities.  The green, vertical dashed-line in each panel represents the $5\sigma = 18.37 \mu$Jy limit of the VLA-COSMOS data. }
\label{fig:fluxes}
\end{figure*}

Since our focus in this paper is on sources below the nominal detection threshold we use the $7\times 7$ pixel box and for sources above $ 0.5$ mJy we use \cite{Smolcic-2017} flux densities. Fig~\ref{fig:fluxes} shows the extracted flux densities for all of the stellar-mass selected sources in each redshift bin. We note that the 3-GHz flux densities follow a Gaussian distribution with an offset from zero and a tail to brighter flux densities, as one would expect.

\section{Bayesian Framework for measuring the RLF}\label{sec:method}
Our ``stacking" analysis is based on a Bayesian formalism that can measure the RLF below the 3-GHz VLA-COSMOS detection threshold, down to sub-$\mu$Jy levels. We make use of a modified version of the software \textsc{bayestack} \footnote{https://github.com/jtlz2/bayestack}
\citep{Zwart-2015}. The idea is to start with a model for the RLF for a given redshift bin. Then translate that into a source-count model and fit to the number of sources per flux-density bin, as extracted from the data as demonstrated by \cite{Malefahlo-2020} for SDSS quasars. Below we review the basics of the method.

\subsection{Bayesian analysis}
The fitting approach uses Bayes' theorem,
\begin{equation}
\centering
\mathcal{P}(\Theta|D,H) =\frac{\mathcal{L}(D|\Theta,H) \Pi(\Theta|H)}{\mathcal{Z}},
\label{eqn:bayes}
\end{equation}
where $\mathcal{P}$ is the posterior distribution of the parameters $\Theta$, given
the data $D$ and model $H$. $\mathcal{L}$ is the likelihood, the probability
distribution of the data given the model, and $\Pi$ is
the prior, the known constraints on the parameters. $\mathcal{Z}$ is the Bayesian evidence, which normalises $\mathcal{P}$ and can be written as an integral of $\mathcal{L}$ and $\Pi$ over the $n$-dimensional parameter space $\Theta$,
\begin{equation} 
\mathcal{Z}  = \int \mathcal{L} \Pi \textrm{d}^n\Theta. 
\label{eqn:Z} 
\end{equation} 
A model has high evidence when a large portion of its prior parameter space is likely (i.e.~large likelihood), and small evidence when a large portion of its parameter space has a small likelihood, irrespective of how peaked the likelihood function is. This therefore automatically encapsulates Occam's razor (e.g.~\citealt{Feroz_Hobson-2008}).

In order to compute this posterior distribution, one needs to sample from it. Sampling has always been one of the most computationally expensive parts of model selection because it involves solving the multidimensional integral in Eq.~\ref{eqn:Z}. Nested sampling (\citealt{Skilling-2004})
was created for its efficiency in calculating the evidence, with an added bonus of producing posterior inferences as a by-product.
\nest\, \citep{Feroz_Hobson-2008,Feroz-2009a,Buchner-2014} is a robust implementation of nested sampling, returning the full posterior distribution from which the uncertainty analysis can be correctly undertaken.

In Bayesian model selection, one compares the evidences of two models, A and B. This is quantified by considering the ratio of their evidences $\mathcal{Z}_A/\mathcal{Z}_B$, or equivalently, the difference of their log-evidence, $\ln(\mathcal{Z}_B) -\ln(\mathcal{Z}_A)$, known as the Bayes factor.  \cite{Jeffreys-1961} introduced a way to interpret how much better Model A is compared to B using the Bayes factor: $\Delta \ln \mathcal{Z} < 1$ is `not significant', $1 < \Delta \ln \mathcal{Z} < 2.5$ is `significant', $2.5 < \Delta \ln \mathcal{Z} < 5$ is `strong', and $\Delta \ln \mathcal{Z} > 5$ is `decisive'. We adopt this scale in our analysis and use it to compare different models for the evolving RLF.
 
\subsection{Likelihood Function}\label{sec:likelihood}

To proceed with our Bayesian analysis we need a likelihood for the data given a model, where the data we have is the extracted flux densities, $S_T$ of individual galaxies in our stellar-mass selected sample. This flux density is a combination of the actual flux density of the galaxy ($S$) and the noise. For this analysis, the noise is assumed to follow a Gaussian distribution, centred at zero with a constant variance $\sigma_n^2$. This assumption is only valid in the central 2\,deg$^2$ of the COSMOS field where the noise of the VLA-COSMOS 3-GHz is relatively homogeneous \citep{Smolcic-2017}. %

The likelihood function that we use requires the binned flux-density distribution, where we can use Poisson statistics. The likelihood of finding $k_{i}$ objects in the $i^{th}$ measured flux-density bin [$S_{T_i}, S_{T_i} + \Delta S_T$] follows a Poisson distribution,
\begin{equation} 
\mathcal{L}_{i}\left(k_i|\pmb{\Theta}\right) = \frac{I_i^{k_i}\me^{-I_i}}{k_i!}, 
\end{equation} 
where $I_i$ is the theoretically-expected number of sources in the $i^{th}$ measured flux-density bin, given by the modified equation taken from \cite{Mitchell_Wynne-2013},
%
\begin{equation} 
\label{eqn:ii} 
I_i= 
\int_{S_{min}}^{S_{max}} 
dS \frac{dN(S)}{dS} 
\int_{S_{T_i}}^{S_{T_i}+\Delta S_{T_i}} 
dS_T
\frac{1}{\sigma_{n}\sqrt{2\pi}} 
\rm{e}^{-\frac{\left(S-S_T\right)^2}{2\sigma_n^2}}.
\end{equation} 
Here $dN/dS$ is the source-count model (number of sources per flux density bin), $\sigma_n$ is the mean noise of the data and $S$ is again the intrinsic flux-density of the source. This approach naturally takes into account sample variance (at the Poisson level) since it does not fix the total number of predicted sources to the observed number (e.g. other regions of the sky could have a different total number). This will have implications for the allowed minimum and maximum flux-density values of our fits, as we will see later. We expect the fits to have a large variance at the low flux-density level (because of the noise) and at the high flux-density level (because of Poisson fluctuations due to the low number of sources). Solving the second integral, \Eq{eqn:ii} becomes 
\begin{equation} 
\label{eqn:iii} 
\begin{aligned}
I_i= & \int_{S_{min}}^{S_{max}} dS \frac{dN(S)}{dS}\\ 
     &\frac{1}{2} \left\{\mathrm{erf}\left(\frac{S-S_{T_i}}{\sigma_n\sqrt{2}}\right) 
- \mathrm{erf}\left(\frac{S-(S_{T_i}+\Delta S_{T_i})}{\sigma_n\sqrt{2}}\right)\right\}. 
\end{aligned}
\end{equation} 
The total likelihood for the $N$ bins is then given by,
\begin{equation} 
\label{eqn:lhood-tot-bins} 
\mathcal{L}\left(\mathbf{k}|\pmb{\Theta}\right) 
    =\prod_{i=1}^{N} \mathcal{L}_{i}\left(k_i|\pmb{\theta}\right). 
\end{equation}
As we aim to fit models that describe the RLF, we need to convert models describing the RLF to the source counts ($dN/dS$), and compare to the binned flux-densities. 

\subsection{Radio Luminosity Function Models \label{sec:models}}

The luminosity per unit frequency (luminosity density) of a radio source, $L_\nu$, can be related to the observed flux density at the same frequency, $S_\nu$, through 
\begin{equation}
L_{\rm 1.4 GHz} = 4 \pi D_L^2( 1+z)^{-\alpha-1} (1.4/3.0)^{\alpha} S_{\rm 3.0 GHz},
\end{equation}
where $D_L$ is the luminosity distance, $\alpha$ is the spectral index of the source (in this work we assume $\alpha=-0.7$, which is typical of SFGs), and $z$ is the redshift of the source.

The RLF, $\rho(L_\nu)$, is the number density of sources per luminosity density bin, e.g. $\rho(L_\nu)=dN/(dL dV)$ (where $dV$ is comoving volume).
Another common definition of the RLF ($\Phi$), which we use here, normalises the radio luminosity function per magnitude (as opposed to using per $\log_{10}{L}$), where $m-m_0=-2.5 \log_{10}(L/L_0)$. The relationship between these two definitions is then
\begin{equation}
\Phi (L_\nu) = \frac{dN}{dV dm} = \frac{dN}{dV dL_\nu}\frac{dL_\nu}{dm} = \ln(10^{0.4}) L_\nu \rho (L_\nu).
\end{equation}
We define parametric models for the RLF consisting of two functions, one for the luminous sources and the other for faint sources (using subscripts 1 and 2 respectively). The RLF at higher luminosities is dominated by AGN and has been shown to follow a double power-law (e.g.~\citealt{Willott-2001, Mauch-Sandler-2007, Prescott-2016}), so we parameterise the luminous part of the RLF as a double power-law for all the models considered here. The shape of the RLF at low luminosities is dominated by SFGs but also contains radio-quiet AGN \citep{Jarvis-2004,White-2015,White-2017}, so for that we consider two models: a double-power-law and a modified Schechter function (log-normal power-law). 

{Model A} has a double power-law for both the high- and low-luminosity sources:
\begin{equation}
\Phi(L)_A = \frac{\Phi_1^*}{(L/L_1^*)^{\alpha_1} + (L/L_1^*)^{\beta_1}} + \frac{\Phi_2^*}{(L/L_2^*)^{\alpha_2} + (L/L_2^*)^{\beta_2}}.
\label{eqn:dpl_dpl}
\end{equation}

{Model B} has a double power-law for the luminous sources and a log-normal power-law for low-luminosity sources (e.g. \citealt{Tammann-1979}):
\begin{equation}
\begin{aligned}
\Phi(L)_B =& \frac{\Phi_1^*}{(L/L_1^*)^{\alpha_1} + (L/L_1*)^{\beta_1}} \\
        &+\Phi_2^* \left( \frac{L}{L_2^*} \right)^{1-{\delta}} \exp \left[-\frac{1}{2\sigma_{LF}^2} \log_{10}^2 \left(1 + \frac{L}{L_2^*}\right) \right].
\end{aligned}
\label{eqn:dpl_lognorm}
\end{equation}

Models A and B are both fit to each individual redshift bin, rather than assuming a fixed shape and adopting an evolution term to fit across redshift bins.

In order to explore a model of fixed functional form that evolves with redshift, and to facilitate comparison with previous work \citep[e.g.][]{McAlpine-2013,Novak-2017,Novak-2018}, we also adopt an additional model, `Model C', which has a total RLF of fixed shape, defined by combining local SFG and AGN RLFs, but allowed to evolve with redshift. We use the local AGN RLF model and parameters from \cite{Mauch-Sandler-2007}, where they constrain both the bright and faint ends of the AGN population.
They fit their RLF with a double power-law (first function of Eq.~\ref{eqn:dpl_lognorm}), with best-fit parameters, $\phi_1^*= 10^{-5.5} \rm{Mpc^{-3} mag^{-1}}$, $L_1^* = 10^{24.59} \rm{W Hz^{-1}}$, $\alpha_1 = 1.27$ and $\beta_1 = 0.49$.

For modelling SFGs, we use the local SFG RLF from \cite{Novak-2017} obtained by fitting a log-normal power-law to combined data from \cite{Condon-2002}, \cite{Best-2005}, and \cite{Mauch-Sandler-2007}, which contains low-resolution and deep high-resolution information to constrain both the faint and bright ends of the SFG RLF. Using an analytical function in the form of a log-normal power-law (second function of Eq.~\ref{eqn:dpl_lognorm}) their best-fit parameters are $\Phi_2^* = 1.42 \times 10^{-3} \rm{Mpc^{-3} mag^{-1}}$ (scaled to our binning), $L_2^* = 1.85 \times 10^{21} \rm{W Hz^{-1}}$, $\delta = 1.22$ and $\sigma_{LF}=0.63$.

The most common ways to quantify evolution in the RLF are through density or luminosity evolution, although we note that the true evolution is probably a mixture of the two \citep[e.g.][]{Yuan-2016}. Density evolution causes a vertical shift in the RLF with redshift, and luminosity evolution causes a horizontal shift with redshift. 
The SFGs and AGN are known to evolve differently, hence we evolve these two populations separately. The combined density- and luminosity-evolution fit is known to have large degeneracies when the knee of the RLF for SFGs is not well constrained, and pure density evolution (PDE) can overestimate sources at low luminosities (e.g \citealt{Novak-2017,Novak-2018}). Therefore, we only consider a pure luminosity evolution (PLE) of the form, 
\begin{equation}
\begin{aligned}
    \Phi(L,z) =&\Phi_0^{SF} \left[\frac{L}{(1+z)^{\alpha_L^{SF} + z\beta_L^{SF}}} \right] \\
               &+ \Phi_0^{AGN} \left[\frac{L}{(1+z)^{\alpha_L^{AGN} + z\beta_L^{AGN}}} \right],
\end{aligned}
\end{equation}
where $\Phi_0^{SF}$ is the local SFG RLF, $\Phi_0^{AGN}$ is the local AGN RLF, and $\alpha_L^{SF,AGN}, \beta_L^{SF,AGN}$ are the evolution parameters.

Finally, we note that each of the model functions will be bounded: $L_{{\rm min}_1} \le L \le L_{{\rm max}_1}$ for the high-luminosity end and $L_{{\rm min}_2} \le L \le L_{{\rm max}_2}$ for the low-luminosity end. The boundaries are allowed to overlap since there might be a contribution from both populations.

The likelihood (\Eq{eqn:iii}) is computed in flux-density space, which means that our RLF models, $\Phi(L)$, have to be converted into source-count models, $dN/dS$:
\begin{equation}
\begin{aligned}
\frac{dN}{dS}   &= \frac{dN}{dL} \frac{dL}{dS}\\
                &= \rho(L)  V_i 4\pi D_L^2(1+z_i)^{-\alpha-1}\\
                &=  \frac{\Phi(L) V_i}{L\ln(10^{0.4})}
                4\pi D_L^2(1+z_i)^{-\alpha-1},\\
\end{aligned}
\end{equation}
where $V_i$ is the volume of the survey for the redshift bin $i$ and $z_i$ is the median of the redshift bin.

\subsection{Priors \label{sec:priors}}
Priors play an important role in Bayesian inference as they define the sampled parameter space. A uniform prior is the simplest form, providing an equal weighting of the parameter space. We assign a uniform prior to the power-law slopes $\alpha_{1,2}$, $\beta_{1,2}$ and $\delta$. The parameter $\sigma_{LF}$ is assigned a Gaussian prior. To avoid degeneracies in the slopes for the double power-law, we also impose $\alpha_{1,2} \ge \beta_{1,2}$. The parameters $L_{1,2}^*$, $L_{{\rm min}_{1,2}}$, $L_{{\rm max}_{1,2}}$ and $\phi_{1,2}^*$ all have uniform priors in log-space. We impose an additional prior on the AGN break ($L_{1}^*$), in that it must never be less than 0.5 dex above the detection threshold. We have this prior because the bright end of our RLF is not always well constrained by our data. Furthermore, the prior is justified because the AGN RLF is well explored in the literature and the break is found at luminosities well above the luminosity corresponding to $5\sigma$ \citep[e.g][]{Smolcic-2017c, Ceraj-2018}.
The priors are summarised in Table ~\ref{table:prior}.
 

\begin{table} 
\centering 
\caption{Assumed priors. $L_{5\sigma}$ is the luminosity corresponding to the $5\sigma_n$ flux-density cut for a given redshift.}
\label{table:prior} 
\begin{tabular}{ll} 
\hline 
Parameter  & Prior   \\ 
\hline 
$\alpha_1,\beta_1,\alpha_2, \beta_2,\delta$           & uniform $\in \left[-5,5\right]$  \\ 
$\sigma_{LF}$                                         & Gaussian$ \sim \left(\mu=0.6,\sigma=0.1\right)$  \\ 
$\log_{10}[L_{\rm{min_{\{1,2\}}}}/(\rm{WHz}^{-1})]$          & uniform $\in \left[18,30\right]$ \\ 
$\log_{10}[L_{\rm{max_{\{1,2\}}}}/(\rm{WHz}^{-1})] $         & uniform $\in \left[18,30\right]$ \\ 
$\log_{10}[\phi_{\{1,2\}}^*/(\rm{Mpc}^{-3} \rm{mag}^{-1})]$ & uniform $\in \left[-12,-2\right]$ \\
$\log_{10}[L_1^*/(\rm{WHz}^{-1})]$                   & uniform $\in \left[\log_{10}(L_{5\sigma})+0.5,30\right]$ \\
$\log_{10}[L_2^*/(\rm{WHz}^{-1})]$                    & uniform $\in \left[18, \log_{10}(L_{5\sigma}) + 1\right]$ \\ 
\hline 
\end{tabular} 
\end{table} 

\begin{figure*}
    \centering
    \includegraphics[width=1\textwidth]{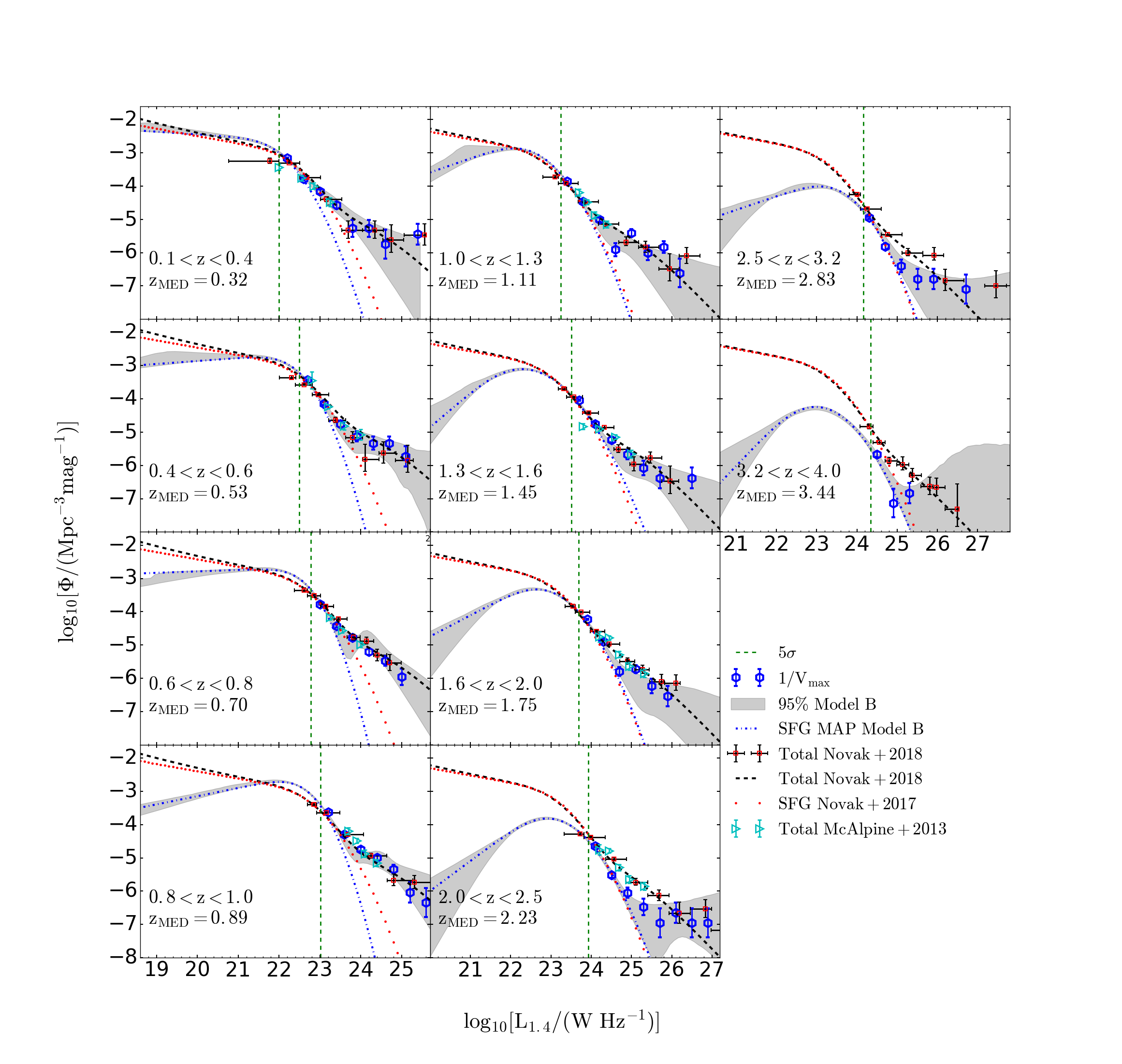}
    \caption{The rest-frame 1.4-GHz RLF of both AGN and SFGs in the COSMOS field. The blue dash-dotted is the SF RLF reconstructed using MAP parameters from the lognormal power-law (Model B) fit to each redshift bin. The grey region represents the 95-per-cent confidence interval of the distribution of reconstructions of models in the posterior. The blue hexagons represent $1/V_{\rm{max}}$ estimations for our detected sources. The red squares are radio-selected RLF data-points from \protect\cite{Novak-2018}, with the curved black, dashed line showing a pure luminosity evolution fit to them. The cyan triangles represent the total RLF from \protect\cite{McAlpine-2013}. The red dots show a PLE fit to SFGs from \protect\cite{Novak-2017}. The vertical, green dashed lines correspond to the detection threshold ($\sigma$) computed using the median redshift for each redshift bin.}
    \label{fig:RLF}
\end{figure*}

\begin{table*}
    \centering
    \begin{tabular}{cccccc}
&$\triangle \log_{10}\mathcal{Z}$ & $\triangle \log_{10}\mathcal{Z}$ & $\triangle \log_{10}\mathcal{Z}$ & $\triangle \log_{10}\mathcal{Z}$ & $\triangle \log_{10}\mathcal{Z}$ \\  
         \hline
Model & $0.10 <z< 0.40$  &      $0.40 <z< 0.60$ &      $0.60 <z< 0.80$ &      $0.80 <z< 1.00$ & $1.00 <z< 1.30$ \\ \hline 
A & $\tb{19.4 \pm 0.26}$ &   $\tb{12.7 \pm 0.27}$  & $\tb{46.3 \pm 0.28}$ & ${27.5 \pm 0.28}$ & ${0.0 \pm 0.00}$ \\
B & ${12.1 \pm 0.25}$ & ${10.6 \pm 0.27}$ & ${41.5 \pm 0.28}$ & $\tb{31.6 \pm 0.28}$ & ${4.7 \pm 0.31}$ \\
C & ${0.0 \pm 0.00}$ & ${0.0 \pm 0.00}$ & ${0.0 \pm 0.00}$ & ${0.0 \pm 0.00}$ & $\tb{8.8 \pm 0.28}$\\
\hline
 & $1.30 <z< 1.60$ & $1.60 <z< 2.00$ & $2.00 <z< 2.50$ & $2.50 <z< 3.20$ & $3.20 <z< 4.00$\\
\hline
A & ${28.5 \pm 0.27}$ & ${87.3 \pm 0.27}$ & ${56.9 \pm 0.27}$ & ${93.6 \pm 0.27}$ & ${12.7 \pm 0.24}$ \\
B & $\tb{36.5 \pm 0.26}$ & $\tb{95.1 \pm 0.26}$ & $\tb{61.1 \pm 0.26}$ & $\tb{95.8 \pm 0.26}$ & $\tb{15.5 \pm 0.23}$ \\
C & ${0.0 \pm 0.00}$ & ${0.0 \pm 0.00}$ & ${0.0 \pm 0.00}$ & ${0.0 \pm 0.00}$ & ${0.0 \pm 0.00}$\\
\hline
    \end{tabular}
    \caption{The relative evidence for the different models (Sec~\ref{sec:models}) in each redshift bin of the NIR-selected radio data. In each redshift bin the reference evidence is from the model with the lowest log-evidence and the winning model is in bold.} 
    \label{table:evidence}
\end{table*}

\section{Results}\label{sec:results}
In this section we provide a binned RLF for the radio-detected sources in our mass-selected sample, based on the $1/V_{\rm{max}}$ statistic, and then present the results of our RLF modelling described in Section~\ref{sec:models}.

\subsection{The binned RLF}\label{sec:binnedRLF}
The RLF for sources with high signal-to-noise (described as `detected') can be calculated directly by converting flux density to luminosity (neglecting noise) and binning the number of sources in luminosity. 

We use the $1/V_{\rm{max}}$ method \citep{Schmidt-1968} given by,
\begin{equation}
\Phi(L_\nu) = \frac{1}{\Delta m}\sum^N_{i=1}\left(\frac{1}{V_{\rm{max}}}\right)_i, 
\end{equation}
with an uncertainty
\begin{equation}
\sigma(\Phi) = \frac{1}{\Delta m}\left[ \sum^N_{i=1}\left(\frac{1}{V_{\rm{max}}}\right)_i^2 \right]^{1/2}, 
\end{equation}
where $V_{\rm{max}}$ is the maximum comoving volume at which the source can be detected given the depth of the data. We assume that the radio sources  have a detection in the NIR data, a photometric redshift, and that the value of $V_{\rm{max}}$ is determined either by the upper limit of the particular redshift bin or by the radio luminosity of the source. 

Fig~\ref{fig:RLF} includes the $1/V_{\rm{max}}$ measurements for our stellar-mass-selected sources above the nominal $5\sigma$ detection threshold. Due to our stellar-mass selection our RLFs are not expected to be exactly the same as the RLF determined using a purely radio-selected sample. 
 
However, we note that the VLA-COSMOS 3-GHz sources all have optical/NIR counterparts up to $z\sim 1.5$, and $\sim 95$ per cent completeness at $z\sim 4$ \citep{Smolcic-2017b}. Given that the main goal of this paper is to measure the RLF for the fainter population of SFGs and how they evolve, this does not affect our results.

Our $1/V_{\rm{max}}$ data-points (dark-blue data-points in Fig~\ref{fig:RLF}) are in good agreement with \cite{McAlpine-2013} and \cite{Novak-2018} measurements for $z < 2$. At $z>2$  our $1/V_{\rm{max}}$ points lie below the volume density found in these studies at the intermediate luminosities between $24 <\log_{10}[L_{1.4}/{\rm W\,Hz}^{-1}] < 26$. This is mainly due to our mass-selection, rather than using the full optical/NIR data and the associated photometric redshifts. However, as we note above, this has little effect on our main results.

\subsection{The free RLF models}\label{sec:freemodel}
We use \textsc{bayestack} to determine the best-fit parameters for the RLF of our mass-selected sample using Models A and B (Section~\ref{sec:models}) in each redshift bin. For each redshift bin we record the Bayesian evidence, posterior distributions for each parameter along with the median, maximum-likelihood and maximum-a-posteriori (MAP) values for each parameter (shown in Table~\ref{table:parameters}). The Bayes factors for each redshift bin are shown in Table~\ref{table:evidence}, where the reference evidence is for the model with the lowest evidence and the model with the highest evidence is in bold text. We find that the data mostly prefers Model~B, the model with a log-normal power-law describing the faint sources (dominated by SFGs) and a double power-law describing the bright-end sources.

In Fig~\ref{fig:RLF} we show the stellar-mass-selected RLF, reconstructed using the MAP parameters from Model B along with the 95 per cent confidence interval. The 95 per cent region is calculated by reconstructing the RLF in a chosen set of luminosity bins, using all the models in the posterior, and determining the 95 per cent limits in each luminosity bin independently. The MAP reconstruction follows the $1/V_{\rm{max}}$ data-points very well and also follows the \cite{Novak-2018} extrapolated evolution fit well, for at least an order of magnitude below the detection threshold, to $z \sim 1.6$. However, the faint-end of the reconstructed RLF underestimates the extrapolated evolution fit from \cite{Novak-2018} at higher redshifts ($z>1.6$). This is due to the \cite{Novak-2018} RLF having a fixed faint-end slope that extends below their detection threshold (and is essentially fixed by the low-redshift data). Instead, we are using a mass-selected sample with the aim of probing this regime, and so in Models A and B, we allow the faint-end slope of the SFG RLF to vary freely.

We also see that the reconstructed RLF, at much lower luminosities (two or more orders of magnitude below the detection threshold, noticeably above $z \sim 0.4$), drops off steeply. This is due to the mass selection, in that we are approaching the point in the RLF where there are not any galaxies at low stellar mass to populate this part of the RLF, due to the relationship between galaxy mass and SFR \citep[][]{Noeske-2007,Daddi-2007,Whitaker-2014, Johnston-2015}. We have checked this by including all NIR-detected sources (rather than using the mass-limited sample) and find that the luminosity where the drop-off occurs moves to lower luminosities, as expected. This shows that it is not a feature of the RLF, but a feature of the parent sample, due to the lack of low-stellar-mass sources in our sample, and this stellar mass limit obviously increases with redshift due to the flux limit of the NIR data. 

\subsection{The fixed RLF model \label{sec:fixed}}
Our main goal is to measure the RLF to low radio luminosities to obtain a measurement of the cosmic SFRD from a stellar-mass-selected sample. Through the \textsc{bayestack} technique we are able to constrain the RLF to luminosities below the nominal $5\sigma$ threshold. However, as shown in Section~\ref{sec:freemodel}, our mass selection causes the free-fitting models to drop off towards lower radio luminosities. This is not an underlying feature of the SFG RLF, and will therefore affect the cosmic SFRD estimation. To address this we follow the work of \cite{McAlpine-2013} and \cite{Novak-2017,Novak-2018} in fixing the shape of the RLF to that of the local RLF (Section~\ref{sec:models}). 

\begin{figure*}
    \centering
    \includegraphics[width=\textwidth]{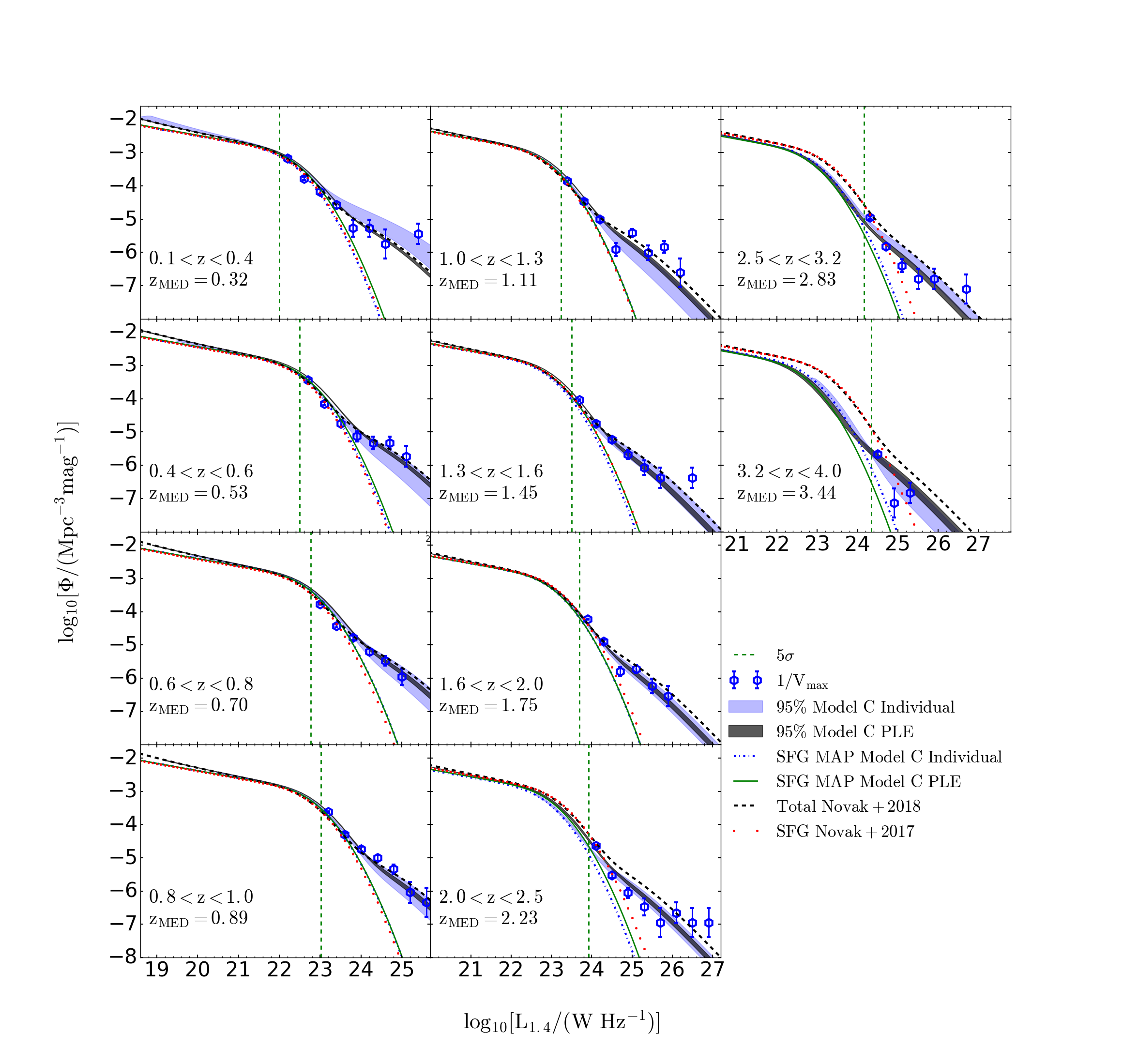}
    \caption{The rest-frame RLF of both AGN and SFGs in the COSMOS field. The blue hexagons represent $1/V_{\rm{max}}$ estimations for the UltraVISTA sources with VLA-COSMOS 3-GHz detected sources. 
    The dark grey and blue regions represents the 95-per-cent confidence interval of the distribution of reconstructions for Model C PLE fit and Model C Individual fit to each redshift bin respectively. The green 
    and blue dashed-dotted lines represent the SFG components of the total RLF of the Model C PLE fit and Model C Individual fit to each redshift bin respectively. The curved, black, dashed line representing the PLE fit to the radio-selected RLF from \protect\cite{Novak-2018}. The red dots are a PLE fit to the SFGs from \protect\citep{Novak-2017}. The vertical, green dashed lines correspond to the detection threshold ($\sigma$) computed using the median redshift for each redshift bin.}
    \label{fig:RLF_v2}
\end{figure*}

We start by modelling the individual redshift bins using the fixed model (Model C), with $\beta_{LF}^{SF,AGN}=0$ (i.e. only allowing a re-normalisation of the RLF in each redshift bin, with a single luminosity evolution term). The resulting RLF is shown in Fig.~\ref{fig:RLF_v2}. \T{table:evidence} shows that Model C is (almost) always the least preferred model (having the lowest log-evidence). This is because Model C {\it forces} a fixed faint-end slope. For our mass-selected sample the fall-off in the number of sources at the low-mass end, and therefore at low radio-luminosity, means that this fixed slope will struggle to produce a fit as good as the models with more freedom, as it assumes that the lower-mass galaxies are in the sample. Thus, formally it is the worst fitting model, even though it may accurately represent the underlying RLF.
 
 We also run \textsc{bayestack} simultaneously over all of the redshift bins using the pure luminosity evolution model.
 The PLE for our AGN and SFG galaxies is given by the following MAP values and 95 per cent confidence limits,
\begin{equation*} 
L_{\rm AGN} \propto (1+z)^{1.83\pm 0.22 - (0.47 \pm 0.10)z} 
\end{equation*}
and
\begin{equation*}
L_{\rm SFG} \propto (1+z)^{3.88\pm 0.04 - (0.82 \pm 0.03)z}
\end{equation*}

In Fig.~\ref{fig:RLF_v2} we show the RLF fits and 95 per cent confidence intervals for the individual redshift bins, alongside the PLE RLF model fits, both with the fixed RLF shape. The PLE RLF model agrees with the $1/V_{\rm{max}}$ data-points across all redshifts up to $z\sim 2$.
At the highest redshifts ($z>2$) we find some differences between our model (plus our binned data) and models of \cite{Novak-2017,Novak-2018}. At the high radio luminosities, small number statistics, coupled with slight differences in the photometric redshifts used by us and \cite{Novak-2017}, offer some explanation as to why our RLF lies below theirs. Furthermore, we note that we are also becoming increasingly incomplete at these redshifts, and sources that are relatively bright at radio wavelengths could have lower-mass/faint host galaxies \citep[e.g.][]{Jarvis-2009, Norris-2011,Collier-2014}. However, more relevant to the focus of this work are the differences in the evolution of the lower-luminosity component of the RLF, which we assume to be dominated by SFGs.

The SFG component of our RLF model evolves with a similar strength to that of \cite{Novak-2017} up to $z\sim 1.5$, with the degeneracy between $\alpha_{L}^{SF}$ and $\beta_{L}^{SF}$ across this redshift range explaining the apparent difference in the evolutionary parameters (Table~\ref{tab:evol}).  
However, beyond $z\sim 2$, the \cite{Novak-2017} RLF continues to evolve, whereas we find that the SFG RLF from our model reaches a steady state and then begins to decline (for the \cite{Novak-2017} PLE model, this decline does not take effect until $z > 3.5$).
  
We note that this decline coincides with the decrease in low stellar-mass sources in our stellar-mass-limited sample at these redshifts. However, it is also worth mentioning that the total RLF (AGN + SFGs) does continue to be a reasonable fit to the binned $1/V_{\rm max}$ points of \cite{Novak-2018} out to $z\sim 2.5$, suggesting that some of the deficit in the low-luminosity RLF is compensated for by the evolving high-luminosity RLF that we associate with AGN. 

\begin{table*}
\begin{center}
\caption{Comparison with determinations in the literature of the pure luminosity evolution of the radio luminosity function.}
\renewcommand{\arraystretch}{1.4}
\begin{tabular}{cccccc}
\hline
 Reference & Description & $\alpha_L^{\text{SF}}$ & $\beta_L^{\text{SF}}$ & $\alpha_L^{\text{AGN}}$ & $\beta_L^{\text{AGN}}$ \\
\hline
 This work& Total RLF fit & $3.88\pm 0.04$ & $-0.82 \pm 0.03$ & $1.83 \pm 0.22$ & $-0.54 \pm 0.10 $\\
  \citealt{Novak-2018}&Total RLF fit & 2.95 $\pm$  0.04 & $-$0.29 $\pm$  0.02 & 2.86 $\pm$  0.16 & $-$0.70 $\pm$  0.06 \\
 \citealt{Novak-2017} and \citealt{Smolcic-2017c}& Individual SF and AGN fit & 3.16 $\pm$  0.04 & $-$0.32 $\pm$  0.02 & 2.88 $\pm$  0.17 & $-$0.84 $\pm$  0.07 \\
 \citealt{McAlpine-2013}& Total RLF fit & \multicolumn{2}{c}{2.47 $\pm$ 0.12$^*$} & \multicolumn{2}{c}{1.18 $\pm 0.21^*$} \\
\hline
\multicolumn{3}{l}{$^*$ \cite{McAlpine-2013} fit only one evolutionary term each for the SF and AGN}
\end{tabular}
\label{tab:evol}
\end{center}
\end{table*}

\section{Cosmic history of star formation}\label{sec:CSFRD}
The RLF for SFGs derived from our stellar-mass-selected sample, obtained from Model C, provides a good estimate of the RLF of SFGs (based on \F{fig:RLF_v2}), and from this we can obtain the SFRD by integrating under it, i.e.
\begin{equation}
SFRD = \int^{\rm Lmax}_{\rm Lmin} SFR(L_{1.4}) \Phi(L_{1.4}) dL,
\label{eq:sfrd}
\end{equation}
where $\Phi(L_{1.4})$ is our RLF for SFGs and $SFR(L_{1.4})$ is the SFR associated with 1.4-GHz radio luminosities. Using the \cite{Kennicutt-1998} calibration, the total infrared luminosity ($L_{\rm{TIR}}$) is related to the SFR by,
\begin{equation}
    \frac{SFR}{\rm{M}_\odot \rm{yr}^{-1}} = 4.5 \times 10^{-37} \frac{L_{\rm{TIR}}}{W}.
    \label{eq:SFRD_IR}
\end{equation}
where $L_{TIR}$ is the total infrared luminosity. The radio luminosity can be converted into the total infrared luminosity and linked to SFR using the infrared--radio correlation (IRRC; e.g. \citealt{Delhaize-2017}),
\begin{equation}
    \frac{SFR}{\rm{M}_\odot \rm{yr}^{-1}} = \mathit{f}_{\rm{IMF}} \times 10^{q_{\rm{TIR}}-24} \frac{L_{1.4 \rm{GHz}}}{\rm{WHz^{-1}}},
\end{equation}  
where $\mathit{f}_{\rm{IMF}}$ is the IMF (equal to 1 for a Chabrier IMF; \citealt{Chabrier-2003}) and $q_{\rm{TIR}}$ is a parameter that quantifies the IRRC given by,
\begin{equation}
    q_{\rm{TIR}} = \log \left(\frac{L_{\rm{TIR}}}{3.75\times 10^{12} \rm{W}} \right) - \log\left(\frac{L_{1.4 \rm{GHz}}}{\rm{WHz^{-1}}} \right).
\label{eqn:qvalue}
\end{equation}
We adopt a $q_{\rm{TIR}}$ value that evolves with redshift, given by $q_{\rm{TIR}}(z)= 2.78 \pm 0.02 (1+z)^{-0.14\pm0.01}$ \citep{Novak-2017}. Although we note that the evolution may be due to a mass dependence of the IRRC \citep[e.g.][]{Gurkan2018, Delvecchio-2020, Smith2020}. We then obtain the SFRD by numerically integrating the product of the RLF and the SFR over 1.4-GHz radio luminosities (\Eq{eq:sfrd}). For this, the integral should cover all radio luminosities 
and not just the range dictated by our fitted values of $L_{\rm{min}}$ and $L_{\rm{max}}$. Although we note that this makes little difference in the derived cosmic SFRD, due to the shallowness of the faint end slope for low SFRs, and the steep exponential decline at high SFRs. With this in mind we use $L_{\rm{min}}=10^{21}$~W~Hz$^{-1}$ in all redshift bins and SFG RLF models.

In \F{fig:sfrd} we present the cosmic SFRDs obtained using our various models. As would be expected, the different models used in fitting the RLF result in different determinations of the cosmic SFRD. Here we present cosmic SFRDs obtained using Model B (blue data points) and Model C, both from fitting individual redshift bins (blue shading\footnote{The 95 per-cent region is calculated using the 95 per-cent confidence interval from the RLF.The conversion error associated with the $q$ value \citep{Novak-2017} has not being unaccounted for in this calculation.}) and from the PLE fit (black line). 

The SFRD from Model C for the individual redshift bins steadily increases with redshift out to $z\sim 1.2$, flattens  and then steadily decreases towards higher redshifts. The SFRD based on the Model B individual redshift bins is in good agreement with that based on the individual Model C bins below $z\sim1.7$. However, between $1 < z < 1.7$ Model B has  slightly higher SFRD due to the fact that the faint end slope remains high to around 1 dex below the knee in the SFG RLF. Above $z\sim 2$, Model B gives a lower SFRDs compared to Model C individual SFRDs. This is because of the downturn in the faint end of the SFG RLF caused by the rising incompleteness due to the stellar-mass-selection. The SFRD based on the Model C PLE RLF behaves similar to Model B but gives a higher SFRD than the other models between $1.6 <z<2.5$, where the stellar-mass selection still enables the knee in the SFG RLF to be well constrained, and the fixed faint-end slope ensures that the SFRD remains high. At $z>2.5$ the stellar-mass limit starts imposing on our ability to constrain the position of the knee in the SFG RLF, and the best-fit evolutionary terms force the position of the knee to lower radio luminosities in order to fit the incomplete parent sample. All our SFRDs estimations start to steadily decline between $1.5 <z<2$ because of the rising stellar-mass limit with redshift. 

\begin{figure*}
    \centering
    \includegraphics[width=0.8\textwidth]{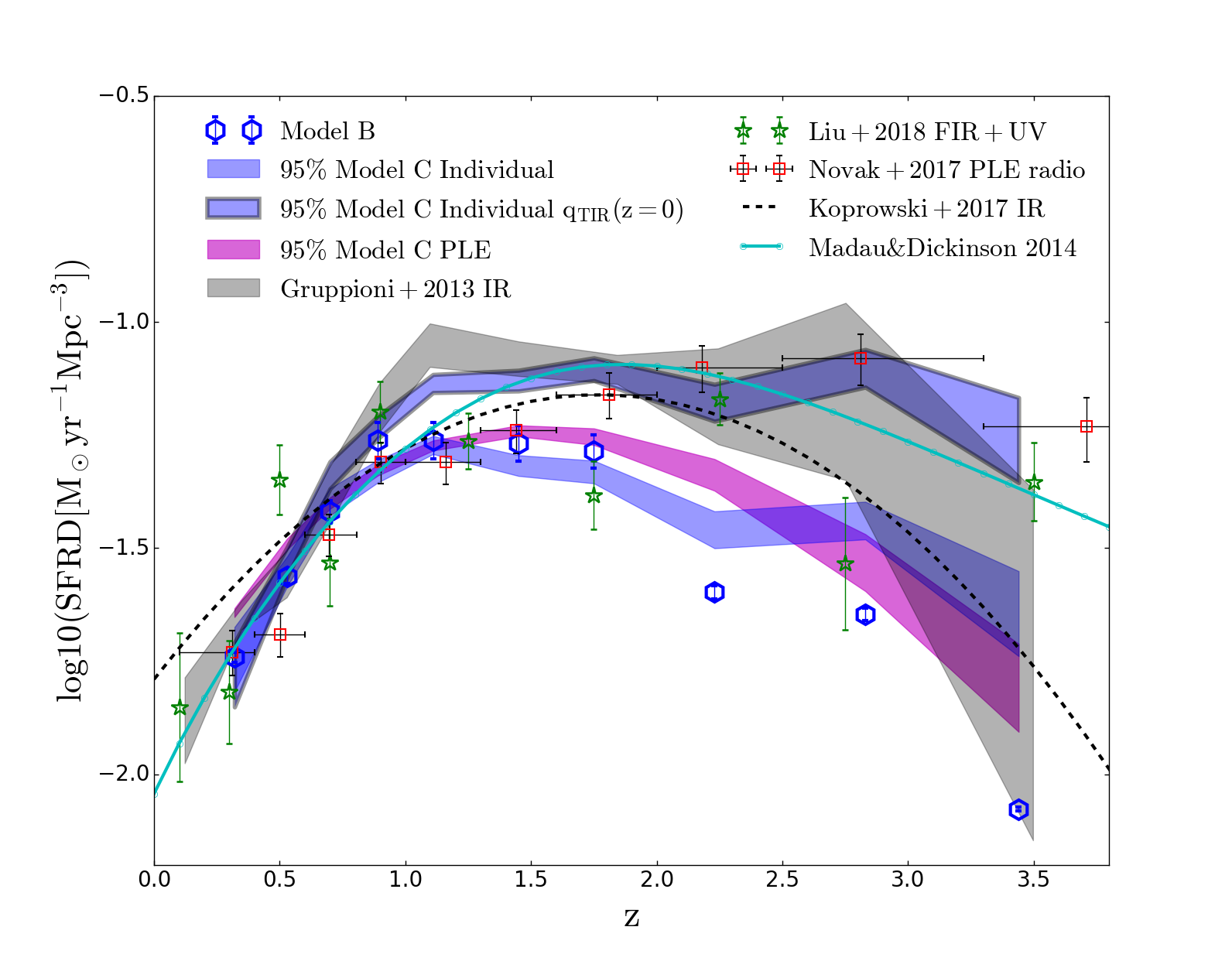}
    \caption{The cosmic star-formation rate density (SFRD). The blue hexagons are generated from the SFG MAP values for the Model B fit to each individual redshift bin. The blue shading corresponds to the 95-per-cent confidence region of the SFG component of the Model C fit to each redshift bin (individually). The blue region with borders also corresponds to the 95-percent confidence region of the individual Model C fit but is calculated assuming a non-evolving $q_{TIR}(z=0)=2.64\pm -0.02$ \protect\citep{Bell-2003}.
    The magenta shading corresponds to to the 95-per-cent confidence region of the SFG component of the Model C PLE fit to the combined redshift bins. 
    The green stars are the combined uncorrected IR and UV data from \protect\cite{Liu-2018} and the red squares are from the SFG RLF of \protect\cite{Novak-2017}, using a PLE fit. The black, dashed line is from \protect\cite{Koprowski-2017} and the connected, cyan curve represent a fit by  \protect\cite{Madau_Dickinson-2014} to various cosmic SFRD measurements in the literature. The gray, shaded region is the cosmic SFRD generated from the infrared LF of \protect\cite{Gruppioni-2013}.}
    \label{fig:sfrd}
\end{figure*}

\subsection{Comparison to the literature}
In this subsection we compare our cosmic SFRD determinations, which are based on NIR-selected RLFs of SFGs (constrained below the nominal detection threshold using the \textsc{bayestack} technique), to literature measurements of the cosmic SFRD using a variety of SFR tracers.

\subsubsection{Comparison with the radio-selected cosmic SFRD}
We first compare our cosmic SFRDs to the cosmic SFRD determined by \cite{Novak-2017}, which is based on COSMOS2015 photometry and spectral energy distribution (SED) fits \citep{Laigle-2016, Delvecchio-2017} and VLA-COSMOS 3-GHz data. Our results are in good agreement with \cite{Novak-2017} at  $z < 1.6$, and our lower uncertainties are because we better constrain the RLF using sources that lie below 5\,$\sigma$ in the radio data. However, our SFRDs deviate away from \cite{Novak-2017} at $z >1.6$ as a result of our stellar-mass selection and possibly the assumed extrapolation to faint luminosities and completeness corrections implemented in \cite{Novak-2017}. The impact of this is most apparent in Fig.~\ref{fig:RLF_v2}, where the discrepancy between the SF component(s) of our `Model C' RLF(s) and that of the \cite{Novak-2017} RLF increases with redshift. Furthermore, although not shown in Fig.~\ref{fig:sfrd}, at $z >1.6$ our results are in broad agreement with other radio-based estimates of the cosmic SFRD in the literature \citep[e.g.][]{Smolcic-2009,Karim-2011,Ocran-2020}. Thus, our results should be considered as complementary to those that use radio-selection to measure the RLF. In our case, the incompleteness arises from the stellar-mass selection only, but we are able to directly constrain the faint-end slope to higher redshifts than the pure radio-selection. Whereas completeness corrections for radio-selected samples are required for both the radio data (e.g. Eddington bias, which is relatively straightforward to account for) {\em and} in terms of the ability to identify a host galaxy and measure a redshift (which is less of a problem for fields with excellent ancillary data, such as COSMOS).

\subsubsection{Comparison to other studies}
Next we compare our cosmic SFRDs to the dust-obscured cosmic SFRD from \cite{Gruppioni-2013}. For this they use a total-IR LF based on deep {\it Herschel} data, from the PACS Evolutionary Probe
(PEP; \citealt{Lutz-2011}) and the complementary {\it Herschel} Multi-tiered Extragalactic Survey (HerMES; \citealt{Oliver-2012}), out to $z \sim 4$. We convert their $L_{TIR}$ density (where the $L_{TIR}$ is obtained from an integral over the whole thermal IR spectrum) to the SFRD using eq.~\ref{eq:SFRD_IR}. In Fig.~\ref{fig:sfrd} we also plot the IR-based cosmic SFRD from \cite{Koprowski-2017}, who used {\it Herschel} far-infrared (FIR) flux-densities for their LFs, extracted at the positions of sub-mm sources identified using the James Clerk Maxwell Telescope's SCUBA-2 Cosmology Legacy Survey (S2CLS; \citealt{Geach-2017}) and the Atacama Large Millimeter/sub-millimeter Array (ALMA; \citealt{Dunlop-2017}) in the COSMOS and UKIDSS-UDS fields. Our results are in good agreement with \cite{Gruppioni-2013} below $z\sim 1$, but then deviate towards higher redshifts where the IR SFRD continues to increase (before flattening and then falling around $z\sim 3$). It should be noted that there are a lot of uncertainties in measuring $L_{TIR}$ from a few data points. There are also k-correction effects, since, as one goes to higher redshifts we move away from the peak of the thermal emission at $\sim 100\mu$m in the rest frame. At these high redshifts the $L_{TIR}$ becomes dominated by hotter dust systems, which are more likely to have AGN contributions. 
This implies that converting from $L_{TIR}$ to SFRD for these systems may lead to an overestimate of the cosmic SFRD. 
The SFRD by \cite{Koprowski-2017} is higher than both of our SFRD determinations at most redshifts, except around $z\sim 1$ where our results overlap. We note that \cite{Gruppioni-2019} attributed the discrepancies between the two IF-SFRD functions \citep{Gruppioni-2013,Koprowski-2017} to selection bias, incompleteness effects, and the choice of SED in the SCUBA-selected data from \cite{Koprowski-2017}, which reinforces some of the issues we mention above. 

 We also compare our results 
  with the cosmic SFRD from  \cite{Liu-2018} which represent the total cosmic SFRDs (a combination of the dust-obscured and unobscured cosmic SFRD measurements). \cite{Liu-2018} derived their SFRD using super-deblended FIR to sub-mm {\it Herschel} photometry from confused galaxies in the northern field of the Great Observatories Origins Deep Survey (GOODS). The FIR/sub-mm photometry is extracted based on fitting SEDs to sources selected from deep {\em Spitzer Space Telescope} Multiband Imaging Photometer (MIPS; \citealt{Rieke-2004}) and 1.4GHz~VLA \citep{Morrison-2010,Owen-2018} data. 
The derived SFRD from \cite{Liu-2018} is in good agreement with our results until $z\sim 3$ with minor deviations. They agree with our decline above $z\sim 1.6$, this is possibly because their sample is also limited by stellar mass, due to their optical/NIR detections. 

We also show the \cite{Madau_Dickinson-2014} cosmic SFRD, which is a fit to various cosmic SFRDs in the literature. Our results are again in good agreement below $z\sim 1$. However, our results deviate at  $z > 1$,  which is certainly influenced by our stellar-mass selection. However, it also should be noted that we have assumed the IRRC form that evolves negatively with increasing redshift, meaning that for a given radio luminosity, the SFR would be lower at high redshift, than at low redshift. This could obviously result in a false decrease in the SFRD if the real IRRC did not evolve with redshift, which other studies have suggested, depending on how the galaxies have been selected \citep[e.g.][]{Molnar2018}. For example, one aspect of this is that \cite{Gurkan2018, Delvecchio-2020, Smith2020} all find that the IRRC has a dependence on the stellar mass of the galaxy, and this may be responsible for the observed evolution of the IRRC, as higher-redshift samples are inevitably dominated by more massive galaxies due to the nature of flux-limited samples. However, mass is unlikely to be the only extra parameter that needs to be considered when using the IRRC to convert a radio luminosity to star-formation rate, with \cite{Smith-2014} and \cite{Read2018} showing that dust temperature, and how you include sensible k-corrections for a range of dust temperatures at different redshifts, can be crucial to measure the SFR.

Furthermore, as we move beyond $z\sim 1$, inverse Compton scattering of the Cosmic Microwave Background photons may reduce the level of radio emission from star-forming galaxies observed at a given (relatively high) frequency \citep[e.g.][]{Murphy2009}.
All of these issues result in our understanding of any evolution in $q_{\rm TIR}$ being uncertain. In Fig.~\ref{fig:sfrd} we therefore also show how the SFRD evolves when adopting a constant of $q_{\rm TIR} = 2.64$ \citep{Bell-2003}. One can see that this has a dramatic effect on the high-redshift evolution of the SFRD, with the cosmic SFRD derived from the RLFs determined in this paper becoming significantly higher at high redshift ($\sim 30\%$ at $z\sim 3.5$).

\begin{figure*}
    \centering
    \includegraphics[width=\textwidth]{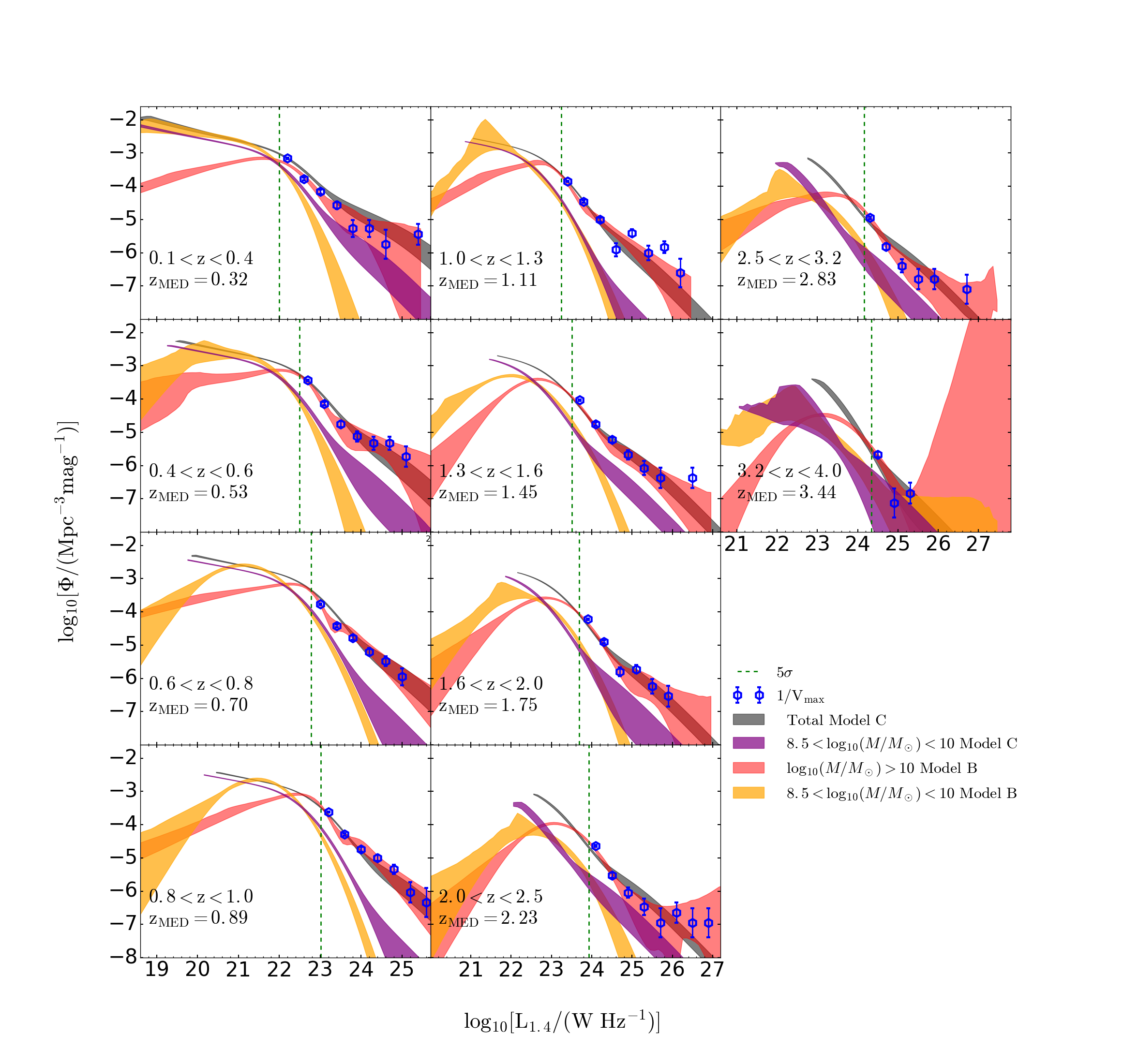}
    \caption{The contribution to the total rest-frame 1.4-GHz RLF from sources with different stellar masses in the COSMOS field. Low stellar mass ($10^{8.5} < M/\mathrm{M}_{\odot} <10^{10} $) are represented by the purple and orange shaded regions, corresponding to the 95 per cent confidence interval of the distribution of reconstructions of models in the posteriors. The contribution from sources with high stellar mass ($M> 10^{10} \mathrm{M}_{\odot}$) are represented by the 95 per cent region. The total RLFs based on Model-C fit to each redshift bin are represented by the grey shading which corresponds to the 95-per-cent region. The blue hexagons represent $1/V_{\rm{max}}$ estimations for our detected sources. The vertical, green dotted lines correspond to the detection threshold ($\sigma$) computed using the median redshift for each redshift bin.}
    \label{fig:RLF_mass}
\end{figure*}

\begin{figure*}
    \centering
    \includegraphics[width=0.8\textwidth]{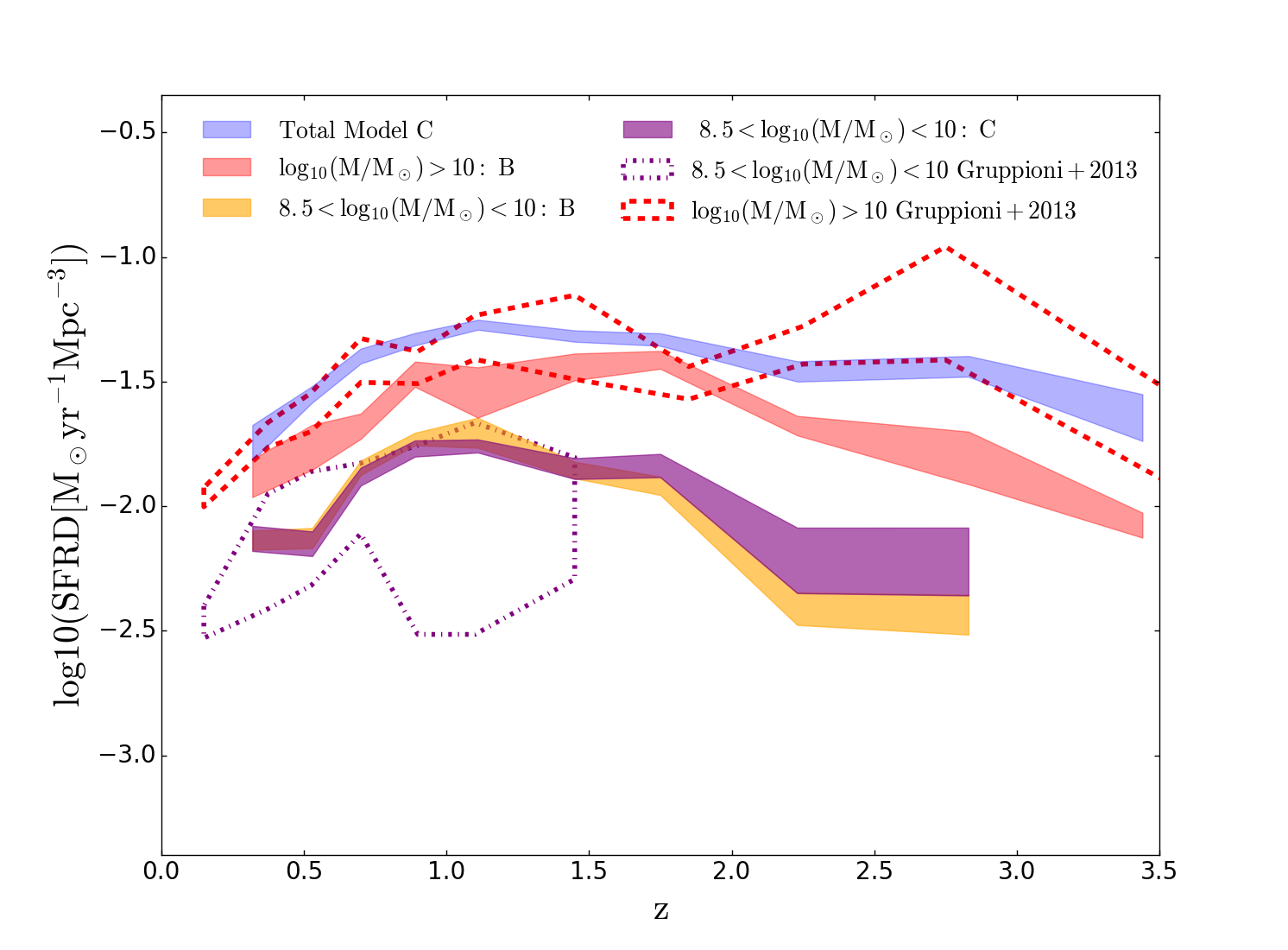}
    \caption{Contribution from different stellar mass populations to the cosmic star-formation rate density (SFRD). The blue shading represents 95-per-cent confidence region associated with the total cosmic SFRD from Model B fit to each redshift bin (individually). The orange and purple shading are the 95-per-cent confidence region from sources with low stellar masses ($10^{8.5} \leq M \leq 10^{10}\mathrm{M}_{\odot}$) from Model C and Model B respectively. The red shading are the 95-per-cent confidence region from sources with high stellar masses ($M>10^{10}\mathrm{M}_{\odot}$). 
    The red dashed and purple dashed-dotted line are the high ($10^{10}\leq M \leq 10^{12}\mathrm{M}_{\odot}$) and low ($M<10^{10}\mathrm{M}_{\odot}$)) stellar mass contributions from \protect\cite{Gruppioni-2013}.}
    \label{fig:sfrd_mass}
\end{figure*}

\subsection{Contribution from different stellar mass populations}
 As noted previously, the stellar-mass selection we have applied to our sample (Sec~\ref{sec:complete}) means that we miss low-stellar mass ($M<10^9 {\rm M}_{\odot}$) sources at high redshift ($z \gtrsim 1.5$). 
 
 To further investigate the effects of stellar mass on the total RLF we divide our sources into low ($10^{8.5} \leq M \leq 10^{10} \mathrm{M}_{\odot}$) and high $(M > 10^{10} \mathrm{M}_{\odot})$ stellar-mass galaxies, shown in Fig~\ref{fig:RLF_mass} for Models B and C. It should be noted that we are splitting at a stellar mass of $M =10^{10} \mathrm{M}_{\odot}$ due to the fact that our sample is complete to $M \sim 10^{10} \mathrm{M}_{\odot}$ in our highest redshift bin (Fig.~\ref{fig:stellar_mass}).  Galaxies with high stellar mass typically have high radio luminosities, as expected, and host a large proportion of the detected radio sources. The low stellar-mass galaxies typically have low radio luminosities and dominate the RLF below the 5-$\sigma$ detection threshold, for $z<1.5$. Above $z> 1.5$, the contribution from the low stellar-mass sources decreases due to our stellar-mass completeness limit (see Fig~\ref{fig:stellar_mass}). It is clear that the bulk of the RLF that we are able to measure at $z>0.4$ is dominated by galaxies with stellar-mass $M>10^{10}\,\mathrm{M}_{\odot}$.
 
 In Fig~\ref{fig:sfrd_mass} we show the contribution from the low ($10^{8.5} \leq M/\mathrm{M}_{\odot} \leq 10^{10}$) and high ($M> 10^{10}\mathrm{M}_{\odot}$) stellar mass sources to the total cosmic SFRD.
 The RLF for low stellar mass sources (orange and purple shades in Fig~\ref{fig:RLF_mass}) shows that they are not the dominant population contributing to the  SFRD at any redshift. However,  they are important to include as they are crucial to determine the position of the knee in the RLF ($L^*$), where the bulk of the SFRD is concentrated, and the steepness of the faint end slope (with a steep slope resulting in a higher contribution to the SFRD from these faint sources). The fact that they are missing in our sample at high redshifts, means that this can affect our cosmic SFRD estimate. 

 The contribution to the SFRD from low stellar mass galaxies increases with redshift up to $z~\sim 1$ (for Model B and $z~\sim 1.5$ for Model C)  where it peaks and drops towards higher redshifts (due to the NIR flux-density limit). The shape of the SFRD derived from just the low stellar mass galaxies is similar to the total SFRD below $z\sim 1$. However, it is the contribution from high stellar mass galaxies that dominates the total SFRD at all redshifts, and hence has a shape almost identical to the total (with minor differences).

We compare our results to the cosmic SFRDs from \cite{Gruppioni-2013}, who divided their sources into three stellar mass (low, mid and high) bins. Their low stellar mass contribution, $M<10^{10}\mathrm{M}_{\odot}$, shows larger error-bars but are fully in agreement with our results. We compare our high stellar mass contribution with the combination of their mid ($10^{10} \leq M/\mathrm{M}_{\odot} \leq 10^{11}$) and high stellar mass ($10^{10} \leq M/\mathrm{M}_{\odot} \leq 10^{12}$), which are largely in agreement below $z\sim2$. Above $z\sim 2$ the \cite{Gruppioni-2013} (high stellar mass) sources result in a higher SFRD, with an increasing contribution from starburst galaxies and SF AGN. This might imply that we are missing these sources in our stellar-mass selection selection, or it is possible that they instead contribute to the total RLF through the high-luminosity part which we do not use to determine the SFRD.

\section{Conclusions}\label{sec:conc}

Our main goal is to constrain the RLF to low radio luminosities and then obtain a measurement of the cosmic star-formation rate density (SFRD) from our stellar mass-limited sample.
Using \textsc{bayestack} we probe the stellar-mass selected RLF orders of magnitude below the nominal $5\sigma$ detection threshold by fitting parametric models to the RLF for both SFGs (low-luminosity radio source) and AGN (high-luminosity radio sources). The reconstructed RLFs follow the $1/V_{\rm{max}}$-points very well above the detection limit. We find that our models also follow the \cite{Novak-2018} extrapolated pure luminosity evolution fit well, for at least an order of magnitude below the detection threshold, to $z\sim 1.6$.  

 However, due to our stellar mass-limit, the (free model; Model B) faint-end slope of our SFG RLF, obtained using \textsc{bayestack}, falls-off towards low radio luminosities, particularly at the higher redshifts ($z>1.5)$. This fall-off is not an underlying feature of the RLF of SFGs but is the result of a lack of fainter radio sources in our parent stellar-mass selected sample. This is due to the known relation between stellar mass and star-formation rates \citep[e.g.][]{Noeske-2007,Daddi-2007,Johnston-2015}, where  our stellar-mass selected sample imposes a natural limit on the level of star formation in galaxies we are able to probe. As our stellar-mass limit increases with redshift, due to the flux limit of the optical/NIR data, this means we do not have the radio-faint SFGs in our sample.
 
 We address this by fixing the shape of the RLF to that of the local RLF and allow it to evolve with redshift. We start by obtaining the RLF in each individual redshift bin, by allowing the knee in both the SFG and AGN RLFs to be a free parameter. We next use a pure luminosity evolution fit (with two luminosity evolution terms) to fit the RLF with a prescribed functional form over all the redshift bins. We find that the best fit PLE model gives $L_{\rm AGN} \propto (1+z)^{1.83\pm 0.2 - (0.47 \pm 0.10)z}$ and $L_{\rm SFG} \propto (1+z)^{3.88\pm 0.04 - (0.82 \pm 0.03)z}$. The evolution strength is similar to that of \cite{Novak-2017} up to $z\sim 1.6$. However, beyond $z\sim 2$, the \cite{Novak-2017} RLF continues to evolve, whereas we find that the RLF does not evolve as strongly beyond $z> 2.5$. The lack of strong evolution coincides with the decrease of low stellar-mass sources in our stellar mass-limited sample at these redshifts. This results in the position of the knee in the RLF moving to lower luminosities for the SFG population, at $z> 2.5$.

We use our RLF models to determine the radio-derived SFRD by numerically integrating the product of the 1.4~GHz RLF of SFGs and the SFR associated with the 1.4~GHz luminosity based on the infrared-radio correlation (IRRC). We found our SFRD to be consistent with the established behaviour at low redshift, where it increases strongly with redshift out to $z\sim 1$ \citep[e.g][]{Gruppioni-2013,Madau_Dickinson-2014,Koprowski-2017,Novak-2017}. Beyond $z\sim 1$ the SFRD determined from radio observations depends strongly on the assumed conversion from radio luminosity to SFR. Assuming an evolving IRRC results in the SFRD decreasing at high redshift. Whereas, if we assume that the IRRC is constant with redshift then the SFRD remains relatively flat out to the limit of our sample at $z = 3.5$. Clearly, if we are to use the radio emission as a tracer of star-formation rate across cosmic time, the relationship between the radio luminosity and star-formation rate needs to be better understood and expanded to include other factors, such as inverse Compton scattering of CMB photons \citep[e.g.][]{Murphy2009}, stellar-mass dependence \citep[e.g.][]{Gurkan2018} and morphology \citep[e.g.][]{Molnar2018}.

We also investigate the effects of stellar mass on the total RLF by splitting our sample into low ($10^{8.5} \leq M/\mathrm{M}_{\odot} \leq 10^{10}$) and high ($M>10^{10}\mathrm{M}_{\odot}$) stellar mass. We find that the low stellar mass sources dominate the faint-end of the RLF and the high stellar mass sources are usually associated with the radio-detected sources, as expected given the relationship between stellar mass and SFR. We find that the SFRD is dominated by sources with high stellar masses ($>10^{10}\mathrm{M}_{\odot}$) at all redshifts, and that the low stellar mass sources excluded from our sample due to the NIR flux limit will not be enough to make up for the decline of our SFRD compared to \cite{Madau_Dickinson-2014} and \cite{Novak-2017} using the evolving IRRC relation. 

Clearly, there is much more work to be done to understand the evolution of the SFRD, with various wavelengths suffering from different selection issues. Here, we have used a new method to determine the evolution of the RLF based on the radio emission from a stellar-mass selected sample in the COSMOS field. However, uncertainties in the conversion from radio luminosity to SFR, and how it may or may not evolve with redshift, means it is difficult to make strong claims about the evolution beyond $z\sim 1$. Extending this study to other radio frequency data may be crucial in overcoming some of these issues. For example, the level of inverse Compton scattering of CMB photons and the contribution from free-free emission from H{\sc ii} regions will impact on the higher frequency emission more than at low frequency. This obviously is more of an issue at high redshifts, where the rest-frame frequency is $>9$\,GHz for sources at $z>2$ in the 3\,GHz data we use here. Thus, undertaking a similar study as we have done here over the deep fields observed by the LOw-Frequency ARray (LOFAR; \citealt{Tasse-2020,Sabater-2020}) at an observed frequency of $150$\,MHz, and the MeerKAT International Giga-Hertz Tiered Extragalactic Exploration  \citep[MIGHTEE; ][]{Jarvis-2016} Survey at 856\,MHz and 1712\,MHz, will provide crucial information necessary to advance our understanding of the cosmic SFRD further.

\subsection*{Acknowledgements}
EDM, MGS, MJJ and SVW acknowledge financial support from the South African Radio Astronomy Observatory (SARAO). EDM and MGS also acknowledge support from the National Research Foundation (Grant No.~84156). We would like to acknowledge the computational resources of the Centre for High Performance Computing. MJJ also acknowledges support from the UK Science and Technology Facilities Council [ST/N000919/1], the Oxford Hintze Centre for Astrophysical Surveys, which is funded through generous support from the Hintze Family Charitable Foundation. NJA acknowledges funding from the STFC Grant Code ST/R505006/1.
RB acknowledges support from the Glasstone Foundation.

\section*{Appendix}
\renewcommand{\thefigure}{A\arabic{figure}}
\setcounter{figure}{0}
\renewcommand{\thetable}{A.\arabic{table}} 
\setcounter{table}{0}
Fig~\ref{fig:perfect} shows the 1-D and 2-D posterior distributions for the PLE Model C to the data at all the redshift bins. The 1-D posterior distribution is the marginalization of each parameter shown at the end of each row. The parameters  have well-defined peaks, and parameters $L_{\rm{min1}}$ and $L_{\rm{max}}$ have two peaks. 

Figs~\ref{fig:Model_B}, \ref{fig:Model_B2}, \ref{fig:Model_B3}, \ref{fig:Model_B4} and \ref{fig:Model_B5} show the 1-D and 2-D posterior distributions for Model B applied to all the redshift bins. The boundary parameters are mostly do not have a well defined peak and also hit the prior range. This does not affect the fit as long as the prior space is large enough. $L^*_1$ (the AGN break) also hit the prior edge however, this prior is motivated by literature data (Sec~\ref{sec:priors}). 

Fig~\ref{fig:Model_C} and \ref{fig:Model_C2} show the posterior distribution for Model C applied to all the redshift bins individually. Table~\ref{table:parameters} show the MAP parameters obtained from the various models applied to the each individual redshift bin. 

Table~\ref{table:parameters2} are the MAP parameters for the various models applied to the low and high stellar mass galaxies in each redshift bin.  

\begin{landscape}
\begin{table}
\caption{The MAP posterior parameters of Models A, B and C for the NIR-selected RLF, in each of the redshift bins and their $\sigma$. The units of the parameters are as shown in \T{table:prior}.}
\begin{tabular}{l|rrrrrrrrrr}
\hline
    Parameter & $0.1 < z < 0.4$ & $0.4 < z < 0.6$ & $0.6 < z < 0.8$ & $0.8 < z < 1.0$ & $1.0 < z < 1.3$ & $1.3 < z < 1.6$ & $1.6 < z < 2.0$ & $2.0 < z < 2.5$ & $2.5 < z < 3.2$ & $3.2 < z < 4.0$ \\ \hline 
    &\multicolumn{10}{c}{{Model A}}\\
    \hline
$\log_{10}[L_{\rm{min_1}}]$ &$ 17.40_{-0.28}^{+0.48}$ & $ 17.55_{-0.38}^{+0.93}$ & $ 18.98_{-0.73}^{+1.00}$ & $ 18.85_{-0.50}^{+0.77}$ & $ 21.00_{-0.30}^{+0.14}$ & $ 18.73_{-0.35}^{+0.50}$ & $ 21.46_{-2.03}^{+0.20}$ & $ 21.91_{-0.29}^{+0.17}$ & $ 20.23_{-0.69}^{+0.83}$ & $ 22.26_{-0.50}^{+0.17}$ \\ [3pt]
$\log_{10}[L_{\rm{max_1}}]$ &$ 24.65_{-0.10}^{+0.10}$ & $ 25.65_{-0.57}^{+0.54}$ & $ 25.40_{-0.13}^{+0.13}$ & $ 26.51_{-0.32}^{+0.31}$ & $ 26.20_{-0.13}^{+0.13}$ & $ 26.53_{-0.33}^{+0.30}$ & $ 26.49_{-0.32}^{+0.32}$ & $ 26.99_{-0.66}^{+0.68}$ & $ 26.79_{-0.47}^{+0.47}$ & $ 26.76_{-0.50}^{+0.48}$ \\ [3pt]
$\log_{10}[L_{\rm{min_2}}]$ &$ 19.54_{-0.19}^{+0.11}$ & $ 19.64_{-0.99}^{+0.90}$ & $ 20.93_{-0.04}^{+0.03}$ & $ 20.45_{-0.39}^{+0.35}$ & $ 22.63_{-0.20}^{+0.12}$ & $ 22.55_{-0.16}^{+0.17}$ & $ 21.57_{-0.83}^{+0.81}$ & $ 21.66_{-0.90}^{+0.91}$ & $ 22.64_{-0.22}^{+0.23}$ & $ 22.65_{-0.23}^{+0.22}$ \\ [3pt]
$\log_{10}[L_{\rm{max_6}}]$ &$ 25.30_{-0.14}^{+0.10}$ & $ 26.13_{-0.60}^{+0.18}$ & $ 26.05_{-0.55}^{+0.33}$ & $ 27.07_{-0.44}^{+0.29}$ & $ 26.26_{-0.16}^{+0.15}$ & $ 26.86_{-0.14}^{+0.10}$ & $ 26.55_{-0.35}^{+0.31}$ & $ 26.09_{-0.06}^{+0.07}$ & $ 26.38_{-0.33}^{+0.31}$ & $ 26.79_{-0.49}^{+0.48}$ \\ [3pt]
    $\log_{10}[\Phi_1^*]$ &$ -4.50_{-0.29}^{+0.15}$ & $ -4.85_{-0.40}^{+0.29}$ & $ -5.58_{-0.28}^{+0.30}$ & $ -4.41_{-0.36}^{+0.26}$ & $ -5.50_{-0.55}^{+0.58}$ & $ -5.89_{-0.59}^{+0.43}$ & $ -5.58_{-0.59}^{+0.34}$ & $ -6.08_{-0.98}^{+0.73}$ & $ -6.81_{-1.35}^{+0.94}$ & $ -6.78_{-0.61}^{+0.56}$ \\ [3pt]
       $\log_{10}[L_1^*]$ &$ 24.21_{-0.16}^{+0.32}$ & $ 24.81_{-0.63}^{+0.67}$ & $ 25.13_{-0.27}^{+0.25}$ & $ 24.44_{-0.31}^{+0.45}$ & $ 25.41_{-0.61}^{+0.51}$ & $ 25.62_{-0.44}^{+0.74}$ & $ 25.19_{-0.59}^{+0.56}$ & $ 25.00_{-0.65}^{+0.78}$ & $ 24.65_{-1.31}^{+1.15}$ & $ 25.51_{-0.70}^{+0.74}$ \\ [3pt]
               $\alpha_1$ &$  1.68_{-0.49}^{+1.01}$ & $  1.78_{-1.21}^{+2.12}$ & $  3.05_{-1.18}^{+1.07}$ & $  1.75_{-0.55}^{+1.33}$ & $  1.91_{-0.77}^{+1.94}$ & $  0.78_{-0.85}^{+0.78}$ & $  1.25_{-0.81}^{+1.74}$ & $  1.32_{-1.12}^{+2.35}$ & $  1.87_{-1.73}^{+2.56}$ & $  1.18_{-0.37}^{+0.39}$ \\ [3pt]
                $\beta_1$ &$  0.68_{-0.02}^{+0.01}$ & $ -0.29_{-1.17}^{+0.52}$ & $  0.97_{-0.03}^{+0.03}$ & $  0.06_{-0.57}^{+0.26}$ & $  0.58_{-0.51}^{+0.23}$ & $ -1.43_{-1.61}^{+1.21}$ & $ -1.96_{-1.97}^{+1.64}$ & $ -2.03_{-1.96}^{+1.72}$ & $ -2.13_{-1.85}^{+1.77}$ & $ -0.01_{-0.45}^{+0.45}$ \\ [3pt]
    $\log_{10}[\Phi_2^*]$ &$ -2.14_{-0.09}^{+0.10}$ & $ -2.32_{-0.06}^{+0.04}$ & $ -2.47_{-0.07}^{+0.06}$ & $ -2.24_{-0.06}^{+0.05}$ & $ -2.86_{-0.27}^{+0.31}$ & $ -2.53_{-0.04}^{+0.04}$ & $ -3.03_{-0.23}^{+0.27}$ & $ -3.54_{-0.36}^{+0.18}$ & $ -3.52_{-0.08}^{+0.06}$ & $ -4.46_{-0.65}^{+1.00}$ \\ [3pt]
       $\log_{10}[L_2^*]$ &$ 21.75_{-0.16}^{+0.16}$ & $ 22.32_{-0.06}^{+0.06}$ & $ 22.65_{-0.06}^{+0.05}$ & $ 22.65_{-0.06}^{+0.06}$ & $ 23.16_{-0.25}^{+0.16}$ & $ 22.89_{-0.07}^{+0.06}$ & $ 23.47_{-0.27}^{+0.21}$ & $ 23.74_{-0.17}^{+0.28}$ & $ 23.71_{-0.09}^{+0.10}$ & $ 24.14_{-0.96}^{+0.39}$ \\ [3pt]
               $\alpha_2$ &$  2.24_{-0.60}^{+1.23}$ & $  1.87_{-0.13}^{+0.16}$ & $  4.31_{-0.68}^{+0.48}$ & $  2.04_{-0.19}^{+0.22}$ & $  2.38_{-0.56}^{+0.89}$ & $  1.51_{-0.07}^{+0.08}$ & $  1.88_{-0.25}^{+0.31}$ & $  1.98_{-0.24}^{+0.88}$ & $  1.69_{-0.12}^{+0.16}$ & $  2.29_{-0.80}^{+1.44}$ \\ [3pt]
                $\beta_2$ &$ -2.74_{-1.45}^{+1.26}$ & $ -0.05_{-0.03}^{+0.07}$ & $ -3.59_{-0.99}^{+1.38}$ & $ -0.15_{-0.04}^{+0.09}$ & $  0.36_{-0.28}^{+0.20}$ & $ -0.38_{-0.05}^{+0.04}$ & $  0.12_{-0.47}^{+0.23}$ & $  0.16_{-0.26}^{+0.30}$ & $ -0.20_{-0.06}^{+0.12}$ & $  0.59_{-1.30}^{+0.35}$ \\ [3pt]

\hline 
&\multicolumn{10}{c}{Model B}\\
\hline
$\log_{10}[L_{\rm{min_1}}]$ &$ 18.20_{-0.15}^{+0.22}$ & $ 17.87_{-0.67}^{+0.69}$ & $ 18.25_{-0.18}^{+0.69}$ & $ 18.26_{-0.18}^{+0.28}$ & $ 19.70_{-0.45}^{+0.83}$ & $ 19.69_{-0.55}^{+0.63}$ & $ 19.42_{-0.30}^{+0.78}$ & $ 19.42_{-0.29}^{+0.93}$ & $ 18.61_{-0.35}^{+0.58}$ & $ 17.46_{-1.65}^{+2.21}$ \\ [3pt]
$\log_{10}[L_{\rm{max_1}}]$ &$ 24.66_{-0.10}^{+0.10}$ & $ 24.70_{-0.84}^{+0.86}$ & $ 25.59_{-0.25}^{+0.27}$ & $ 27.98_{-1.27}^{+1.33}$ & $ 28.06_{-1.29}^{+1.24}$ & $ 27.98_{-1.27}^{+1.29}$ & $ 27.97_{-1.26}^{+1.29}$ & $ 27.91_{-1.17}^{+1.34}$ & $ 27.71_{-1.10}^{+0.98}$ & $ 28.07_{-1.33}^{+1.24}$ \\ [3pt]
$\log_{10}[L_{\rm{min_2}}]$ &$ 19.06_{-0.64}^{+0.63}$ & $ 19.00_{-0.67}^{+0.64}$ & $ 19.41_{-0.82}^{+0.78}$ & $ 20.29_{-0.91}^{+0.67}$ & $ 20.71_{-0.89}^{+0.81}$ & $ 20.63_{-0.84}^{+0.78}$ & $ 20.57_{-0.80}^{+0.79}$ & $ 20.69_{-0.86}^{+0.84}$ & $ 20.22_{-0.61}^{+0.55}$ & $ 22.91_{-0.37}^{+0.38}$ \\ [3pt]
$\log_{10}[L_{\rm{max_6}}]$ &$ 25.18_{-0.23}^{+0.26}$ & $ 25.33_{-0.26}^{+0.34}$ & $ 29.09_{-0.24}^{+0.64}$ & $ 29.45_{-0.31}^{+0.23}$ & $ 29.33_{-0.92}^{+0.44}$ & $ 29.47_{-0.84}^{+0.37}$ & $ 29.62_{-0.96}^{+0.28}$ & $ 29.58_{-0.83}^{+0.24}$ & $ 29.27_{-0.74}^{+0.44}$ & $ 28.68_{-1.18}^{+0.90}$ \\ [3pt]
    $\log_{10}[\Phi_1^*]$ &$ -3.80_{-0.33}^{+0.20}$ & $ -4.58_{-0.36}^{+0.30}$ & $ -4.15_{-0.21}^{+0.16}$ & $ -4.50_{-0.60}^{+0.23}$ & $ -4.64_{-0.46}^{+0.26}$ & $ -5.33_{-0.42}^{+0.23}$ & $ -5.58_{-0.39}^{+0.33}$ & $ -6.14_{-0.48}^{+0.37}$ & $ -6.49_{-0.60}^{+0.36}$ & $ -6.93_{-1.26}^{+1.10}$ \\ [3pt]
       $\log_{10}[L_1^*]$ &$ 23.19_{-0.15}^{+0.34}$ & $ 23.71_{-0.53}^{+1.09}$ & $ 24.10_{-0.07}^{+0.17}$ & $ 24.62_{-0.17}^{+1.08}$ & $ 24.22_{-0.33}^{+0.65}$ & $ 25.37_{-0.27}^{+0.71}$ & $ 25.14_{-0.60}^{+0.77}$ & $ 25.58_{-0.96}^{+0.90}$ & $ 24.71_{-0.80}^{+0.64}$ & $ 25.80_{-1.18}^{+1.00}$ \\ [3pt]
               $\alpha_1$ &$  1.31_{-0.36}^{+0.33}$ & $  0.73_{-0.38}^{+1.33}$ & $  1.67_{-0.34}^{+0.23}$ & $  1.55_{-0.54}^{+0.32}$ & $  1.05_{-0.36}^{+0.48}$ & $  1.27_{-0.50}^{+0.45}$ & $  0.86_{-0.40}^{+0.49}$ & $  0.22_{-0.44}^{+0.72}$ & $  0.24_{-0.35}^{+0.28}$ & $  0.71_{-0.80}^{+0.77}$ \\ [3pt]
                $\beta_1$ &$ -0.52_{-1.15}^{+0.55}$ & $ -0.94_{-2.56}^{+1.05}$ & $ -1.23_{-1.90}^{+1.08}$ & $  0.04_{-0.73}^{+0.33}$ & $ -0.99_{-1.29}^{+1.03}$ & $ -0.11_{-1.00}^{+0.39}$ & $ -0.67_{-0.85}^{+0.69}$ & $ -1.73_{-1.95}^{+1.28}$ & $ -2.67_{-1.48}^{+1.24}$ & $ -1.40_{-1.03}^{+1.09}$ \\ [3pt]
    $\log_{10}[\Phi_2^*]$ &$ -2.18_{-0.05}^{+0.05}$ & $ -2.28_{-0.03}^{+0.03}$ & $ -2.24_{-0.02}^{+0.02}$ & $ -2.21_{-0.02}^{+0.01}$ & $ -2.45_{-0.05}^{+0.03}$ & $ -3.16_{-0.41}^{+0.29}$ & $ -3.16_{-0.46}^{+0.15}$ & $ -4.01_{-0.79}^{+0.39}$ & $ -3.62_{-0.23}^{+0.03}$ & $ -4.99_{-0.60}^{+0.73}$ \\ [3pt]
       $\log_{10}[L_2^*]$ &$ 21.39_{-0.17}^{+0.16}$ & $ 21.69_{-0.15}^{+0.13}$ & $ 21.93_{-0.15}^{+0.13}$ & $ 22.00_{-0.14}^{+0.12}$ & $ 21.78_{-0.27}^{+0.31}$ & $ 21.21_{-0.34}^{+0.43}$ & $ 21.77_{-0.49}^{+0.31}$ & $ 21.77_{-0.50}^{+0.44}$ & $ 22.47_{-0.57}^{+0.22}$ & $ 21.49_{-0.33}^{+0.55}$ \\ [3pt]
               $\alpha_2$ &$  1.11_{-0.03}^{+0.04}$ & $  0.92_{-0.04}^{+0.07}$ & $  0.87_{-0.03}^{+0.04}$ & $  0.74_{-0.04}^{+0.03}$ & $  0.61_{-0.13}^{+0.26}$ & $ -0.02_{-0.34}^{+0.33}$ & $  0.16_{-0.42}^{+0.22}$ & $ -0.29_{-0.57}^{+0.43}$ & $  0.49_{-0.32}^{+0.10}$ & $ -0.64_{-0.39}^{+0.62}$ \\ [3pt]
            $\sigma_{LF}$ &$  0.57_{-0.08}^{+0.08}$ & $  0.52_{-0.05}^{+0.05}$ & $  0.47_{-0.04}^{+0.05}$ & $  0.47_{-0.03}^{+0.04}$ & $  0.59_{-0.05}^{+0.04}$ & $  0.65_{-0.02}^{+0.02}$ & $  0.62_{-0.03}^{+0.02}$ & $  0.59_{-0.02}^{+0.02}$ & $  0.60_{-0.06}^{+0.04}$ & $  0.61_{-0.03}^{+0.03}$ \\ [3pt]

\hline 
&\multicolumn{10}{c}{Model C}\\
\hline

$\log_{10}[L_{\rm{min}}]$ &$ 18.68_{-0.09}^{+0.07}$ & $ 19.52_{-0.02}^{+0.03}$ & $ 19.89_{-0.02}^{+0.02}$ & $ 20.46_{-0.01}^{+0.01}$ & $ 21.00_{-0.01}^{+0.01}$ & $ 21.64_{-0.01}^{+0.01}$ & $ 22.02_{-0.01}^{+0.01}$ & $ 22.44_{-0.01}^{+0.01}$ & $ 22.41_{-0.02}^{+0.02}$ & $ 22.58_{-0.04}^{+0.03}$ \\ [3pt]

$\log_{10}[L_{\rm{max}}]$ &$ 29.22_{-1.36}^{+0.43}$ & $ 29.50_{-0.32}^{+0.16}$ & $ 26.73_{-0.42}^{+0.06}$ & $ 28.83_{-1.20}^{+0.63}$ & $ 29.03_{-0.42}^{+0.08}$ & $ 29.31_{-3.46}^{+0.54}$ & $ 29.17_{-2.93}^{+0.53}$ & $ 29.65_{-0.09}^{+0.12}$ & $ 26.69_{-0.12}^{+0.26}$ & $ 28.17_{-0.91}^{+1.16}$ \\ [3pt]

           $\alpha_{SF}$ &$  2.77_{-0.29}^{+0.32}$ & $  3.12_{-0.09}^{+0.09}$ & $  3.31_{-0.07}^{+0.07}$ & $  3.13_{-0.04}^{+0.04}$ & $  2.96_{-0.03}^{+0.03}$ & $  2.49_{-0.03}^{+0.03}$ & $  2.26_{-0.03}^{+0.03}$ & $  1.55_{-0.04}^{+0.04}$ & $  1.13_{-0.05}^{+0.04}$ & $  0.90_{-0.07}^{+0.06}$ \\ [3pt]
           
           $\alpha_{AGN}$ &$  6.89_{-1.90}^{+1.56}$ & $  0.68_{-0.91}^{+0.90}$ & $  1.35_{-0.76}^{+0.72}$ & $  1.33_{-0.55}^{+0.49}$ & $  0.67_{-0.44}^{+0.41}$ & $  1.39_{-0.29}^{+0.25}$ & $  1.44_{-0.18}^{+0.19}$ & $  1.15_{-0.15}^{+0.15}$ & $  1.29_{-0.08}^{+0.06}$ & $  0.04_{-0.12}^{+0.09}$ \\ [3pt]

\hline
\end{tabular}
\label{table:parameters}
\end{table}
\end{landscape}

\begin{landscape}
\begin{table}
\caption{The MAP posterior parameters of the low (using Model B and Model C) and high (using Model B) stellar mass contribution to NIR-selected RLF, in each of the redshift bins and their $\sigma$. Model B and Model C  The units of the parameters are as shown in \T{table:prior}.}
\begin{tabular}{l|rrrrrrrrrr}
\hline
    Parameter & $0.1 < z < 0.4$ & $0.4 < z < 0.6$ & $0.6 < z < 0.8$ & $0.8 < z < 1.0$ & $1.0 < z < 1.3$ & $1.3 < z < 1.6$ & $1.6 < z < 2.0$ & $2.0 < z < 2.5$ & $2.5 < z < 3.2$ & $3.2 < z < 4.0$ \\ \hline 
    &\multicolumn{10}{c}{{High stellar mass Model B fit}}\\
    \hline

$\log_{10}[L_{\rm{min_1}}]$ &$ 17.46_{-0.20}^{+0.65}$ & $ 17.44_{-0.30}^{+0.74}$ & $ 18.61_{-0.45}^{+0.61}$ & $ 18.79_{-0.50}^{+0.94}$ & $ 19.24_{-0.70}^{+1.27}$ & $ 18.97_{-0.49}^{+1.39}$ & $ 19.31_{-0.96}^{+1.30}$ & $ 18.98_{-0.66}^{+1.73}$ & $ 20.37_{-0.81}^{+0.81}$ & $ 17.46_{-1.65}^{+2.21}$ \\ [3pt]
$\log_{10}[L_{\rm{max_1}}]$ &$ 24.66_{-0.10}^{+0.09}$ & $ 25.65_{-0.53}^{+0.54}$ & $ 25.40_{-0.13}^{+0.13}$ & $ 26.49_{-0.33}^{+0.34}$ & $ 26.20_{-0.13}^{+0.13}$ & $ 26.51_{-0.32}^{+0.31}$ & $ 26.50_{-0.32}^{+0.32}$ & $ 26.96_{-0.62}^{+0.67}$ & $ 26.76_{-0.47}^{+0.48}$ & $ 28.07_{-1.33}^{+1.24}$ \\ [3pt]
$\log_{10}[L_{\rm{min_2}}]$ &$ 19.78_{-1.00}^{+0.93}$ & $ 19.50_{-0.82}^{+0.91}$ & $ 20.91_{-0.47}^{+0.46}$ & $ 20.42_{-0.40}^{+0.37}$ & $ 22.58_{-0.18}^{+0.14}$ & $ 22.56_{-0.16}^{+0.15}$ & $ 21.63_{-0.83}^{+0.76}$ & $ 21.73_{-0.87}^{+0.83}$ & $ 22.67_{-0.23}^{+0.21}$ & $ 22.91_{-0.37}^{+0.38}$ \\ [3pt]
$\log_{10}[L_{\rm{max_6}}]$ &$ 25.36_{-0.20}^{+0.10}$ & $ 26.15_{-0.34}^{+0.16}$ & $ 26.04_{-0.71}^{+0.32}$ & $ 27.06_{-0.53}^{+0.32}$ & $ 26.31_{-0.20}^{+0.14}$ & $ 26.84_{-0.19}^{+0.11}$ & $ 26.60_{-0.36}^{+0.29}$ & $ 26.26_{-0.20}^{+1.57}$ & $ 26.39_{-0.30}^{+0.31}$ & $ 28.68_{-1.18}^{+0.90}$ \\ [3pt]
    $\log_{10}[\Phi_1^*]$ &$ -4.45_{-0.30}^{+0.27}$ & $ -4.86_{-0.45}^{+0.46}$ & $ -3.75_{-0.09}^{+0.09}$ & $ -3.79_{-0.10}^{+0.10}$ & $ -4.27_{-1.28}^{+0.18}$ & $ -4.95_{-0.35}^{+0.19}$ & $ -5.37_{-0.36}^{+0.24}$ & $ -5.92_{-0.54}^{+0.39}$ & $ -5.42_{-0.66}^{+0.63}$ & $ -6.93_{-1.26}^{+1.10}$ \\ [3pt]
       $\log_{10}[L_1^*]$ &$ 23.98_{-0.33}^{+0.58}$ & $ 24.78_{-0.82}^{+0.68}$ & $ 23.34_{-0.15}^{+0.13}$ & $ 23.57_{-0.12}^{+0.15}$ & $ 23.92_{-0.24}^{+1.49}$ & $ 24.61_{-0.33}^{+0.61}$ & $ 25.04_{-0.49}^{+0.60}$ & $ 25.18_{-0.57}^{+1.17}$ & $ 24.54_{-0.83}^{+0.88}$ & $ 25.80_{-1.18}^{+1.00}$ \\ [3pt]
               $\alpha_1$ &$  1.64_{-0.87}^{+1.15}$ & $  1.63_{-0.78}^{+1.96}$ & $  1.11_{-0.16}^{+0.21}$ & $  1.13_{-0.18}^{+0.24}$ & $  1.13_{-0.21}^{+0.74}$ & $  1.02_{-0.29}^{+0.38}$ & $  1.19_{-0.56}^{+1.22}$ & $  0.85_{-0.91}^{+1.41}$ & $  0.94_{-0.48}^{+0.68}$ & $  0.71_{-0.80}^{+0.77}$ \\ [3pt]
                $\beta_1$ &$ -0.29_{-1.27}^{+0.42}$ & $  0.23_{-0.51}^{+0.15}$ & $ -2.92_{-1.43}^{+1.75}$ & $ -2.86_{-1.50}^{+1.76}$ & $ -1.44_{-2.35}^{+2.18}$ & $ -0.98_{-1.87}^{+1.33}$ & $ -1.15_{-2.20}^{+1.17}$ & $ -1.52_{-2.10}^{+1.26}$ & $ -1.80_{-2.01}^{+1.60}$ & $ -1.40_{-1.03}^{+1.09}$ \\ [3pt]
    $\log_{10}[\Phi_2^*]$ &$ -2.71_{-0.02}^{+0.02}$ & $ -2.61_{-0.02}^{+0.03}$ & $ -2.54_{-0.04}^{+0.04}$ & $ -2.47_{-0.04}^{+0.05}$ & $ -2.76_{-0.04}^{+0.04}$ & $ -3.73_{-0.29}^{+0.33}$ & $ -3.56_{-0.37}^{+0.23}$ & $ -4.30_{-0.44}^{+0.43}$ & $ -3.75_{-0.07}^{+0.05}$ & $ -4.99_{-0.60}^{+0.73}$ \\ [3pt]
       $\log_{10}[L_2^*]$ &$ 21.40_{-0.09}^{+0.11}$ & $ 22.19_{-0.16}^{+0.18}$ & $ 23.02_{-0.22}^{+0.15}$ & $ 22.67_{-0.21}^{+0.21}$ & $ 22.56_{-0.15}^{+0.21}$ & $ 21.55_{-0.18}^{+0.27}$ & $ 21.96_{-0.32}^{+0.32}$ & $ 21.94_{-0.27}^{+0.38}$ & $ 23.29_{-0.44}^{+0.43}$ & $ 21.49_{-0.33}^{+0.55}$ \\ [3pt]
               $\alpha_2$ &$  0.67_{-0.04}^{+0.04}$ & $  0.73_{-0.12}^{+0.04}$ & $  0.76_{-0.03}^{+0.04}$ & $  0.56_{-0.05}^{+0.06}$ & $  0.46_{-0.07}^{+0.11}$ & $ -0.49_{-0.24}^{+0.32}$ & $ -0.11_{-0.35}^{+0.27}$ & $ -0.50_{-0.35}^{+0.42}$ & $  0.59_{-0.18}^{+0.11}$ & $ -0.64_{-0.39}^{+0.62}$ \\ [3pt]
            $\sigma_{LF}$ &$  0.57_{-0.04}^{+0.02}$ & $  0.36_{-0.07}^{+0.06}$ & $  0.15_{-0.03}^{+0.06}$ & $  0.28_{-0.06}^{+0.06}$ & $  0.37_{-0.06}^{+0.04}$ & $  0.55_{-0.02}^{+0.02}$ & $  0.57_{-0.03}^{+0.02}$ & $  0.55_{-0.02}^{+0.02}$ & $  0.44_{-0.14}^{+0.11}$ & $  0.61_{-0.03}^{+0.03}$ \\ [3pt]

\hline
    &\multicolumn{10}{c}{{low stellar mass Model $B'$ fit}}\\
    \hline

 $\log_{10}[L_{\rm{min}}]$ &$ 18.63_{-0.56}^{+0.10}$ & $ 18.47_{-0.93}^{+0.95}$ & $ 18.84_{-0.73}^{+0.39}$ & $ 18.77_{-0.47}^{+0.45}$ & $ 21.24_{-0.12}^{+0.06}$ & $ 19.30_{-0.21}^{+0.91}$ & $ 20.32_{-0.88}^{+0.90}$ & $ 20.39_{-0.92}^{+0.92}$ & $ 19.89_{-1.26}^{+1.34}$ & $ 19.89_{-1.26}^{+1.34}$ \\ [3pt]
  $\log_{10}[L_{\rm{max}]}$ &$ 25.44_{-0.38}^{+0.04}$ & $ 25.41_{-0.25}^{+0.25}$ & $ 26.24_{-0.19}^{+0.20}$ & $ 26.70_{-0.40}^{+0.22}$ & $ 26.91_{-0.57}^{+0.71}$ & $ 26.72_{-0.08}^{+0.22}$ & $ 26.77_{-0.49}^{+0.47}$ & $ 27.00_{-0.66}^{+0.66}$ & $ 27.26_{-0.16}^{+0.15}$ & $ 27.26_{-0.16}^{+0.15}$ \\ [3pt]
    $\log_{10}[\Phi_1^*]$ &$ -2.26_{-0.09}^{+0.09}$ & $ -2.25_{-0.13}^{+0.05}$ & $ -2.99_{-0.66}^{+0.52}$ & $ -2.83_{-0.54}^{+0.39}$ & $ -3.05_{-0.53}^{+0.52}$ & $ -3.31_{-0.31}^{+0.18}$ & $ -3.46_{-0.50}^{+0.30}$ & $ -4.26_{-0.91}^{+0.38}$ & $ -4.27_{-0.72}^{+0.35}$ & $ -4.27_{-0.72}^{+0.35}$ \\ [3pt]
       $\log_{10}[L_1^*]$ &$ 21.27_{-0.20}^{+0.16}$ & $ 20.95_{-0.65}^{+0.32}$ & $ 19.77_{-0.42}^{+0.52}$ & $ 20.38_{-0.35}^{+0.42}$ & $ 22.50_{-0.74}^{+0.39}$ & $ 21.01_{-0.27}^{+0.25}$ & $ 21.47_{-0.41}^{+0.48}$ & $ 21.65_{-0.65}^{+0.62}$ & $ 21.54_{-0.61}^{+0.77}$ & $ 21.54_{-0.61}^{+0.77}$ \\ [3pt]
               $\alpha_1$ &$  1.23_{-0.12}^{+0.05}$ & $  0.77_{-0.31}^{+0.27}$ & $ -0.31_{-0.44}^{+0.49}$ & $ -0.32_{-0.45}^{+0.45}$ & $  1.99_{-0.49}^{+0.19}$ & $ -0.06_{-0.28}^{+0.22}$ & $ -0.04_{-0.50}^{+0.58}$ & $ -0.12_{-0.80}^{+0.66}$ & $  0.05_{-0.70}^{+0.67}$ & $  0.05_{-0.70}^{+0.67}$ \\ [3pt]
            $\sigma_{LF}$ &$  0.51_{-0.05}^{+0.06}$ & $  0.55_{-0.06}^{+0.08}$ & $  0.65_{-0.02}^{+0.02}$ & $  0.57_{-0.02}^{+0.02}$ & $  0.61_{-0.07}^{+0.08}$ & $  0.61_{-0.02}^{+0.02}$ & $  0.56_{-0.03}^{+0.02}$ & $  0.57_{-0.04}^{+0.04}$ & $  0.62_{-0.04}^{+0.04}$ & $  0.62_{-0.04}^{+0.04}$ \\ [3pt]

\hline
 &\multicolumn{10}{c}{{Low stellar mass Model C fit}}\\
    \hline
$\log_{10}[L_{\rm{min}}]$ &$ 18.08_{-0.03}^{+0.05}$ & $ 19.31_{-0.02}^{+0.02}$ & $ 19.73_{-0.02}^{+0.02}$ & $ 20.12_{-0.02}^{+0.02}$ & $ 20.85_{-0.01}^{+0.01}$ & $ 21.47_{-0.02}^{+0.02}$ & $ 21.85_{-0.02}^{+0.02}$ & $ 22.18_{-0.05}^{+0.05}$ & $ 22.18_{-0.07}^{+0.05}$ & $ 22.38_{-0.14}^{+0.09}$ \\ [3pt]
$\log_{10}[L_{\rm{max}}]$ &$ 26.54_{-0.68}^{+0.39}$ & $ 26.25_{-0.49}^{+0.52}$ & $ 26.69_{-0.83}^{+0.20}$ & $ 26.45_{-0.66}^{+0.53}$ & $ 26.25_{-0.52}^{+0.50}$ & $ 26.25_{-0.52}^{+0.50}$ & $ 26.27_{-0.51}^{+0.50}$ & $ 26.50_{-0.34}^{+0.33}$ & $ 26.50_{-0.35}^{+0.34}$ & $ 26.24_{-0.51}^{+0.52}$ \\ [3pt]

$\alpha_{SF}$ &$ -0.15_{-0.19}^{+0.18}$ & $  0.05_{-0.12}^{+0.12}$ & $  1.29_{-0.07}^{+0.07}$ & $  1.59_{-0.06}^{+0.06}$ & $  1.50_{-0.04}^{+0.04}$ & $  1.16_{-0.06}^{+0.05}$ & $  1.13_{-0.05}^{+0.06}$ & $  0.39_{-0.13}^{+0.12}$ & $  0.44_{-0.13}^{+0.10}$ & $  0.27_{-0.26}^{+0.17}$ \\ [3pt]
           $\alpha_{AGN}$ &$ -7.87_{-1.54}^{+2.35}$ & $ -8.39_{-1.13}^{+2.06}$ & $ -7.25_{-1.84}^{+1.85}$ & $ -6.84_{-1.97}^{+1.63}$ & $ -7.06_{-1.98}^{+2.14}$ & $ -1.23_{-0.39}^{+0.34}$ & $ -1.40_{-0.50}^{+0.40}$ & $ -1.06_{-0.36}^{+0.29}$ & $ -1.32_{-0.33}^{+0.28}$ & $ -1.67_{-0.96}^{+0.44}$ \\ [3pt]

\hline
\end{tabular}
\label{table:parameters2}
\end{table}
\end{landscape}

\begin{figure*}
    \includegraphics[width=1.1\textwidth,height=1\textwidth]{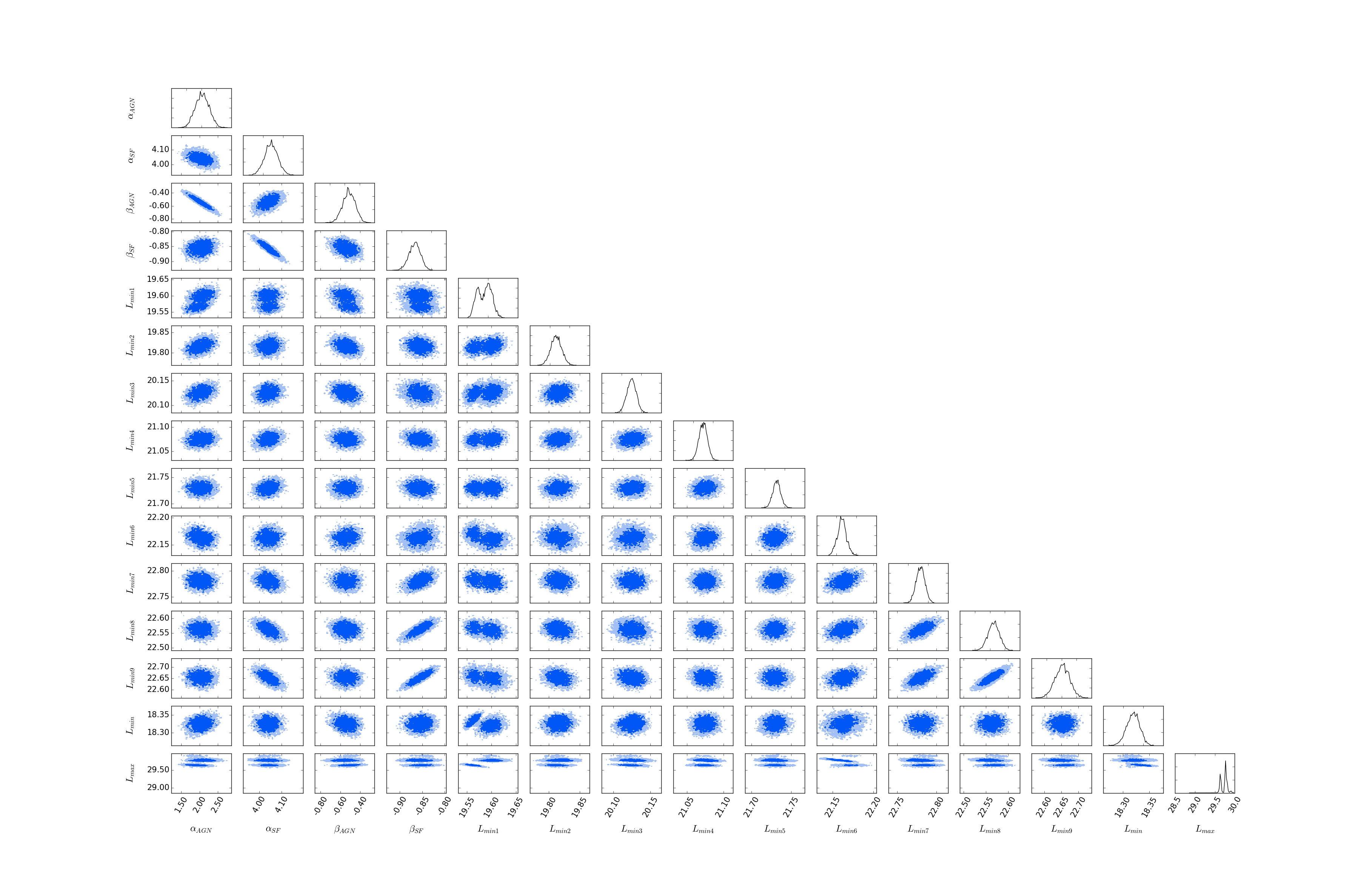}
    \caption{The triangle plot of Model C pure luminosity evolution to all the redshift bins.}
    \label{fig:perfect}
\end{figure*}

\begin{figure*}
\centering
\subfloat[$0.1 <z< 0.4$]{\includegraphics[width=0.85\textwidth]{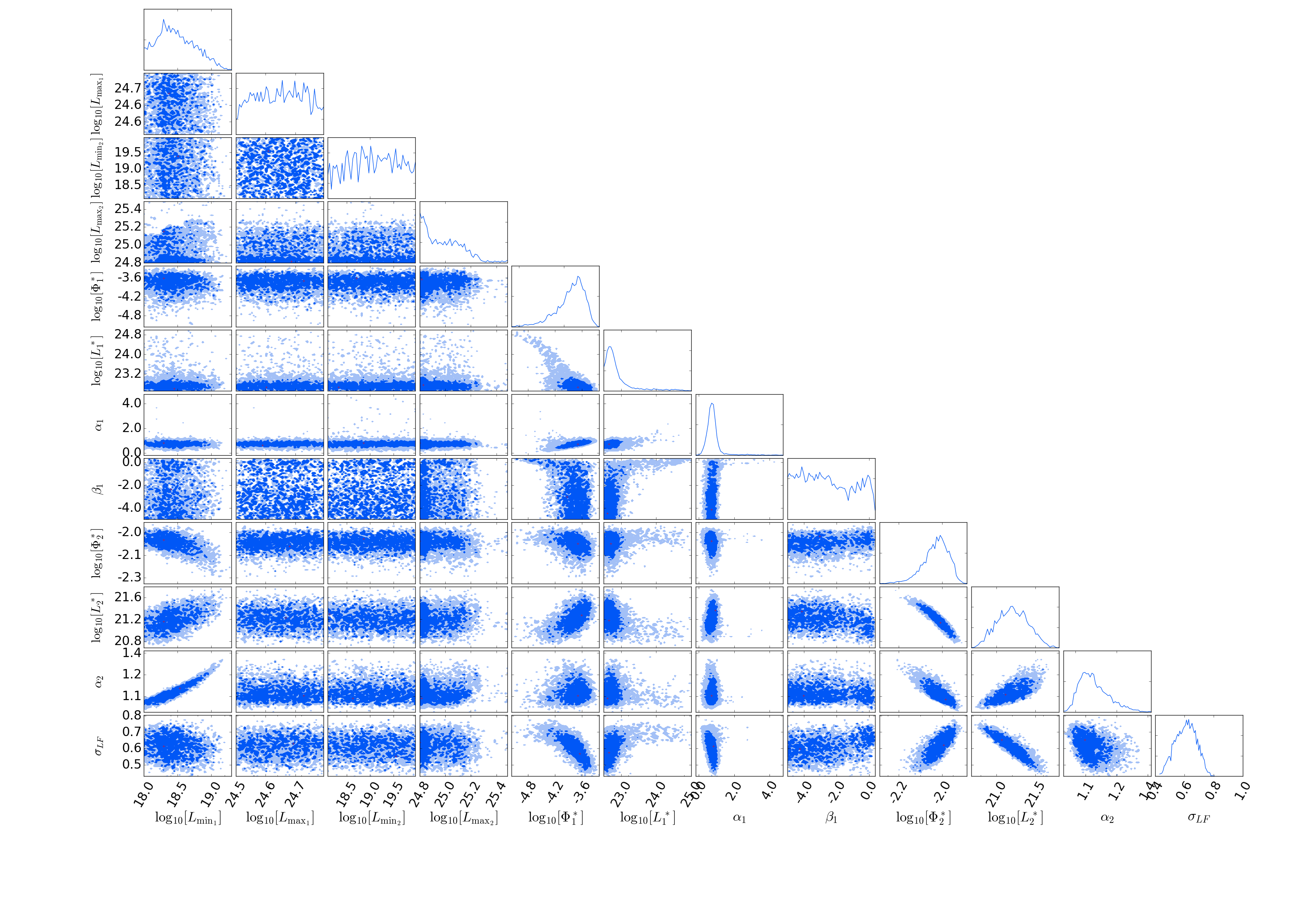}}\\
\subfloat[$0.4 <z< 0.6$]{\includegraphics[width=0.85\textwidth]{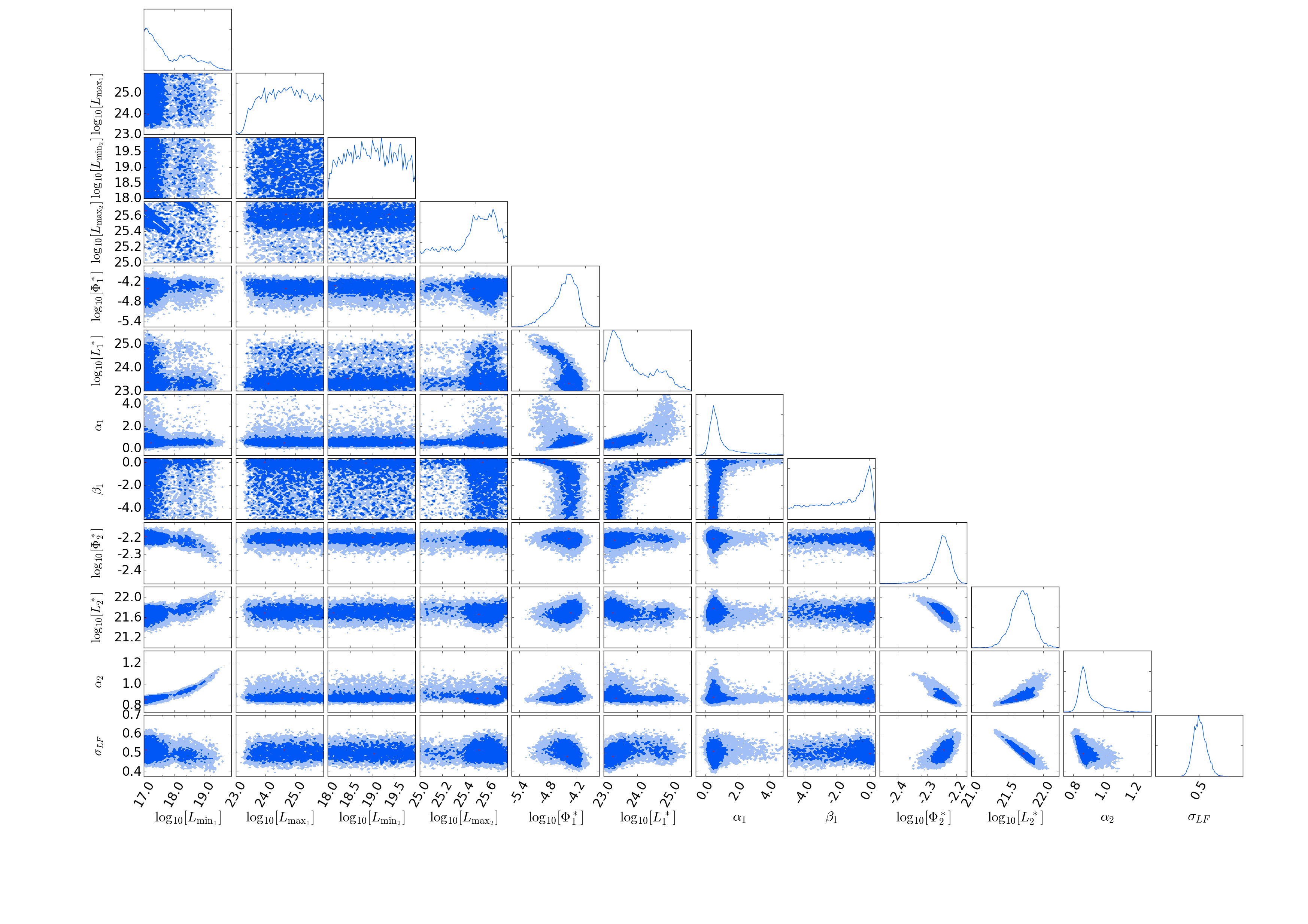}}
\caption{The triangle plots for model B fit to the the individual redshift bins. }
\label{fig:Model_B}
\end{figure*}

\begin{figure*}
\centering
\subfloat[$0.6 <z< 0.8$]{\includegraphics[width=0.85\textwidth]{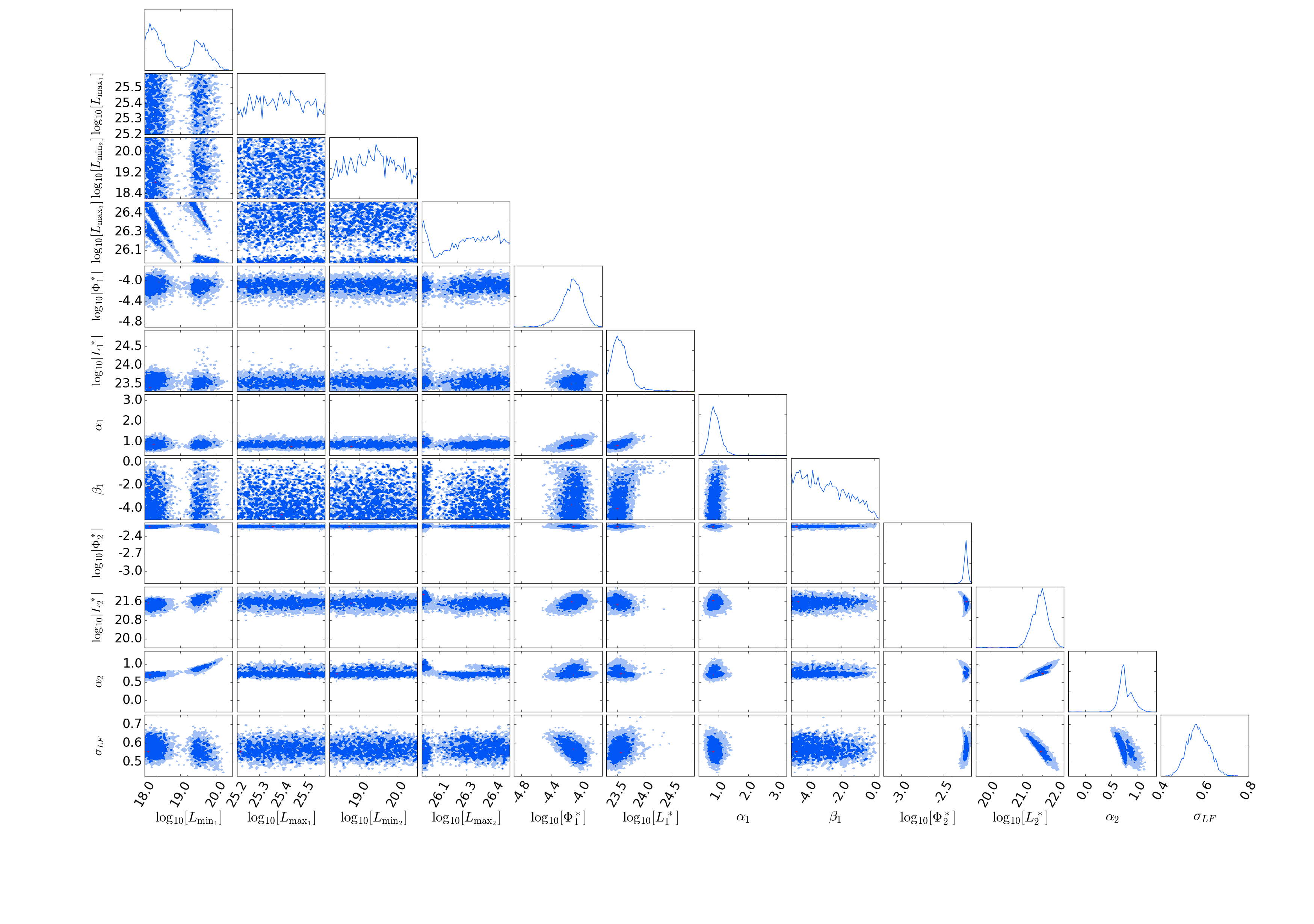}}\\
\subfloat[$0.8 <z< 1.0$]{\includegraphics[width=0.85\textwidth]{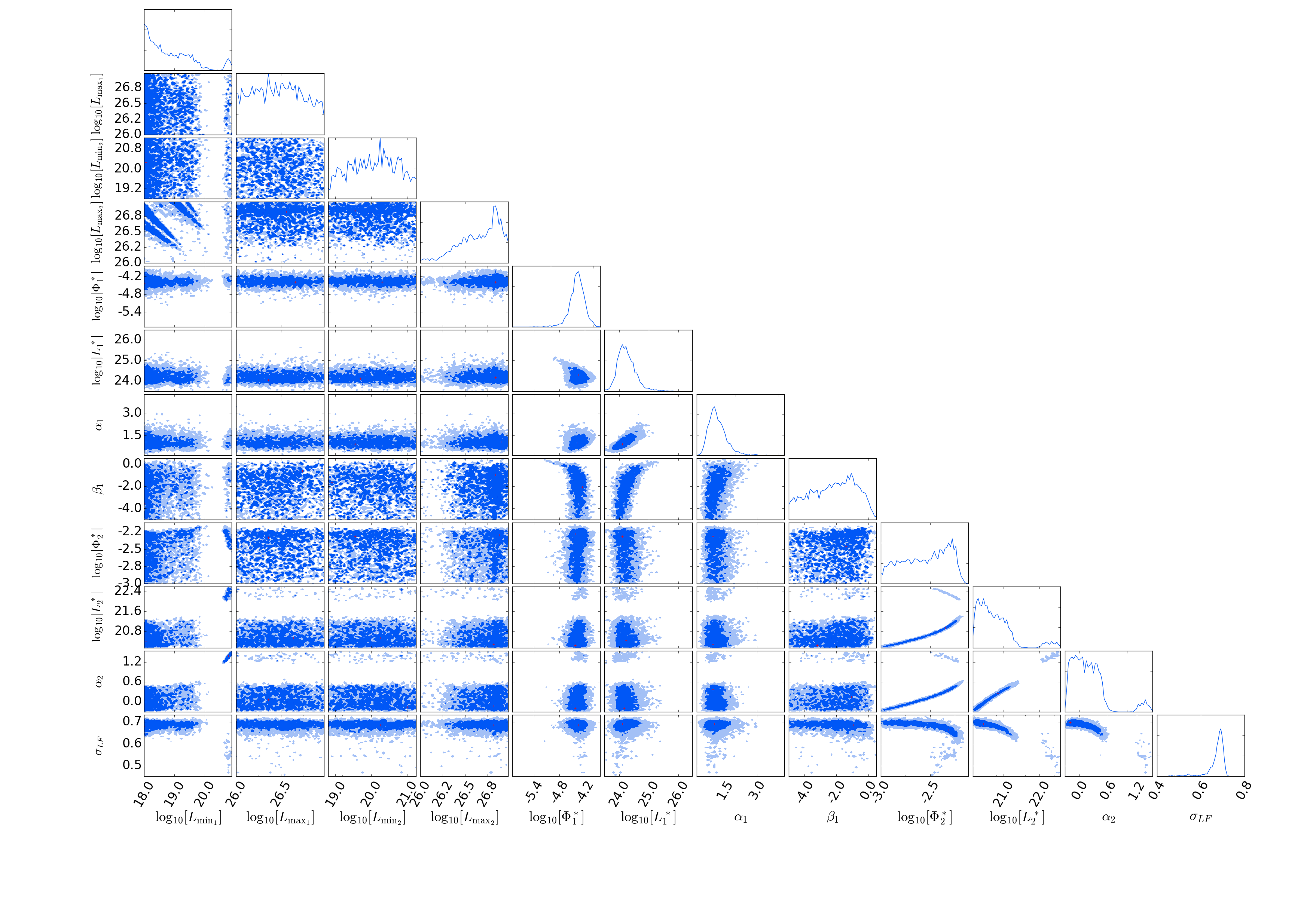}}
\caption{Continued.}
\label{fig:Model_B2}
\end{figure*}

\begin{figure*}
\centering
\subfloat[$1.0 <z< 1.3$]{\includegraphics[width=0.85\textwidth]{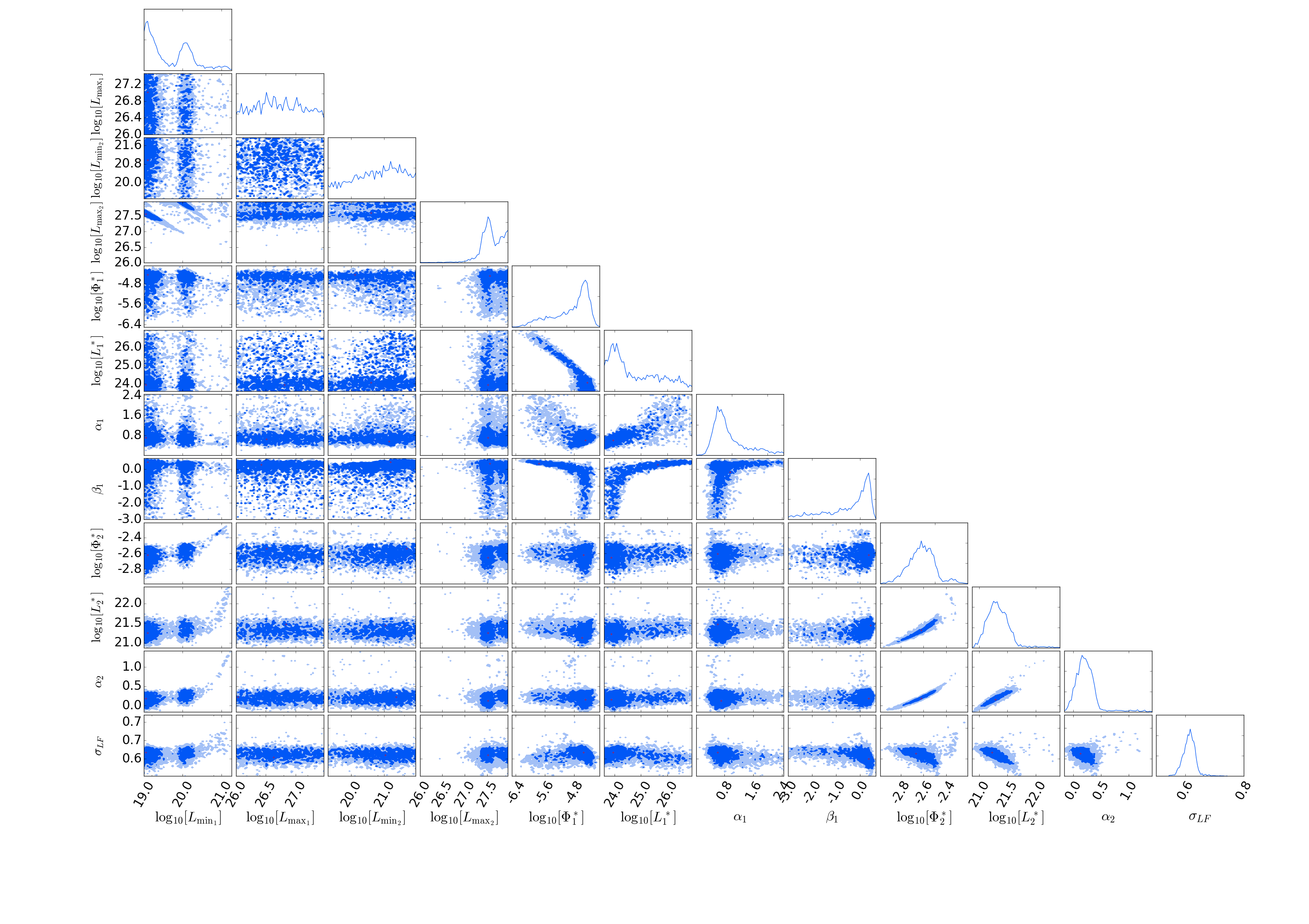}}\\
\subfloat[$1.3 <z< 1.6$]{\includegraphics[width=0.85\textwidth]{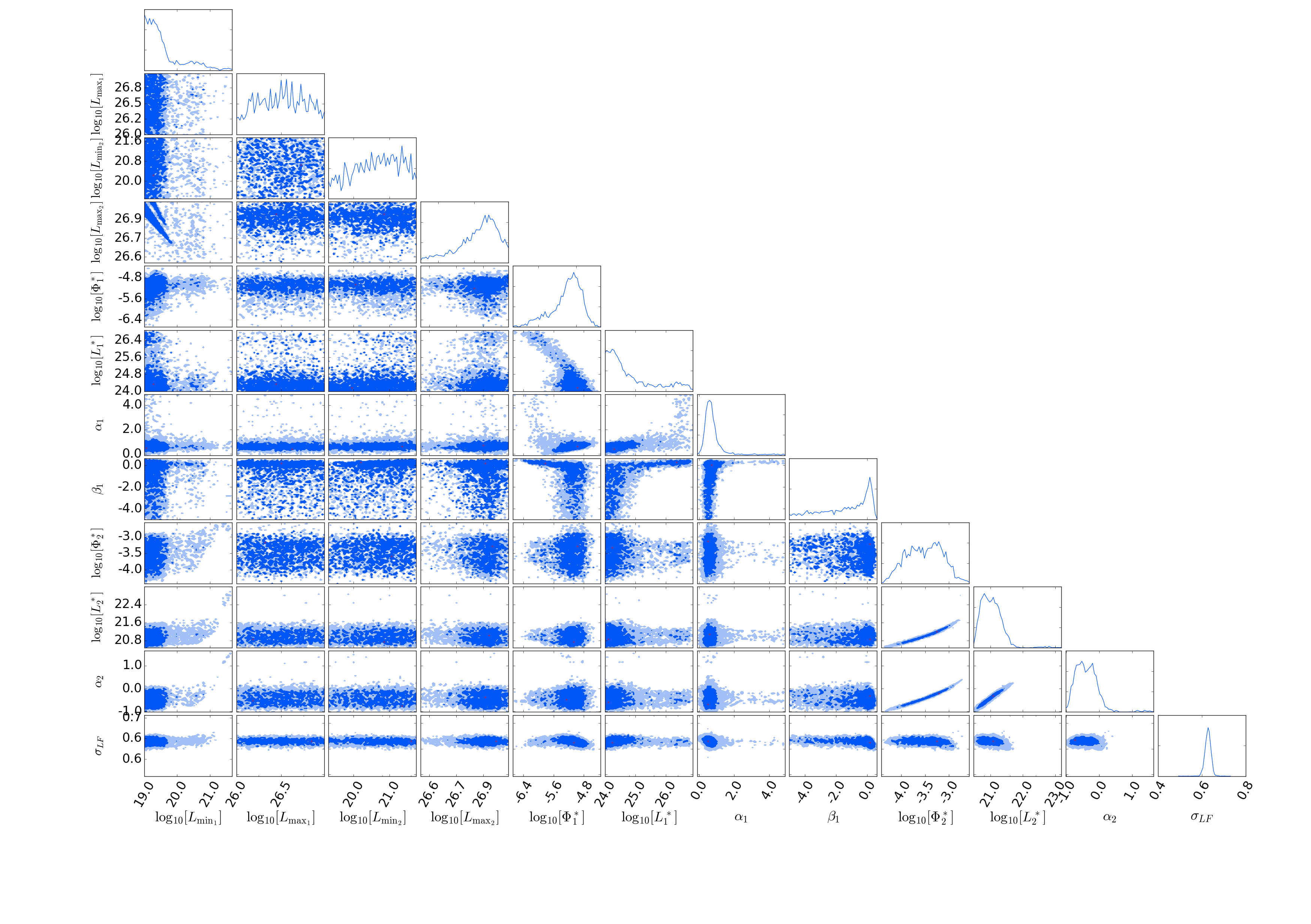}}
\caption{Continued.}
\label{fig:Model_B3}
\end{figure*}

\begin{figure*}
\centering
\subfloat[$1.6 <z< 2.0$]{\includegraphics[width=0.85\textwidth]{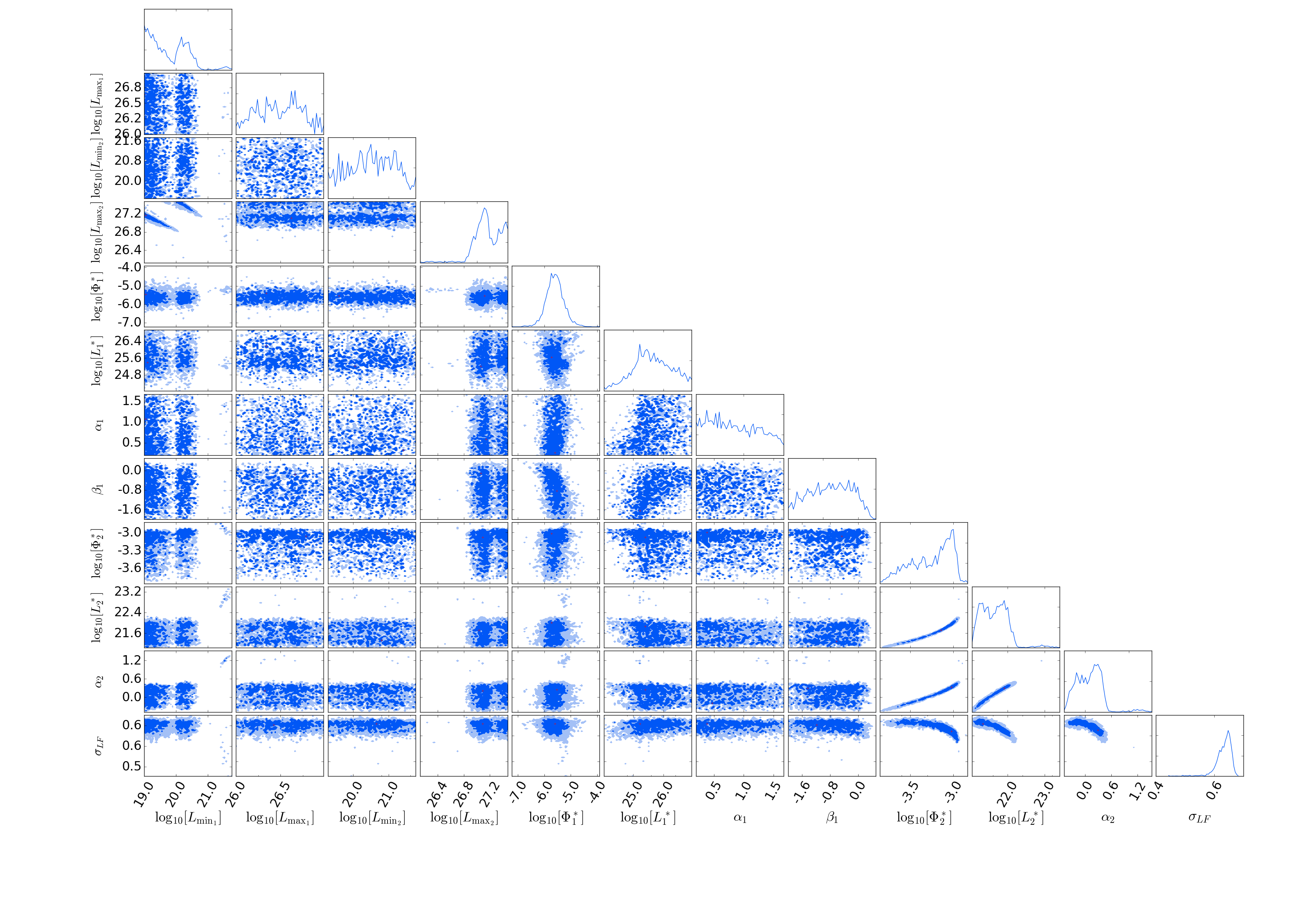}}\\
\subfloat[$2.0 <z< 2.$]{\includegraphics[width=0.85\textwidth]{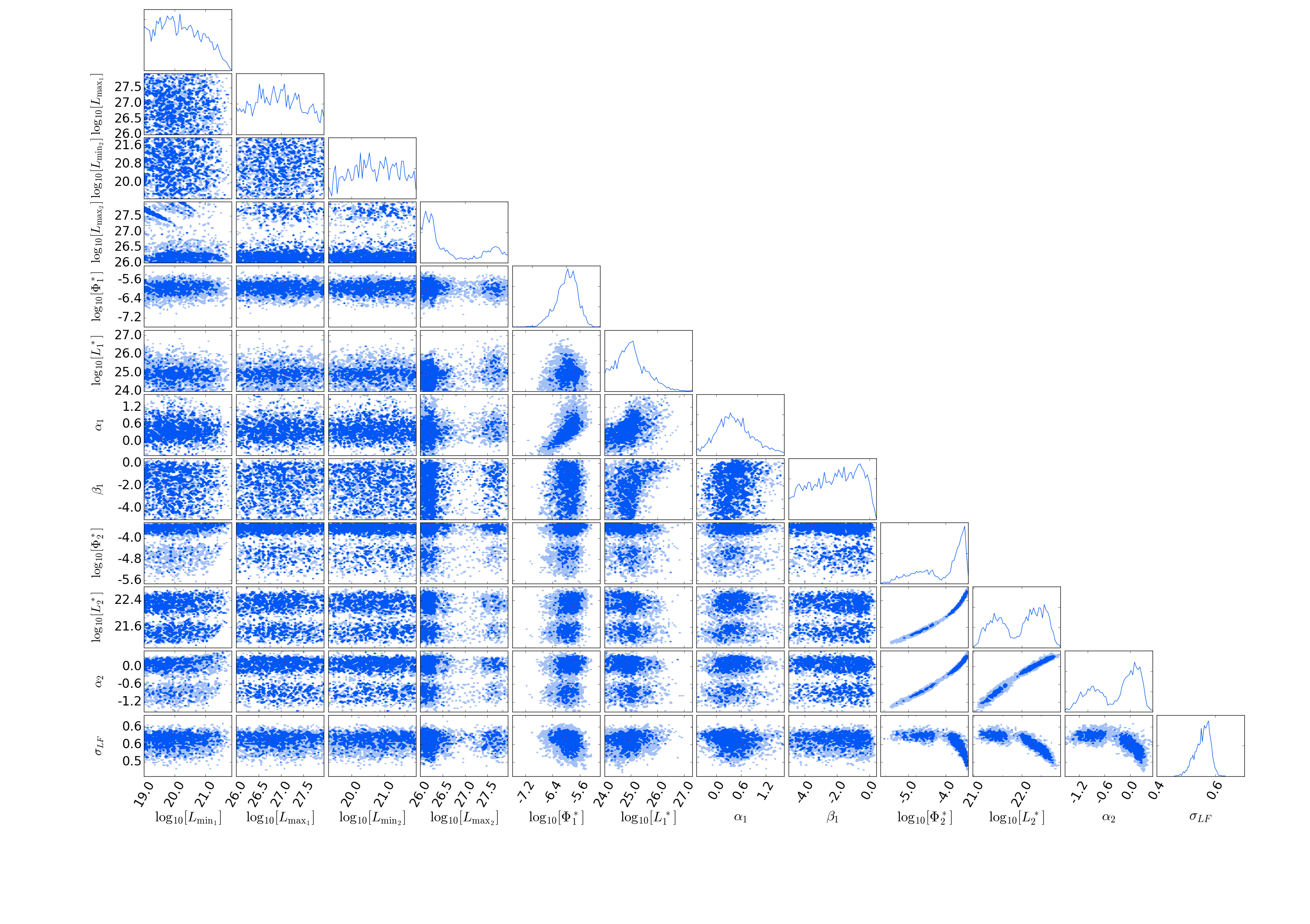}}
\caption{Continued.}
\label{fig:Model_B4}
\end{figure*}

\begin{figure*}
\centering
\subfloat[$2.5 <z< 3.2$]{\includegraphics[width=0.85\textwidth]{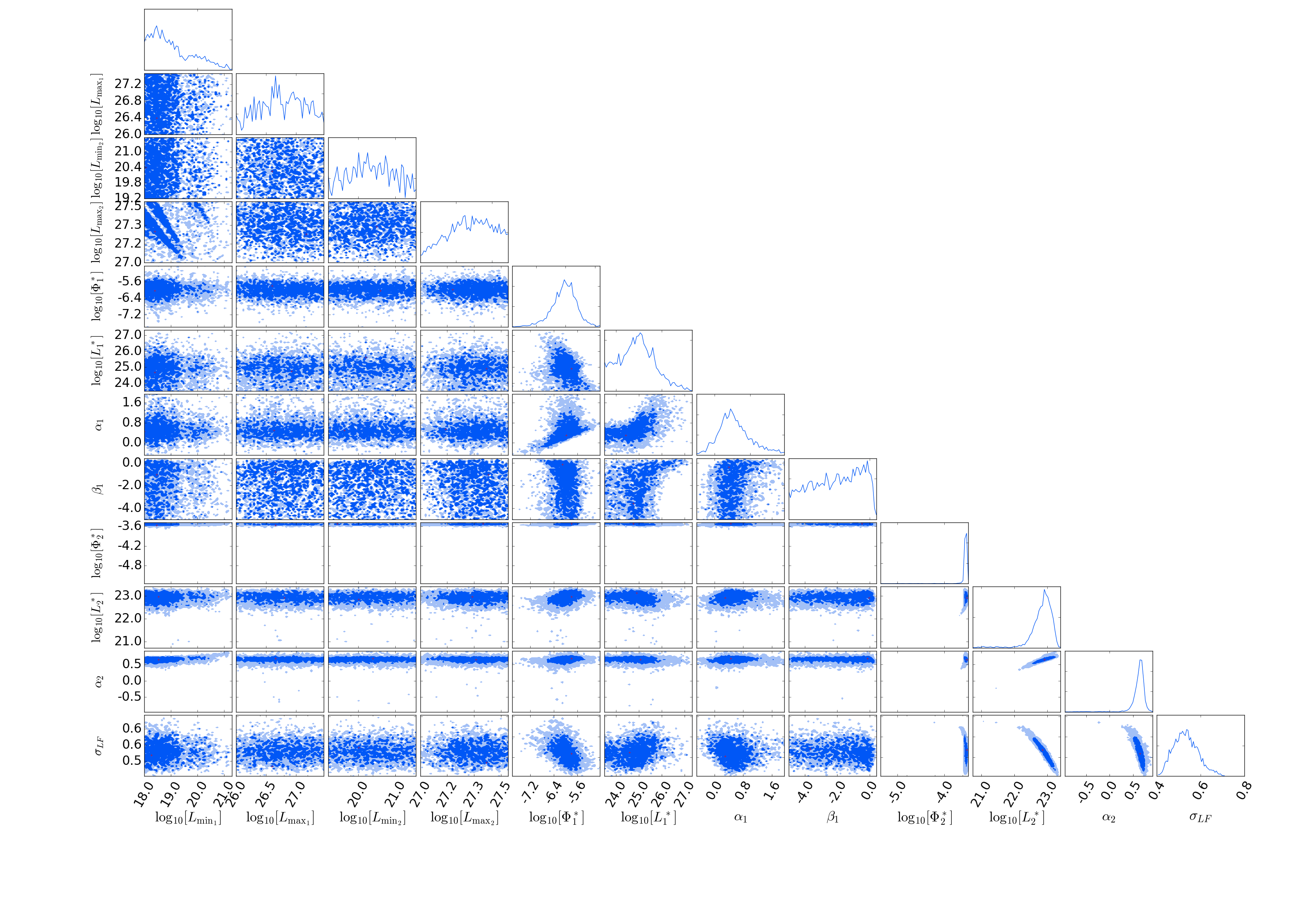}}\\
\subfloat[$3.2 <z< 4.0$]{\includegraphics[width=0.85\textwidth]{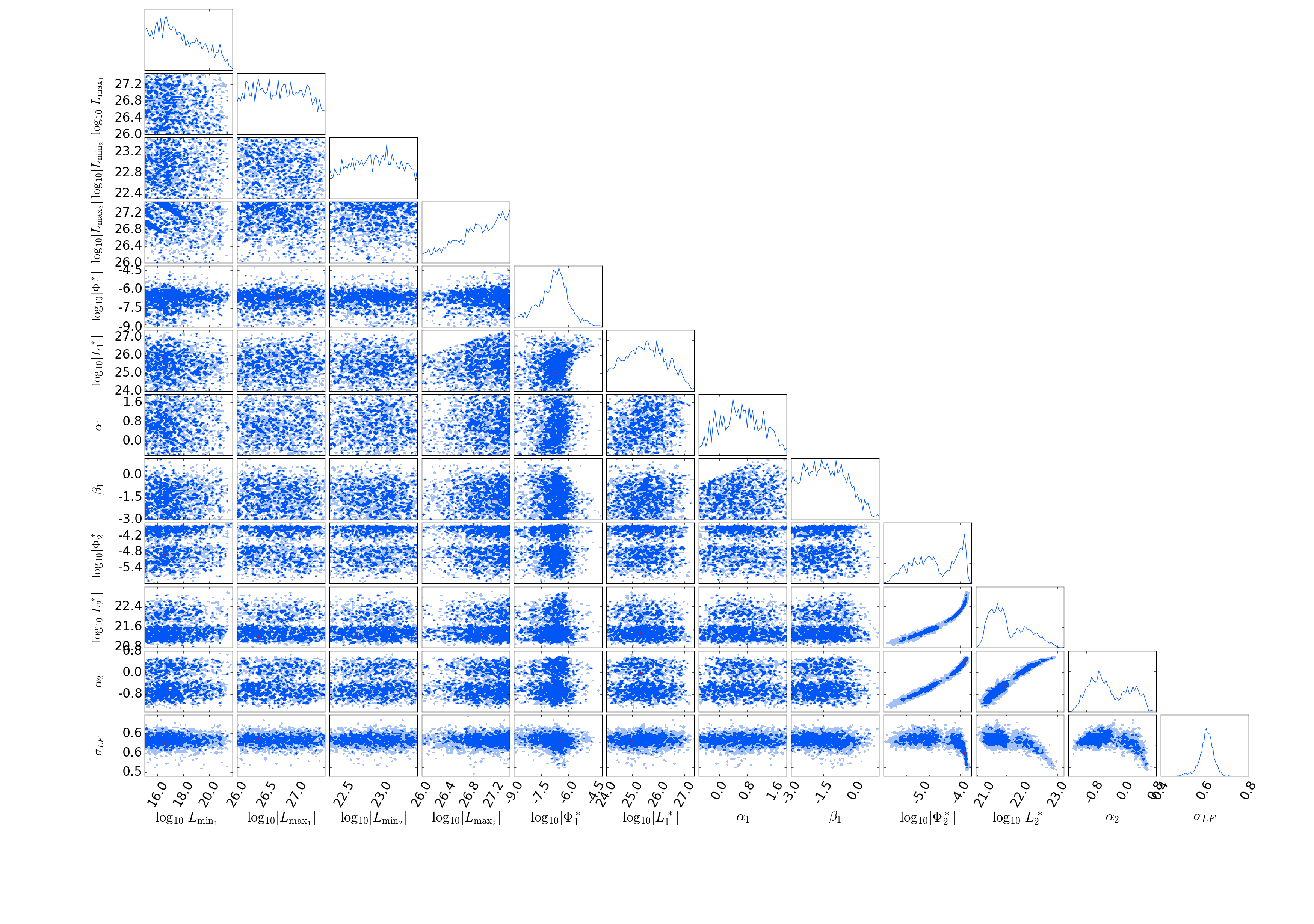}}
\caption{Continued.}
\label{fig:Model_B5}
\end{figure*}

\begin{figure*}
\centering
\subfloat[$0.1 <z< 0.4$]{\includegraphics[width=0.5\textwidth]{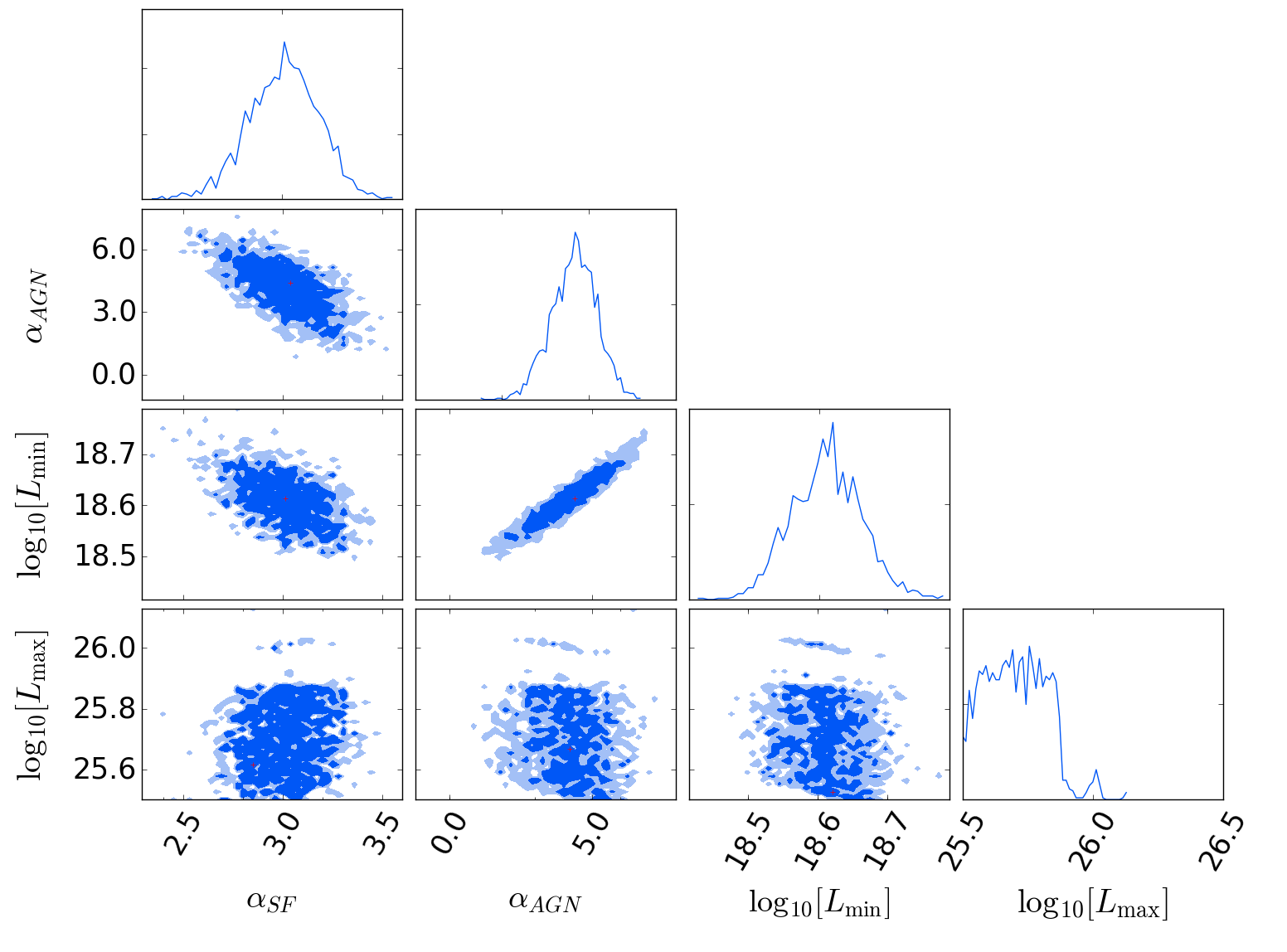}}
\subfloat[$0.4 <z< 0.6$]{\includegraphics[width=0.5\textwidth]{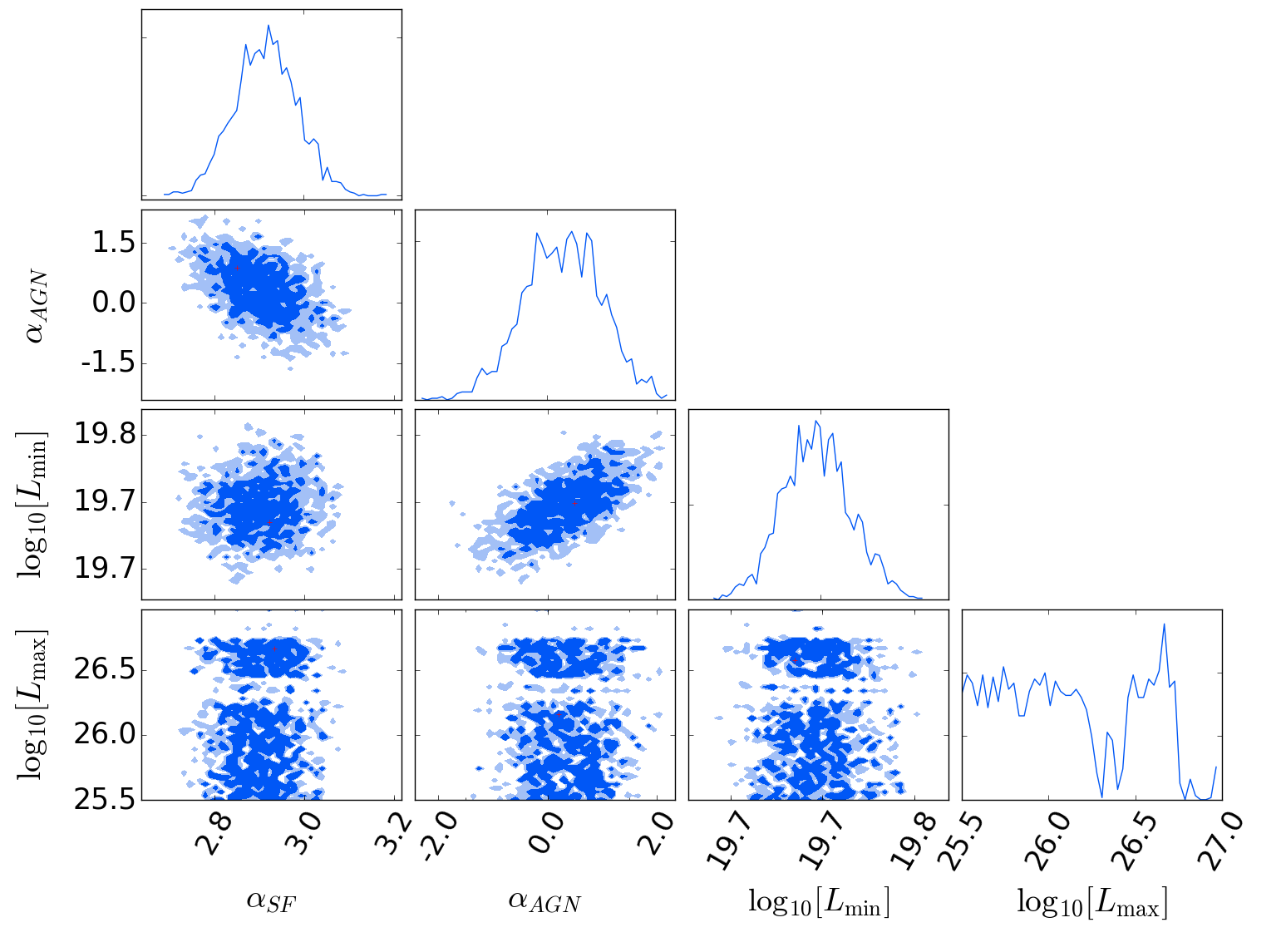}}\\
\subfloat[$0.6 <z< 0.8$]{\includegraphics[width=0.5\textwidth]{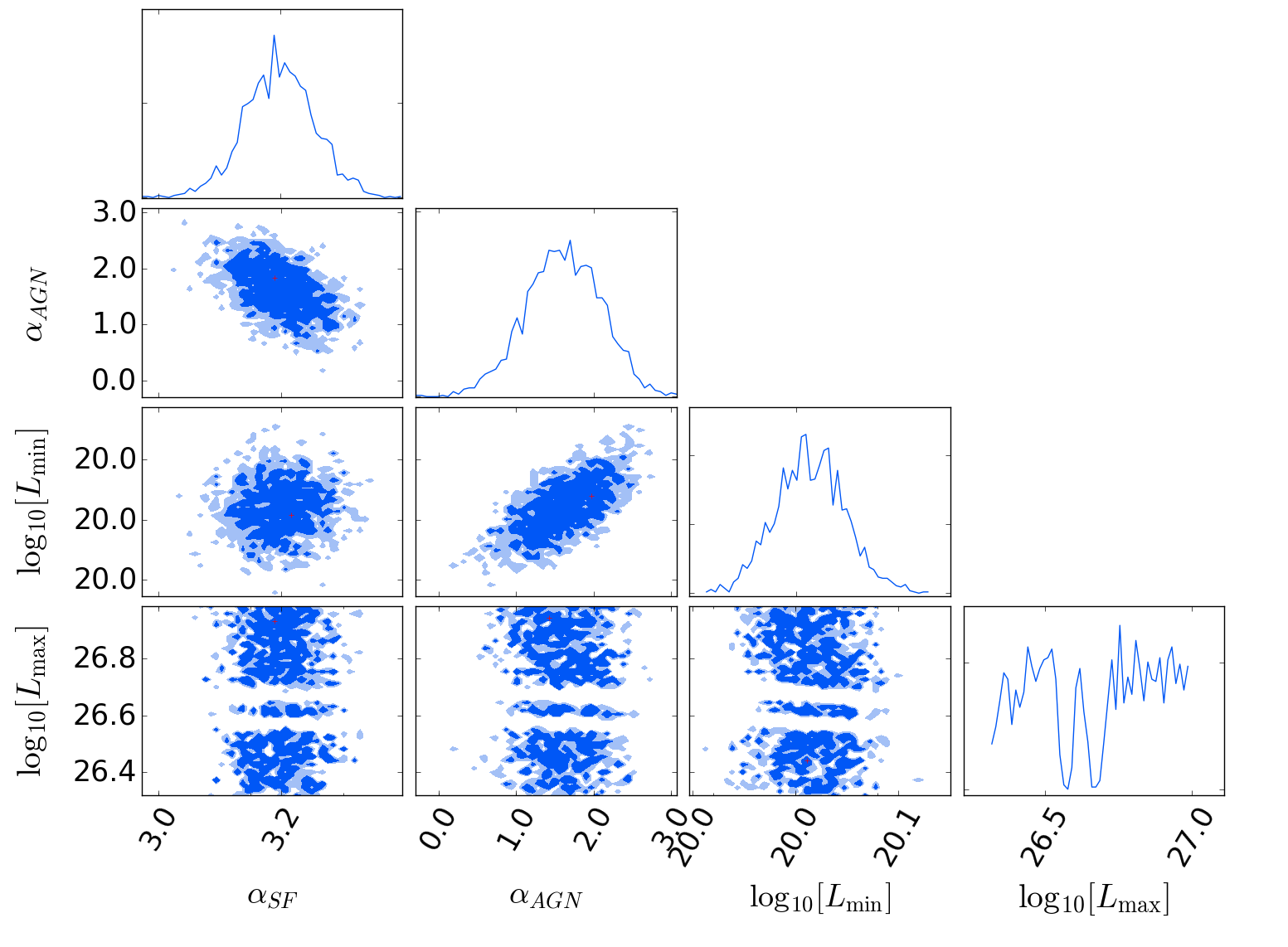}}
\subfloat[$0.8 <z< 1.0$]{\includegraphics[width=0.5\textwidth]{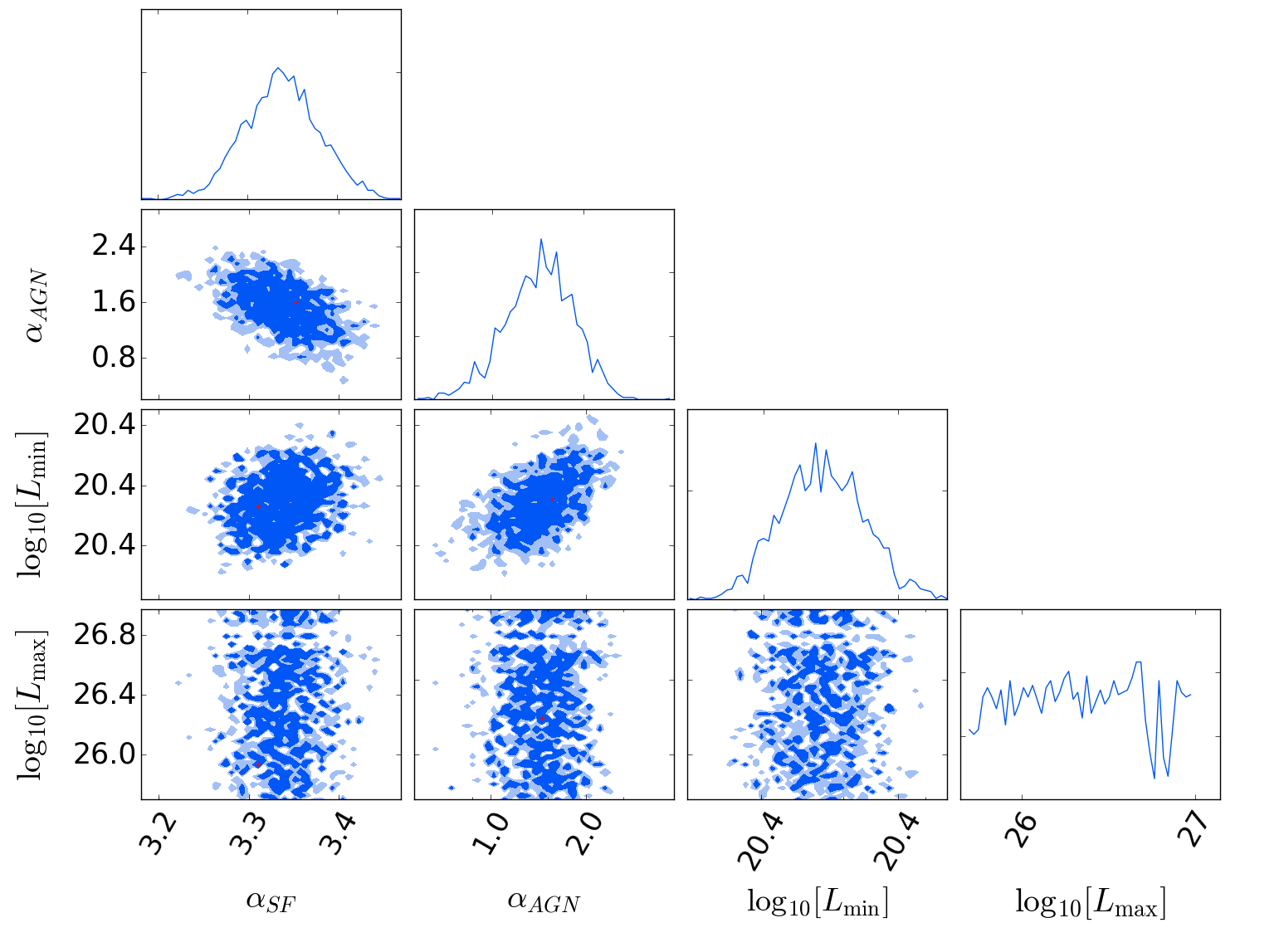}}\\
\subfloat[$1.0 <z< 1.3$]{\includegraphics[width=0.5\textwidth]{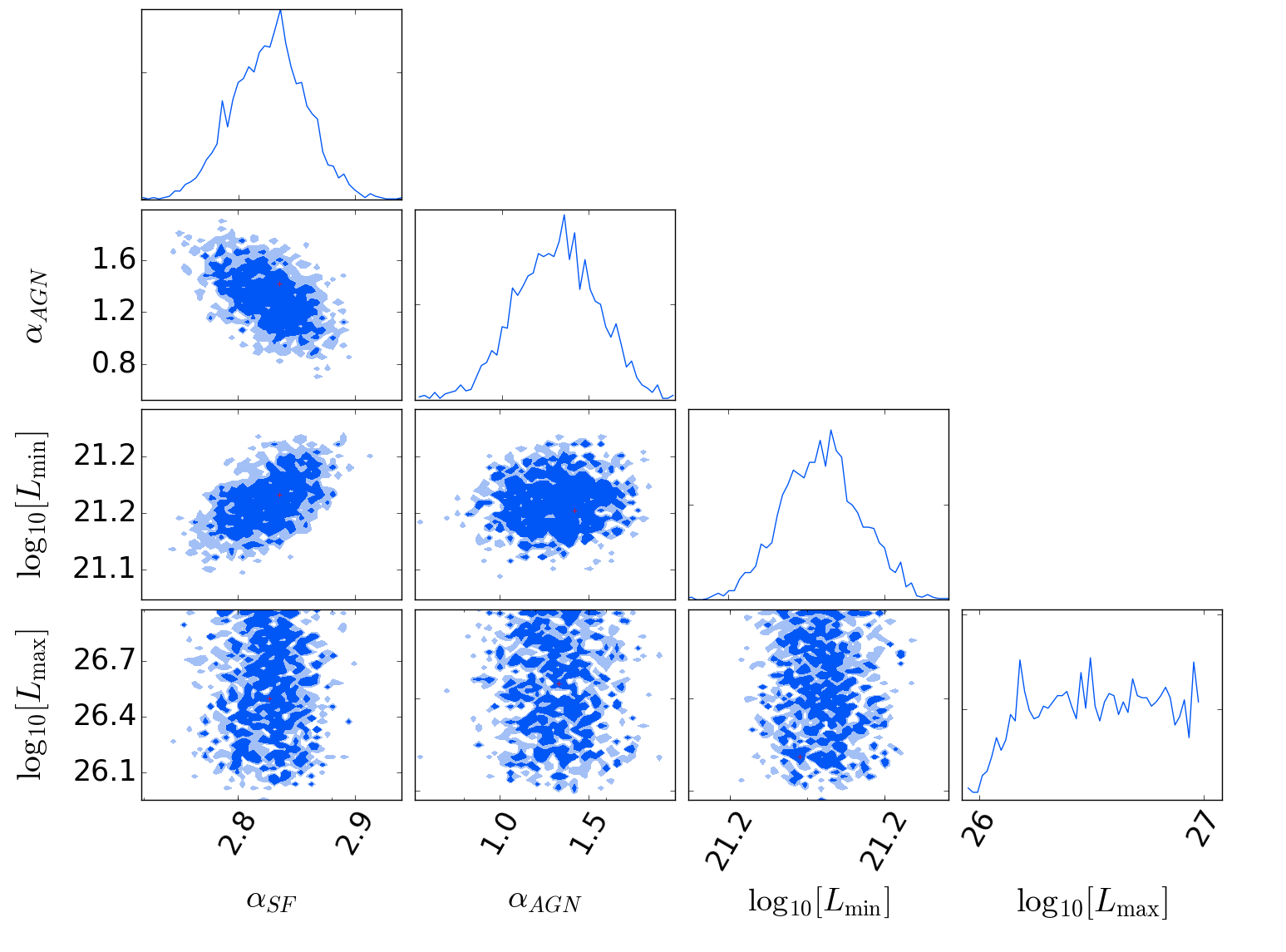}}
\subfloat[$1.3 <z< 1.6$]{\includegraphics[width=0.5\textwidth]{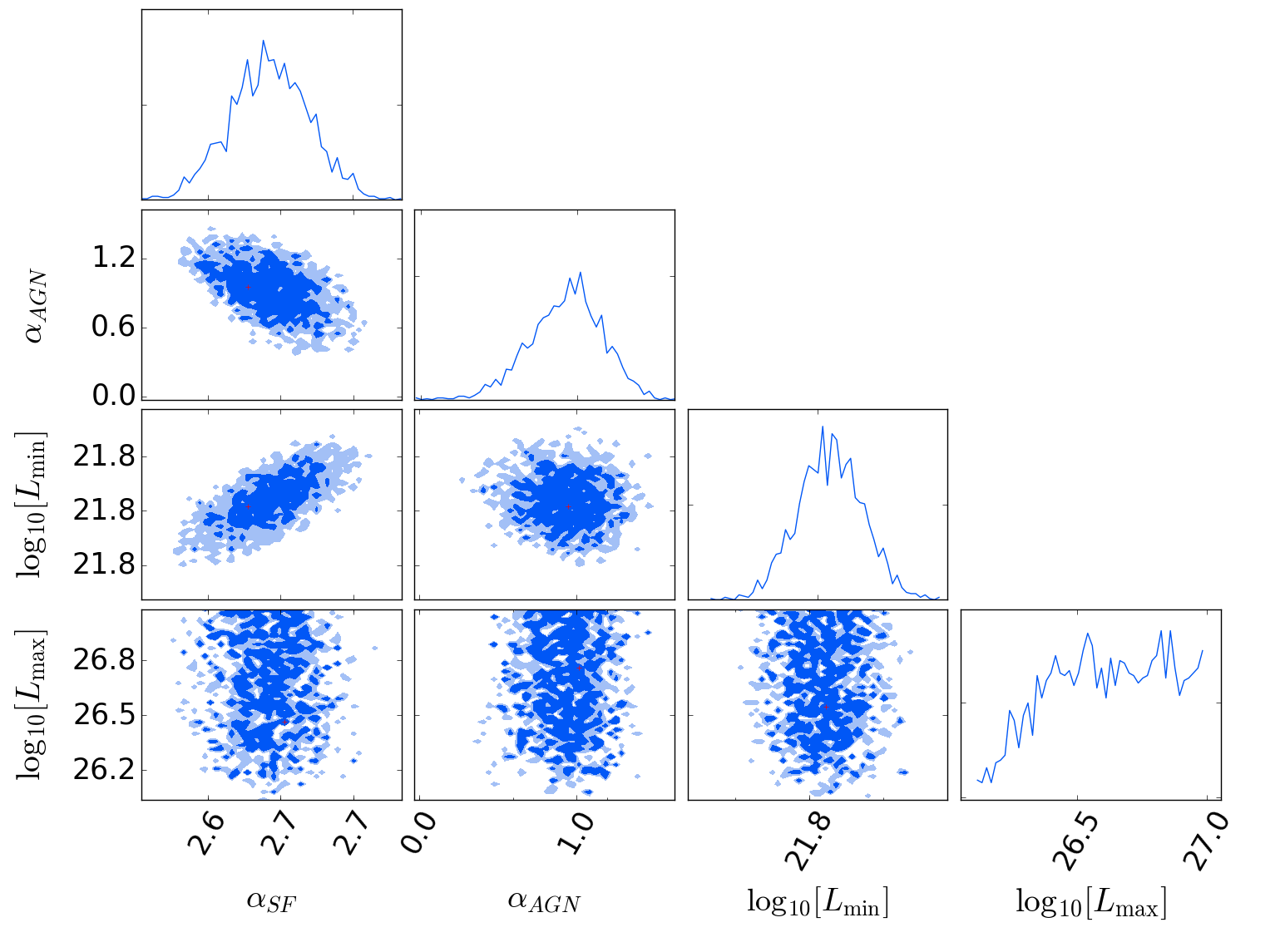}}
\caption{The triangle plots for model C fit to the the individual redshift bins. }
\label{fig:Model_C}
\end{figure*}

\begin{figure*}
\centering
\subfloat[$1.6 <z< 2.0$]{\includegraphics[width=0.5\textwidth]{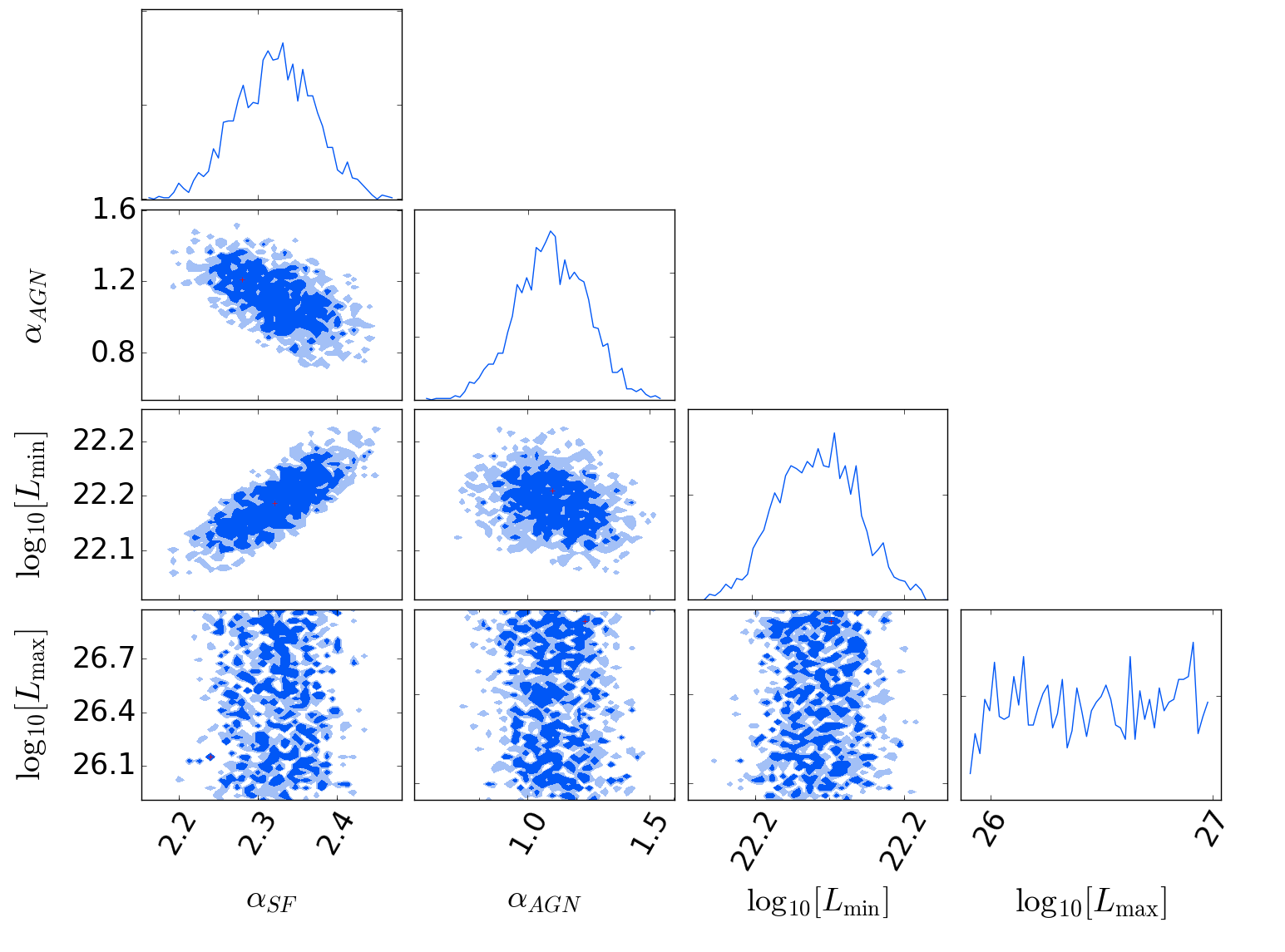}}
\subfloat[$2.0 <z< 2.$]{\includegraphics[width=0.5\textwidth]{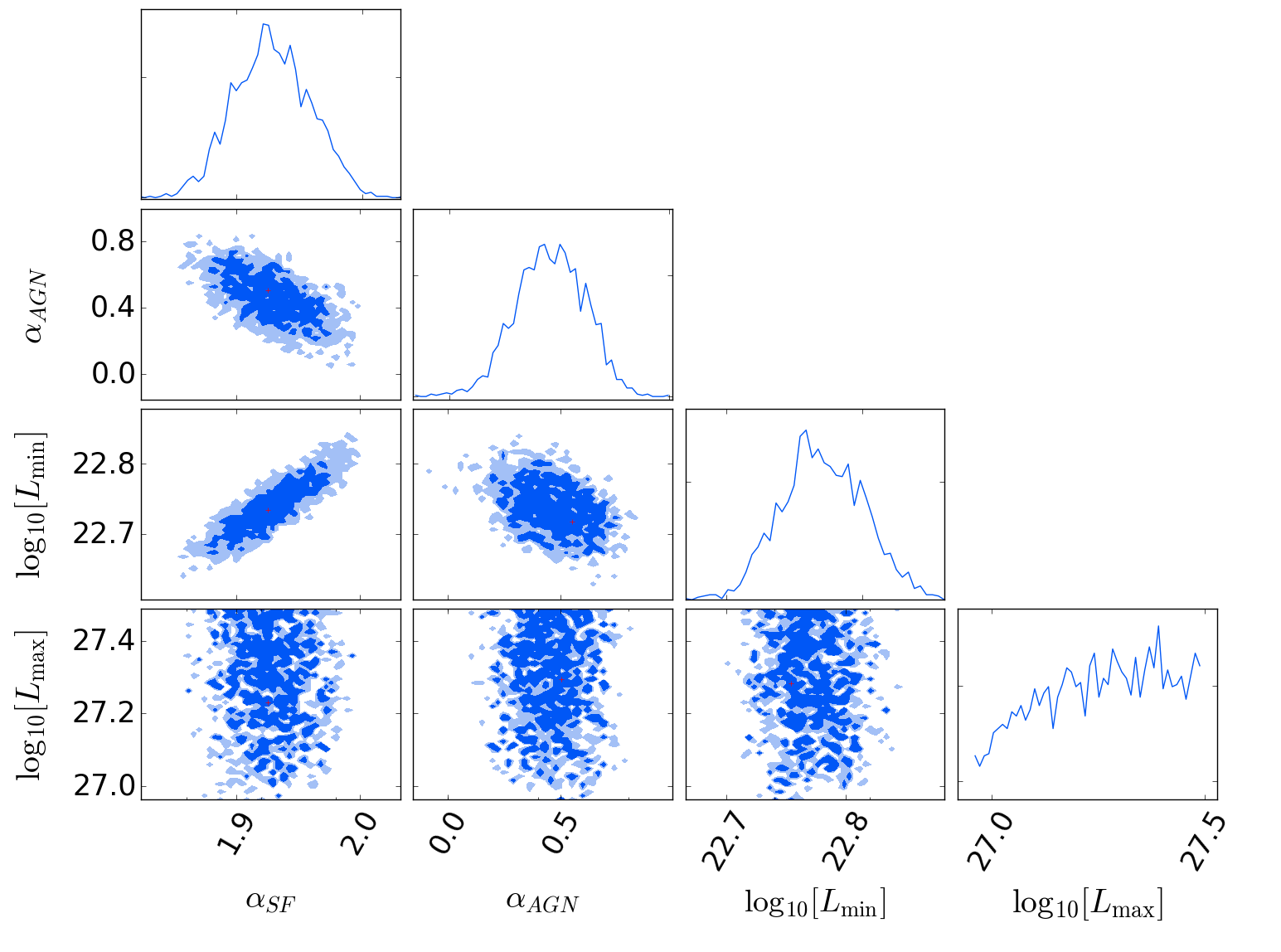}}\\
\subfloat[$2.5 <z< 3.2$]{\includegraphics[width=0.5\textwidth]{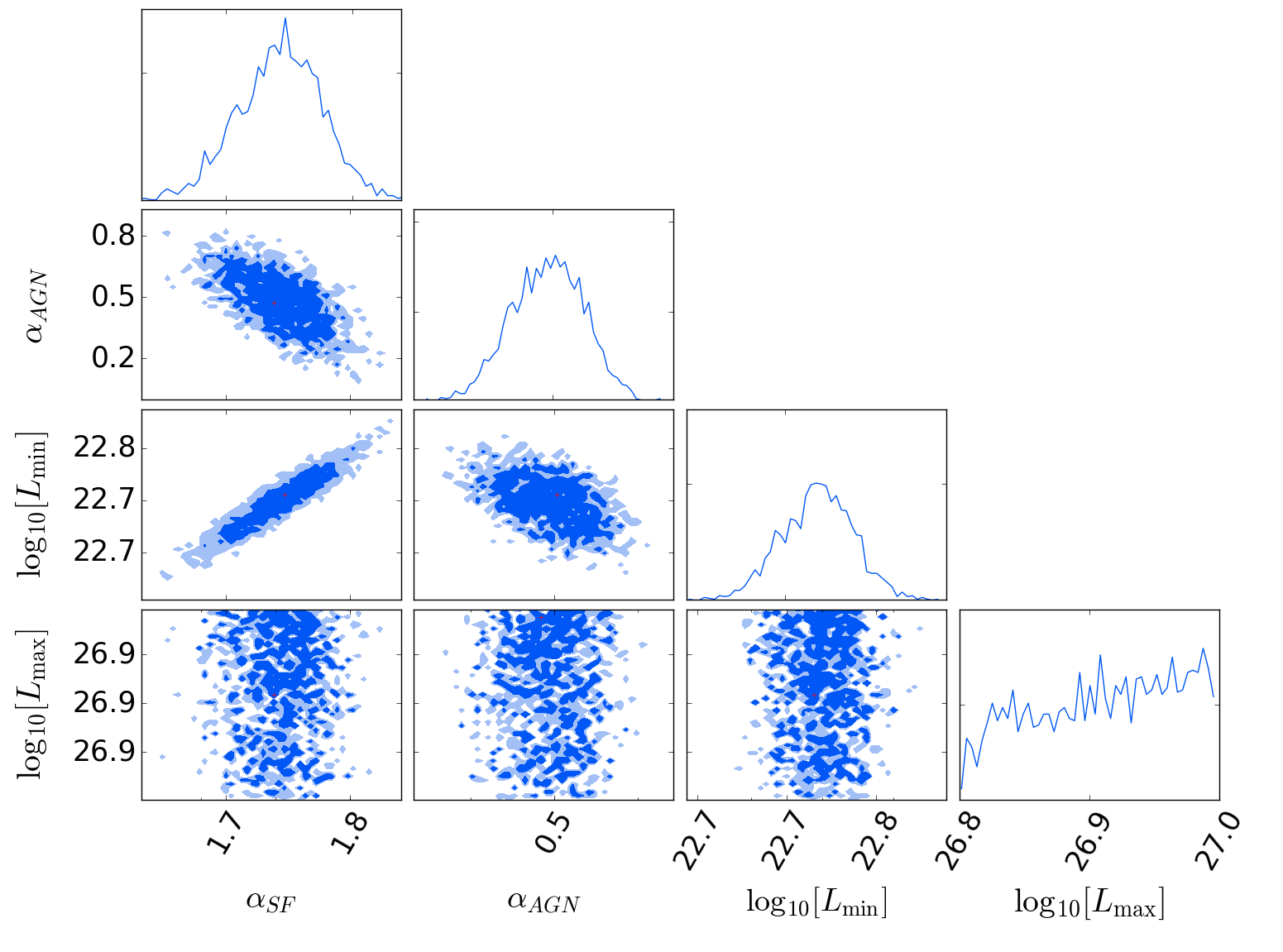}}
\subfloat[$3.2 <z< 4.0$]{\includegraphics[width=0.5\textwidth]{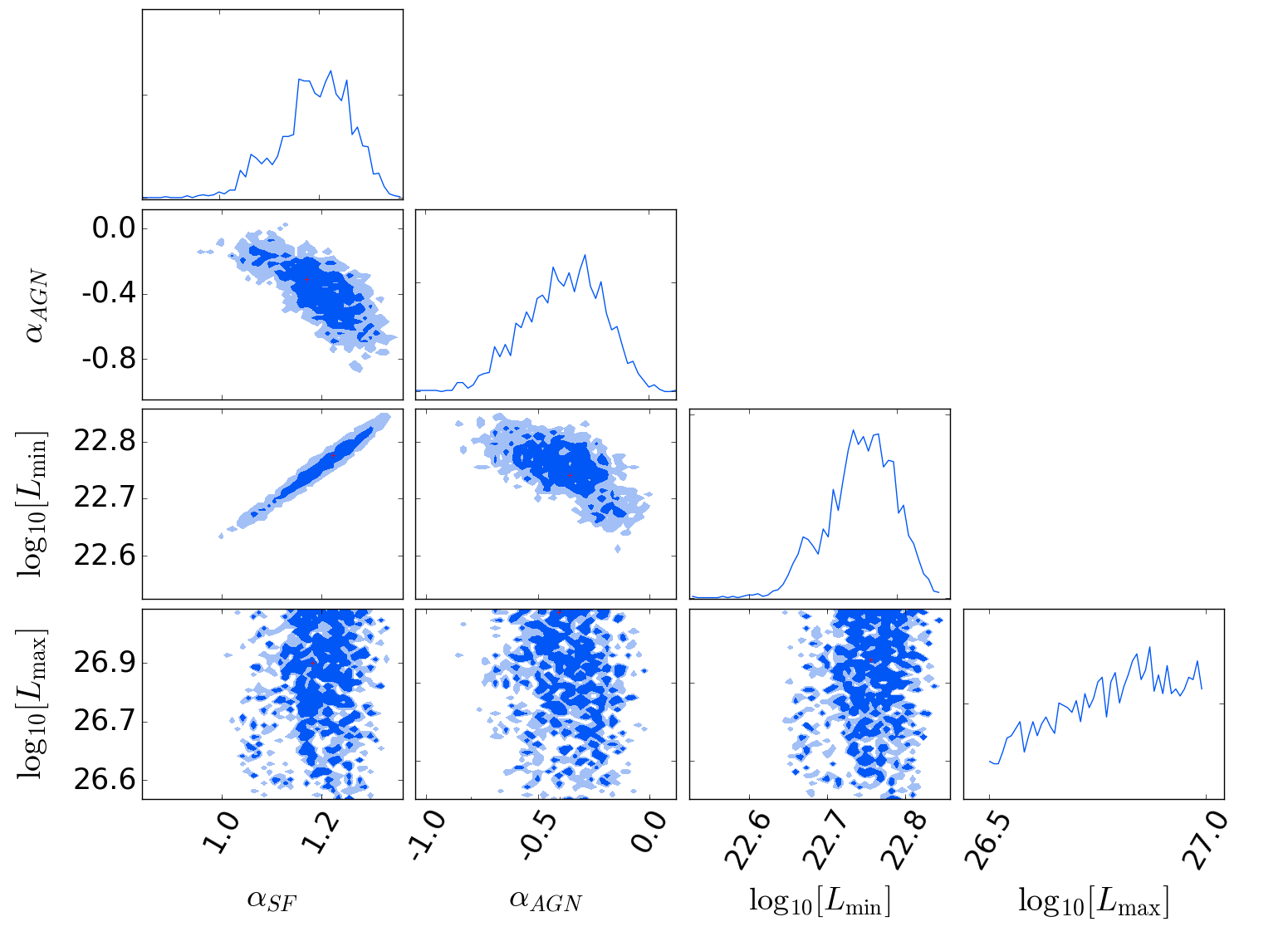}}
\caption{Continued.}
\label{fig:Model_C2}
\end{figure*}

\begin{tiny}
\bibliographystyle{mnras}
{\footnotesize 
\bibliography{literature}
}
\end{tiny}

\end{document}